\documentclass{lmcs}

\keywords{Linear Temporal Logic, Operator-Precedence Languages,
    Model Checking, First-Order Completeness, Visibly Pushdown Languages,
    Input-Driven Languages}

\usepackage[inline]{enumitem}
\usepackage{amsmath}
\usepackage{amssymb}
\usepackage{stmaryrd}
\usepackage{wasysym}
\usepackage{mathrsfs}
\usepackage{array}
\usepackage{wrapfig}

\usepackage{tikz}
\usetikzlibrary{arrows,automata}
\usetikzlibrary{matrix}
\usetikzlibrary{decorations.pathmorphing}
\usetikzlibrary{quotes}

\newcommand*{\lnext}{\ocircle}
\newcommand*{\lanext}{\ocircle_{\chi}}
\newcommand*{\lgnext}[1]{\ocircle^{#1}}
\newcommand*{\lganext}[1]{\chi_{F}^{#1}}
\newcommand*{\lback}{\circleddash}
\newcommand*{\laback}{\circleddash_{\chi}}
\newcommand*{\lgback}[1]{\circleddash^{#1}}
\newcommand*{\lgaback}[1]{\chi_{P}^{#1}}
\newcommand*{\lhynext}{\lgnext{\lessdot}_H}

\newcommand*{\lthrnext}{\mathit{CallThr}}

\newcommand*{\lguntil}[4]{#3 \mathbin{\mathcal{U}^{#1}_{#2}} #4}
\newcommand*{\luntil}[3]{#2 \mathbin{\mathcal{U}^{#1}} #3}
\newcommand*{\lluntil}[2]{\luntil{}{#1}{#2}}
\newcommand*{\lgsince}[4]{#3 \mathbin{\mathcal{S}^{#1}_{#2}} #4}
\newcommand*{\lsince}[3]{#2 \mathbin{\mathcal{S}^{#1}} #3}

\newcommand*{\lhyuntil}[2]{\lguntil{\lessdot}{H}{#1}{#2}}
\newcommand*{\lhysince}[2]{\lgsince{\lessdot}{H}{#1}{#2}}
\newcommand*{\lhtuntil}[2]{\lguntil{\gtrdot}{H}{#1}{#2}}
\newcommand*{\lhtsince}[2]{\lgsince{\gtrdot}{H}{#1}{#2}}
\newcommand*{\lcallsince}{\mathit{Scall}}
\newcommand*{\lfuntil}[3]{#2 \mathbin{\mathcal{U}(#1)} #3}

\newcommand*{\lpglob}[1]{\boxminus^{#1}}
\newcommand*{\llglob}{\square}
\newcommand*{\llpglob}{\boxminus}

\newcommand*{\lhupglob}{\lpglob{u}_H}

\newcommand*{\ldnext}{\lnext^d}
\newcommand*{\lunext}{\lnext^u}
\newcommand*{\ldback}{\lback^d}
\newcommand*{\luback}{\lback^u}
\newcommand*{\lcnext}[1]{\chi_F^{#1}}
\newcommand*{\lcdnext}{\lcnext{d}}
\newcommand*{\lcunext}{\lcnext{u}}
\newcommand*{\lcback}[1]{\chi_P^{#1}}
\newcommand*{\lcdback}{\lcback{d}}
\newcommand*{\lcuback}{\lcback{u}}
\newcommand*{\lcduntil}[2]{{#1} \mathbin{\mathcal{U}_\chi^d} {#2}}
\newcommand*{\lcdsince}[2]{{#1} \mathbin{\mathcal{S}_\chi^d} {#2}}
\newcommand*{\lcuuntil}[2]{{#1} \mathbin{\mathcal{U}_\chi^u} {#2}}
\newcommand*{\lcusince}[2]{{#1} \mathbin{\mathcal{S}_\chi^u} {#2}}
\newcommand*{\lhnext}[1]{\lnext_H^{#1}}
\newcommand*{\lhback}[1]{\lback_H^{#1}}
\newcommand*{\lhdnext}{\lhnext{d}}
\newcommand*{\lhdback}{\lhback{d}}
\newcommand*{\lhduntil}[2]{{#1} \mathbin{\mathcal{U}_H^d} {#2}}
\newcommand*{\lhdsince}[2]{{#1} \mathbin{\mathcal{S}_H^d} {#2}}
\newcommand*{\lhunext}{\lhnext{u}}
\newcommand*{\lhuback}{\lhback{u}}
\newcommand*{\lhuuntil}[2]{{#1} \mathbin{\mathcal{U}_H^u} {#2}}
\newcommand*{\lhusince}[2]{{#1} \mathbin{\mathcal{S}_H^u} {#2}}

\newcommand*{\chain}{\chi}

\newcommand*{\powset}[1]{{\mathcal{P}(#1)}}
\newcommand*{\lsucc}{\operatorname{succ}}
\newcommand*{\prf}{\pi}
\newcommand*{\pr}{\mathrel{\prf}}

\newcommand*{\clos}[1]{\operatorname{Cl}({#1})}

\newcommand*{\atoms}[1]{\operatorname{Atoms}({#1})}
\newcommand*{\fprop}[1]{\mathrm{q}_{(#1)}}

\newcommand*{\ra}{\operatorname{Rcc}}
\newcommand*{\invtau}{\tau^{-1}_{AP}}
\newcommand*{\first}{\operatorname{first}}
\newcommand*{\last}{\operatorname{last}}
\newcommand*{\rchild}{R_\Downarrow}
\newcommand*{\rsibl}{R_\Rightarrow}
\newcommand*{\rparen}{R_\Uparrow}
\newcommand*{\rlsibl}{R_\Leftarrow}
\newcommand*{\uotrees}{\mathcal{T}}

\newcommand*{\xuntil}{$\mathcal{X}_\mathit{until}$}

\newcommand*{\lcall}{\mathbf{call}}
\newcommand*{\lret}{\mathbf{ret}}
\newcommand*{\lhandle}{\mathbf{han}}
\newcommand*{\lthrow}{\mathbf{exc}}

\newcommand{\mrk}[1]{{#1}'}
\newcommand{\oldstack}[3]{%
{\ifthenelse{\equal{#1}{1}}{%
\mrk{#2}
}%
{#2}}_{#3}%
}
\newcommand{\stack}[3]{%
[%
{\ifthenelse{\equal{#1}{1}}{%
\mrk{#2}
}%
{#2}}
{\ifthenelse{\equal{#1}{0}}{\ }{} }
{#3}%
]%
}

\newcommand{\tstack}[2]{%
[#1,\ #2]%
}
\newcommand{\tconfig}[3]{\langle #1, \ #2, \ #3 \rangle}

\newcommand{\transition}[1]{\stackrel {{#1}} \vdash}

\newcommand{\va}[1]{\stackrel{#1}{\longrightarrow}}
\newcommand{\vshift}[1]{\stackrel{#1}{\dashrightarrow}}
\newcommand{\ourpath}[1]{\stackrel{#1}{\leadsto}}
\newcommand{\flush}[1]{\stackrel{#1}{\Longrightarrow}}

\newcommand{\ochain}[3]{{}^{#1}[ #2 ]{}^{#3}}

\newcommand{\config}[3]{\langle #1, \allowbreak #2, \allowbreak #3 \rangle}
\newcommand{\symb}[1]{\mathop{smb}(#1)}
\newcommand{\state}[1]{\mathop{st}(#1)}

\begin{document}

\title[POTL: A FO-complete Temporal Logic for OPL]
{POTL: A First-Order Complete Temporal Logic for Operator Precedence Languages}

\author[M. Chiari]{Michele Chiari}
\address{DEIB, Politecnico di Milano, Italy}
\email{michele.chiari@polimi.it}

\author[D. Mandrioli]{Dino Mandrioli}
\address{DEIB, Politecnico di Milano, Italy}
\email{dino.mandrioli@polimi.it}

\author[M. Pradella]{Matteo Pradella}
\address{DEIB, Politecnico di Milano, Italy and
IEIIT, Consiglio Nazionale delle Ricerche}
\email{matteo.pradella@polimi.it}

\begin{abstract}
  The problem of model checking procedural programs has fostered much research
  towards the definition of temporal logics for reasoning on context-free
  structures. The most notable of such results are temporal logics on Nested
  Words, such as CaRet and NWTL. Recently, the logic OPTL was introduced, based
  on the class of Operator Precedence Languages (OPL), more powerful than Nested
  Words. We define the new OPL-based logic POTL, prove its FO-completeness,
  and provide a model checking procedure for it. POTL improves on
  NWTL by enabling the formulation of requirements involving
  pre/post-conditions, stack inspection, and others in the presence of
  exception-like constructs. It improves on OPTL by being FO-complete, and by
  expressing more easily stack inspection and function-local properties.
\end{abstract}

\maketitle

\section{Introduction}
\label{sec:intro}

Model checking is one of the most successful techniques for the
verification of software programs.  It consists in the exhaustive
verification of the mathematical model of a program against a
specification of its desired behavior.  The kind of properties that
can be proved in this way depends both on the formalism employed to
model the program, and on the one used to express the specification.
The initial and most classical frameworks consist in the use of
operational formalisms, such as Transition Systems
and Finite State Automata (generally B\"uchi automata) for the model, and
temporal logics such as LTL, CTL and CTL* for the specification.  The
success of such logics is due to their ease in reasoning about linear
or branching sequences of events over time, by expressing liveness and
safety properties, their conciseness with respect to automata, and the
complexity of their model checking.

In this paper we consider linear-time temporal domains.
LTL limits its set of expressible properties to the
First-Order Logic (FOL) definable fragment of regular languages.  This
is quite restrictive when compared with the most popular abstract
models of procedural programs, such as Pushdown Systems, Boolean
Programs \cite{BallR00}, and Recursive State Machines~\cite{AlurBEGRY05}.  
All such stack-based formalisms show behaviors which are
expressible by means of Context-Free Languages (CFL), rather than regular ones.
State and configuration reachability, fair
computation problems, and model checking of \emph{regular specifications}
have been thoroughly studied for such formalisms
\cite{BouajjaniEM97,FinkelWW97,BurkartS99,EsparzaHRS00,KupfermanPV02a,PitermanV04,Walukiewicz2001,AlurBEGRY05,GodefroidY13,AlurBE18}.
To expand the expressive power of specification languages too, 
\cite{Bouajjani95,BouajjaniH96} augmented LTL with
Presburger arithmetic constraints on the occurrences of states,
obtaining a logic capable of even some context-sensitive
specifications, but with only restricted decidable fragments.
\cite{KupfermanPV02b} introduced model checking of pushdown tree automata
specifications on regular systems, and Dynamic Logic was extended
to some limited classes of CFL~\cite{HarelKT02}.
Decision procedures for different kinds of regular constraints on
stack contents have been given in~\cite{JensenLT99,EsparzaKS03,ChatterjeeMMZHP04}.

A coherent approach came with the introduction of temporal
logics based on Visibly Pushdown Languages (VPL) \cite{AluMad04},
a.k.a.\ Input-Driven Languages \cite{DBLP:conf/icalp/Mehlhorn80}.
Such logics, namely CaRet \cite{AlurEM04} and the FO-complete
NWTL \cite{lmcs/AlurABEIL08}, model the execution trace of a
procedural program as a Nested Word \cite{jacm/AlurM09}, consisting in
a linear ordering augmented with a one-to-one matching relation
between function calls and returns.  They are the first ones featuring
temporal modalities that explicitly refer to the nesting structure of
CFL~\cite{AlurBE18}.  This enables requirement
specifications to include Hoare-style pre/post-conditions,
stack-inspection properties, and more.  A $\mu$-calculus based on VPL
extends model checking to branching-time semantics in
\cite{DBLP:journals/toplas/AlurCM11}, while \cite{BozzelliS14}
introduces a temporal logic capturing the whole class of VPL.  Timed
extensions of CaRet are given in \cite{BozzelliMP18}.

VPL too have their limitations.  
They are more general than Parenthesis Languages~\cite{McNaughton67},
but their \emph{matching relation} is essentially constrained to be
one-to-one~\cite{MP18}.
This hinders their suitability to model processes in which a single event must be put in
relation with multiple ones.  Unfortunately, computer programs often
present such behaviors: exceptions, continuations, and
context-switches in real-time operating systems are single events that
cause the termination (or re-instantiation) of multiple functions on
the stack.
Colored Nested Words \cite{AlurF16} have been an early and partial attempt at modeling such behaviors.
To be able to reason about them, temporal logics based on Operator Precedence Languages
(OPL) were proposed.  OPL were initially introduced with the purpose of efficient
parsing \cite{Floyd1963}, a field in which they continue to offer
useful applications \cite{BarenghiEtAl2015}.  They are capable of
expressing arithmetic expressions, and other constructs whose
context-free structure is not immediately visible.  The
generality of the structure of their syntax trees is much
greater than that of VPL, which are strictly included in OPL
\cite{CrespiMandrioli12}.  Nevertheless, they retain the same closure
properties that make regular languages and VPL suitable for
automata-theoretic model checking: OPL are closed under
Boolean operations, concatenation, Kleene *, and language emptiness and inclusion are
decidable \cite{LonatiEtAl2015}.  Moreover, they have been
characterized by means of a Monadic Second-Order Logic.

OPTL~\cite{ChiariMP18} is the first linear-time temporal logic for which a model
checking procedure has been given on both finite and $\omega$-words of OPL. It
enables reasoning on procedural programs with exceptions, expressing
properties regarding the possibility of a function to be terminated by
an exception, or to throw one, and also pre/post-conditions.
NWTL can be translated into OPTL in linear time,
thus the latter is capable of expressing all properties of
CaRet and NWTL, and many more.
Unfortunately, we were not able to prove the FO-completeness of OPTL
due to some limitations of its semantics.
In OPTL it is difficult to navigate the syntax tree of a word,
and thus to express certain function-frame local properties.

One of the characterizing features of linear-time temporal logics is
their equivalence to FOL on their respective algebraic structure.
This was the reason for introducing NWTL, since it was not possible to
deduce the position of CaRet in this respect
\cite{lmcs/AlurABEIL08}.  This is also our motivation for presenting
Precedence Oriented Temporal Logic (POTL).  POTL redefines the
semantics of OPTL to be much closer to the ``essence'' of OPL, i.e.\
to the syntax tree structure of words.  In this paper, we prove the FO-completeness of
POTL over both finite and $\omega$ Operator Precedence (OP) words.
The greater theoretical expressive power
benefits POTL also in practice: it is easier to express stack
inspection properties in the presence of uncaught exceptions, as well
as function-frame local properties.
We conjecture some of such properties are not expressible at all in OPTL, although
proving the ``strict containment'' of OPTL in POTL seems to be
arduous, as is that of CaRet in NWTL.  Nevertheless, the
FO-completeness of POTL and the expressibility of OPTL in FOL allow us
to conclude that POTL is at least as expressive as OPTL.
We also give a tableaux-construction procedure for model checking POTL, which
yields nondeterministic automata of size at most singly exponential in formula length,
and is thus not asymptotically greater that that of LTL and NWTL.

The paper is organized as follows: Section~\ref{sec:opl} provides some background on OPL;
Section~\ref{sec:potl-syntax-semantics} presents the syntax and semantics of POTL,
also providing some qualitative demonstration of its expressive power;
Section~\ref{sec:fo-completeness} proves equivalence to FOL on finite words;
Section~\ref{sec:mc} provides a finite model checking procedure;
Appendices~\ref{subsec:expansion-proofs} and \ref{sec:xpath-proofs} contain some proofs
that would not fit into the main text.

\section{Operator Precedence Languages}
\label{sec:opl}

Operator Precedence Languages (OPL) are usually defined through their generating grammars
\cite{Floyd1963}; in this paper, however, we
characterize them through their accepting automata
\cite{LonatiEtAl2015} which are the natural way to state equivalence
properties with logic characterization. We assume some
familiarity with classical language theory concepts such as
context-free grammar, parsing, shift-reduce algorithm, syntax tree (ST) \cite{GruneJacobs:08}.
Readers not familiar with OPL may refer to \cite{MP18} for more informal
explanations on the following basic concepts;
an explanatory example is also given at the end of this section.

Let $\Sigma$ be a finite alphabet, and $\varepsilon$ the empty string.
We use a special symbol $\# \not\in \Sigma$ to mark the beginning and
the end of any string.
%
  An \textit{operator precedence matrix} (OPM) $M$ over $\Sigma$ is a partial function
  $(\Sigma \cup \{\#\})^2 \to \{\lessdot, \allowbreak \doteq, \allowbreak \gtrdot\}$,
  that, for each ordered pair $(a,b)$, defines the \emph{precedence relation} (PR) $M(a,b)$
  holding between $a$ and $b$. If the function is total we say that M is \emph{complete}.
  We call the pair $(\Sigma, M)$ an \emph{operator precedence alphabet}.
  Relations $\lessdot, \doteq, \gtrdot$, are respectively named
  \emph{yields precedence, equal in precedence}, and \emph{takes precedence}.
  By convention, the initial \# can only yield precedence, and other
  symbols can only take precedence on the ending \#.
  If $M(a,b) = \prf$, where $\prf \in \{\lessdot, \doteq, \gtrdot \}$,
  we write $a \pr b$.  For $u,v \in \Sigma^+$ we write $u \pr v$ if
  $u = xa$ and $v = by$ with $a \pr b$.
  The role of PR is to give structure to words:
  they can be seen as special and more concise parentheses, where
  e.g. one ``closing'' $\gtrdot$ can match more than one ``opening'' $\lessdot$.
  Despite their graphical appearance, PR are not ordering relations.
%


\begin{defi}\label{def:OPA}
An  \emph{operator precedence automaton (OPA)} is a tuple
$\mathcal A = (\Sigma, \allowbreak M, \allowbreak Q, \allowbreak I, \allowbreak F, \allowbreak \delta) $ where:
$(\Sigma, M)$ is an operator precedence alphabet,
$Q$ is a finite set of states (disjoint from $\Sigma$),
$I \subseteq Q$ is the set of initial states,
$F \subseteq Q$ is the set of final states,
$\delta \subseteq Q \times ( \Sigma \cup Q) \times Q$ is the transition relation,
which is the union of the three disjoint relations
$\delta_{\mathit{shift}}\subseteq Q \times \Sigma \times Q$,
$\delta_{\mathit{push}}\subseteq Q \times \Sigma \times Q$,
and
$\delta_{\mathit{pop}}\subseteq Q \times Q \times Q$.
An OPA is deterministic iff $I$ is a singleton,
and all three components of $\delta$ are --possibly partial-- functions.
\end{defi}

To define the semantics of OPA, we need some new notations.
Letters $p, q, p_i, \allowbreak q_i, \dots$ denote states in $Q$.
We sometimes use
$q_0 \va{a}{q_1}$ for $(q_0, a, q_1) \in \delta_{\mathit{push}}$,
$q_0 \vshift{a}{q_1}$ for $(q_0, a, q_1) \in \delta_{\mathit{shift}}$,
$q_0 \flush{q_2}{q_1}$  for $(q_0, q_2, q_1) \in  \delta_{\mathit{pop}}$,
and ${q_0} \ourpath{w} {q_1}$, if the automaton can read $w \in \Sigma^*$ going from $q_0$ to $q_1$.
Let $\Gamma$ be $\Sigma \times Q$ and let $\Gamma' = \Gamma \cup  \{\bot\}$
be the \textit{stack alphabet}; 
we denote symbols in $\Gamma'$ as $\tstack aq$ or $\bot$.
We set $\symb {\tstack aq} = a$, $\symb {\bot}=\#$, and
$\state {\tstack aq} = q$.
For a stack content $\gamma = \gamma_n \dots \gamma_1 \bot$,
with $\gamma_i \in \Gamma$ , $n \geq 0$, 
we set $\symb \gamma = \symb{\gamma_n}$ if $n \geq 1$, $\symb \gamma = \#$ if $n = 0$.

A \emph{configuration} of an OPA is a triple $c = \tconfig w q \gamma$,
where $w \in \Sigma^*\#$, $q \in Q$, and $\gamma \in \Gamma^*\bot$.
A \emph{computation} or \emph{run} is a finite sequence
$c_0 \transition{} c_1 \transition{} \dots \transition{} c_n$
of \emph{moves} or \emph{transitions}
$c_i \transition{} c_{i+1}$.
There are three kinds of moves, depending on the PR between the symbol
on top of the stack and the next input symbol:

\noindent {\bf push move:} if $\symb \gamma \lessdot \ a$ then
$
\tconfig {ax} p  \gamma \transition{} \tconfig {x} q {\tstack   a p \gamma }$,
with $(p,a, q) \in \delta_{\mathit{push}}$;

\noindent {\bf shift move:} if $a \doteq b$ then 
$
\tconfig {bx} q { \tstack a p \gamma}  \transition{} \tconfig x  r { \tstack b p \gamma}$,
with $(q,b,r) \in \delta_{\mathit{shift}}$;

\noindent {\bf pop move:} if $a \gtrdot b$
then 
$
\tconfig {bx} q  { \tstack a p \gamma}\transition{} \tconfig {bx} r \gamma $,
with $(q, p, r) \in \delta_{\mathit{pop}}$.

Shift and pop moves are not performed when the stack contains only $\bot$.
%
Push moves put a new element on top of the stack consisting of the input symbol together with the current state of the OPA.
Shift moves update the top element of the stack by \textit{changing its input symbol only}.
Pop moves remove the element on top of the stack,
and update the state of the OPA according to $\delta_{\mathit{pop}}$ on the basis of the current state of the OPA and the state of the removed stack symbol.
They do not consume the input symbol, which is used only to establish the $\gtrdot$ relation, remaining available for the next move.
%
The OPA accepts the language
$
L(\mathcal A) = \left\{ x \in \Sigma^* \mid  \tconfig {x\#} {q_I} {\bot} \vdash ^* 
\tconfig {\#} {q_F}{\bot} , \allowbreak q_I \in I, \allowbreak q_F \in F \right\}.
$

We now introduce the concept of {\em chain}, which makes the connection between OP relations and
context-free structure explicit, through brackets.
\begin{defi}\label{def:chain}
A \emph{simple chain}
$
\ochain {c_0} {c_1 c_2 \dots c_\ell} {c_{\ell+1}}
$
is a string $c_0 c_1 c_2 \dots c_\ell c_{\ell+1}$,
such that:
$c_0, \allowbreak c_{\ell+1} \in \Sigma \cup \{\#\}$,
$c_i \in \Sigma$ for every $i = 1,2, \dots \ell$ ($\ell \geq 1$),
and $c_0 \lessdot c_1 \doteq c_2 \dots c_{\ell-1} \doteq c_\ell
\gtrdot c_{\ell+1}$.
A \emph{composed chain} is a string 
$c_0 s_0 c_1 s_1 c_2  \dots c_\ell s_\ell c_{\ell+1}$, 
where
$\ochain {c_0}{c_1 c_2 \dots c_\ell}{c_{\ell+1}}$ is a simple chain, and
$s_i \in \Sigma^*$ is the empty string 
or is such that $\ochain {c_i} {s_i} {c_{i+1}}$ is a chain (simple or composed),
for every $i = 0,1, \dots, \ell$ ($\ell \geq 1$). 
Such a composed chain will be written as
$\ochain {c_0} {s_0 c_1 s_1 c_2 \dots c_\ell s_\ell} {c_{\ell+1}}$.
$c_0$ (resp.\ $c_{\ell+1}$) is called its \emph{left} (resp.\ \emph{right}) \emph{context}.
\end{defi}

\begin{figure}[tb]
\centering
\begin{tabular}{p{0.3\textwidth} p{0.6\textwidth}}
\begin{minipage}{0.3\textwidth}
\[
\begin{array}{r | c c c c}
         & \lcall   & \lret   & \lhandle & \lthrow \\
\hline
\lcall   & \lessdot & \doteq  & \lessdot & \gtrdot \\
\lret    & \gtrdot  & \gtrdot & \gtrdot & \gtrdot \\
\lhandle & \lessdot & \gtrdot & \lessdot & \doteq \\
\lthrow  & \gtrdot  & \gtrdot & \gtrdot  & \gtrdot \\
\end{array}
\]
\end{minipage}
&
\begin{minipage}{0.6\textwidth}
\[
\# [ \lcall [ [ [ \lhandle
[ \lcall [ \lcall [ \lcall ] ] ]
\lthrow ] \lcall \; \lret ] \lcall \; \lret ] \lret ] \#
\]
\end{minipage}
\end{tabular}
\caption{OPM $M_\lcall$ (left) and a string with chains shown by brackets (right).}
\label{fig:opm-mcall}
\end{figure}

  A finite word $w$ over $\Sigma$ is \emph{compatible} with an OPM $M$ iff
  for each pair of letters $c, d$, consecutive in $w$, $M(c, d)$ is defined and,
  for each substring $x$ of $\# w \#$ which is a chain of the form $^a[y]^b$,
  $M(a, b)$ is defined.
%
E.g., the word
of Fig.~\ref{fig:opm-mcall} is compatible with $M_\lcall$.
An easy way to identify chains is by noting that their bodies
are always enclosed by the $\lessdot$ and $\gtrdot$ relations,
i.e., $\ochain{a}{x}{b}$ iff $a \lessdot x \gtrdot b$.
In Fig.~\ref{fig:opm-mcall}, all the resulting chains are reported, e.g.
$\ochain {\lcall} {\lcall} {\lthrow}$,
$\ochain {\lcall} {\lhandle \, \lthrow} {\lcall}$
are simple chains, while
$\ochain {\lcall} {[ \lhandle [ \lcall [ \lcall [ \lcall ] ] ] \lthrow ] \lcall \; \lret} {\lcall}$,
$\ochain {\lcall} {\lcall [ \lcall ] }  {\lthrow}$
are composed chains.
In Fig.~\ref{fig:sttree-example} we show the syntax tree of this word,
which is isomorphic to the chain structure uniquely determined by the OPM.
Each chain corresponds to a non-terminal (a dot-node in the tree),
and the fringe of the subtree rooted at it is the chain's body.

Let $\mathcal A$ be an OPA.
We call a \emph{support} for the simple chain
$\ochain {c_0} {c_1 c_2 \dots c_\ell} {c_{\ell+1}}$
any path in $\mathcal A$ of the form
$q_0
\va{c_1}{q_1}
\vshift{}{}
\dots
\vshift{}q_{\ell-1}
\vshift{c_{\ell}}{q_\ell}
\flush{q_0} {q_{\ell+1}}$.
The label of the last (and only) pop is exactly $q_0$, i.e.\ the first state of the path;
this pop is executed because of relation $c_\ell \gtrdot c_{\ell+1}$.
We call a \emph{support for the composed chain}
$\ochain {c_0} {s_0 c_1 s_1 c_2 \dots c_\ell s_\ell} {c_{\ell+1}}$
any path in $\mathcal A$ of the form
\(
q_0
\ourpath{s_0}{q'_0}
\va{c_1}{q_1}
\ourpath{s_1}{q'_1}
\vshift{c_2}{}
\dots
\vshift{c_\ell} {q_\ell}
\ourpath{s_\ell}{q'_\ell}
\flush{q'_0}{q_{\ell+1}}
\)
where, for every $i = 0, 1, \dots, \ell$:
if $s_i \neq \epsilon$, then $q_i \ourpath{s_i}{q'_i} $
is a support for the chain $\ochain {c_i} {s_i} {c_{i+1}}$, else $q'_i = q_i$.


Chains fully determine the parsing structure of any
OPA over $(\Sigma, M)$. If the OPA performs the computation
$
\langle sb, q_i, [a, q_j] \gamma \rangle \vdash^*
\langle b,  q_k, \gamma \rangle
$,
then $\ochain asb$
is necessarily a chain over $(\Sigma, \allowbreak M)$, and there exists a support
like the one above with $s = s_0 c_1 \dots c_\ell s_\ell$ and $q_{\ell+1} = q_k$.
This corresponds to the parsing of the string $s_0 c_1 \dots c_\ell s_{\ell}$ within the
context $a$,$b$, which contains all
information needed to build the subtree whose frontier is that string.

Consider the OPA
$\mathcal A(\Sigma, M)$ $=$ $\langle \Sigma, M,$ $\{q\}, \{q\}, \{q\}, \delta_{max}
\rangle $ where  $\delta_{max}(q,q) = q$, and $\delta_{max}(q,c) = q$,
$\forall c \in \Sigma$.
We call it the \emph{OP Max-Automaton} over $\Sigma, M$.
For a max-automaton, each chain has a support.
Since there is a chain $\ochain{\#}{s}{\#}$ for any string $s$
compatible with $M$, a string is accepted by $\mathcal A(\Sigma, M)$
iff it is compatible with $M$.
If $M$ is complete, each string is
accepted by $A(\Sigma, M)$, which defines the
universal language $\Sigma^*$ by assigning to any string the (unique)
structure compatible with the OPM.
With $M_{\lcall}$ of Fig.~\ref{fig:opm-mcall},
if we take e.g. the string
$
\lret \
\lcall \
\lhandle
$, it is accepted by the max-automaton with structure
$
\#[[\lret]
\lcall
[\lhandle]]
\#.
$

In conclusion, given an OP alphabet, the OPM $M$ assigns a unique structure
to any compatible string in $\Sigma^*$;
unlike VPL, such a structure is 
not visible in the string, and must be built by means of a non-trivial parsing 
algorithm.  An OPA defined on the
OP alphabet selects an appropriate subset within the
``universe'' of strings compatible with $M$.
In some sense this property is yet another variation
of the fundamental Chomsky-Sh\"utzenberger theorem.
For a more complete description of the OPL family and of its relations with other CFL
we refer the reader to \cite{MP18}.

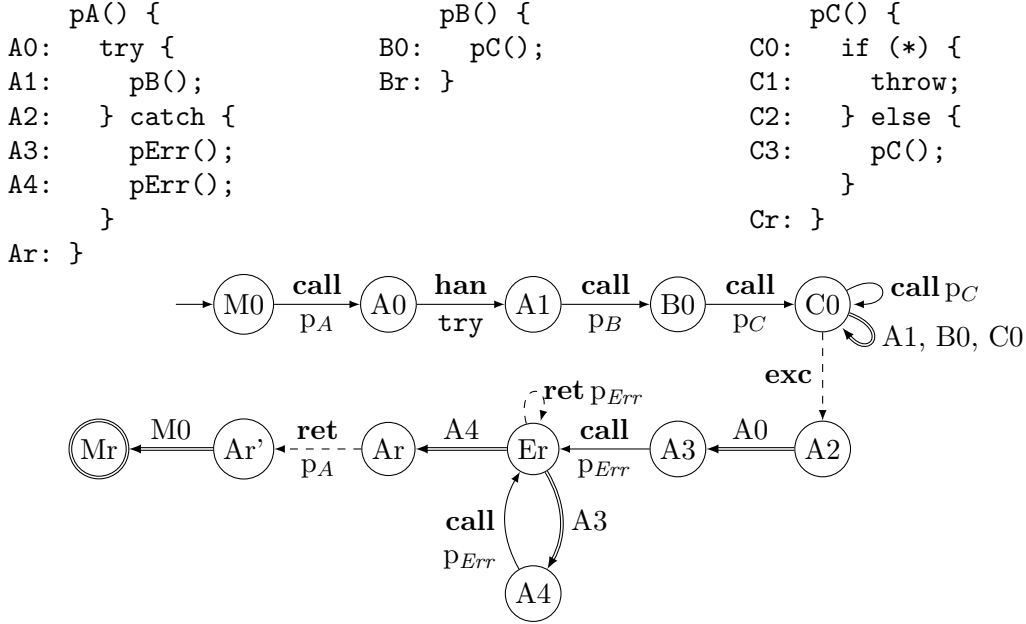
\begin{figure}[tb]
\centering
\begin{tabular}{p{0.3\textwidth} p{0.3\textwidth} p{0.3\textwidth}}
\begin{minipage}[t]{0.3\textwidth}
\begin{verbatim}
    pA() {
A0:   try {
A1:     pB();
A2:   } catch {
A3:     pErr();
A4:     pErr();
      }
Ar: }
\end{verbatim}
\end{minipage}
&
\begin{minipage}[t]{0.3\textwidth}
\begin{verbatim}
    pB() {
B0:   pC();
Br: }
\end{verbatim}
\end{minipage}
&
\begin{minipage}[t]{0.3\textwidth}
\begin{verbatim}
    pC() {
C0:   if (*) {
C1:     throw;
C2:   } else {
C3:     pC();
      }
Cr: }
\end{verbatim}
\end{minipage}
\\
\multicolumn{3}{c}{%
\begin{tikzpicture}
  [node distance=5em, every state/.style={minimum size=0pt, inner sep=2pt}, >=latex]
\node[state, initial by arrow, initial text=] (m0) {M0};
\node[state] (a0) [right of=m0] {A0};
\node[state] (a1) [right of=a0] {A1};
\node[state] (b0) [right of=a1] {B0};
\node[state] (c0) [right of=b0] {C0};
\node[state] (a2) [below of=c0] {A2};
\node[state] (a3) [left of=a2] {A3};
\node[state] (er) [left of=a3] {Er};
\node[state] (ar) [left of=er] {Ar};
\node[state] (arp) [left of=ar] {Ar'};
\node[state] (mr) [accepting, left of=arp] {Mr};
\node[state] (a4) [below of=er] {A4};
\path[->] (m0) edge[above] node[label=below:$\mathrm{p}_A$] {$\lcall$} (a0)
          (a0) edge[above] node[label=below:\texttt{try}] {$\lhandle$} (a1)
          (a1) edge[above] node[label=below:$\mathrm{p}_B$] {$\lcall$} (b0)
          (b0) edge[above] node[label=below:$\mathrm{p}_C$] {$\lcall$} (c0)
          (c0) edge[out=30, in=0, loop, right] node {$\lcall \, \mathrm{p}_C$} (c0)
          (c0) edge[out=-20, in=-50, loop, double, right] node {A1, B0, C0} (c0)
          (c0) edge[dashed, left] node {$\lthrow$} (a2)
          (a2) edge[double, above] node {A0} (a3)
          (a3) edge[above] node[label=below:$\mathrm{p}_{\mathit{Err}}$] {$\lcall$} (er)
          (er) edge[loop above, dashed, right] node {$\lret \, \mathrm{p}_{\mathit{Err}}$} (er)
          (er) edge[bend left, double, right] node {A3} (a4)
          (a4) edge[bend left, left] node[label=below:$\mathrm{p}_{\mathit{Err}}$] {$\lcall$} (er)
          (er) edge[double, above] node {A4} (ar)
          (ar) edge[dashed, above] node[label=below:$\mathrm{p}_{A}$] {$\lret$} (arp)
          (arp) edge[double, above] node {M0} (mr);
\end{tikzpicture}}
\end{tabular}
  \caption{Example procedural program (top) and the derived OPA (bottom).
    Push, shift, pop moves are shown by, resp., solid, dashed and double arrows.}
  \label{fig:example-prog}
\end{figure}

For readers not familiar with OPL, we show how OPA can naturally model
programming languages such as Java and C++.
Given a set $AP$ of
atomic propositions describing events and states of the program, we
use $(\powset{AP}, M_{AP})$ as the OP alphabet.  For
convenience, we consider a partitioning of $AP$ into a set of normal
propositional labels (in round font), and \emph{structural labels}
(SL, in bold).  SL define the OP structure of the word:
$M_{AP}$ is only defined for subsets of $AP$ containing exactly
one SL, so that given two SL $\mathbf{l}_1, \mathbf{l}_2$, for any $a,
a', b, b' \in \powset{AP}$ s.t.\ $\mathbf{l}_1 \in a, a'$ and
$\mathbf{l}_2 \in b, b'$ we have $M_{AP}(a,b) =
M_{AP}(a',b')$.  This way, it is possible to define an OPM on
the entire $\powset{AP}$ by only giving the relations between SL, as
we did for $M_\lcall$.
Fig.~\ref{fig:example-prog} shows how to model a procedural program with
 an OPA.  The OPA simulates the program's behavior with respect to the stack, by expressing its
execution traces with four event kinds: $\lcall$ (resp.\ $\lret$)
marks a procedure call (resp.\ return), $\lhandle$ the
installation of an exception handler by a \texttt{try} statement, and
$\lthrow$ an exception being raised. OPM $M_\lcall$ defines the
context-free structure of the word, which is strictly linked with the
programming language semantics: the $\lessdot$ PR causes
nesting (e.g., $\lcall$s can be nested into other $\lcall$s), and the
$\doteq$ PR implies a one-to-one relation, e.g.\ between a $\lcall$
and the $\lret$ of the same function, and a $\lhandle$ and the
$\lthrow$ it catches. Each OPA state represents a line in
the source code. First, procedure $\mathrm{p}_A$ is called by the
program loader (M0), and $[\{\lcall, \mathrm{p}_A\}, \text{M0}]$ is pushed onto the stack, to track the
program state before the $\lcall$. Then, the \texttt{try} statement at
line \texttt{A0} of $\mathrm{p}_A$ installs a handler. All subsequent
calls to $\mathrm{p}_B$ and $\mathrm{p}_C$ push new stack symbols on
top of the one pushed with $\lhandle$. $\mathrm{p}_C$ may only call
itself recursively, or throw an exception, but never return
normally. This is reflected by $\lthrow$ being the only transition
leading from state C0 to the accepting state Mr, and $\mathrm{p}_B$
and $\mathrm{p}_C$ having no way to a normal $\lret$. The OPA has a
look-ahead of one input symbol, so when it encounters $\lthrow$, it
must pop all symbols in the stack, corresponding to active function
frames, until it finds the one with $\lhandle$ in it, which cannot be
popped because $\lhandle \doteq \lthrow$. Notice that such behavior
cannot be modeled by Visibly Pushdown Automata or Nested Word
Automata, because they need to read an input symbol for each pop
move. Thus, $\lhandle$ protects the parent function from the
exception. Since the state contained in $\lhandle$'s stack symbol is
A0, the execution resumes in the \texttt{catch} clause of
$\mathrm{p}_A$. $\mathrm{p}_A$ then calls twice the library error-handling
function $\mathrm{p}_{\mathit{Err}}$, which ends regularly both times, and returns.
The string of Fig.~\ref{fig:opm-mcall} is accepted by this OPA.

In this example, we only model stack behavior for simplicity, but
other statements, such as assignments, and other behaviors, such as
continuations, could be modeled by a different choice of the OPM, and
other aspects of the program's state by appropriate
abstractions~\cite{JhalaPR18}.

\section{POTL: Syntax and Semantics}
\label{sec:potl-syntax-semantics}
\begin{figure}[tb]
\begin{tikzpicture}
  [edge/.style={}]
\matrix (m) [matrix of math nodes, column sep=-4, row sep=-4]
{
  \#
  & \lessdot & \lcall
  & \lessdot & \lhandle
  & \lessdot & \lcall
  & \lessdot & \lcall
  & \lessdot & \lcall
  & \gtrdot & \lthrow
  & \gtrdot & \lcall
  & \doteq & \lret
  & \gtrdot & \lcall
  & \doteq & \lret
  & \gtrdot & \lret
  & \gtrdot & \# \\
  & & \mathrm{p}_A
  & &
  & & \mathrm{p}_B
  & & \mathrm{p}_C
  & & \mathrm{p}_C
  & &
  & & \mathrm{p}_{\mathit{Err}}
  & & \mathrm{p}_{\mathit{Err}}
  & & \mathrm{p}_{\mathit{Err}}
  & & \mathrm{p}_{\mathit{Err}}
  & & \mathrm{p}_A
  & & \\
  0
  & & 1
  & & 2
  & & 3
  & & 4
  & & 5
  & & 6
  & & 7
  & & 8
  & & 9
  & & 10
  & & 11
  & & 12 \\
};
\draw[edge] (m-1-1) to [out=20, in=160] (m-1-25);
\draw[edge] (m-1-3) to [out=20, in=160] (m-1-23);
\draw[edge] (m-1-3) to [out=20, in=160] (m-1-15);
\draw[edge] (m-1-3) to [out=20, in=160] (m-1-19);
\draw[edge] (m-1-5) to [out=20, in=160] (m-1-13);
\draw[edge] (m-1-7) to [out=20, in=160] (m-1-13);
\draw[edge] (m-1-9) to [out=20, in=160] (m-1-13);
\end{tikzpicture}
\caption{The example string of Fig.~\ref{fig:opm-mcall} as an OP word.
  Chains are highlighted by arrows joining their contexts;
  structural labels are in bold,
  and other atomic propositions are shown below them.
  $\mathrm{p}_l$ means a $\lcall$ or a $\lret$ is related to procedure $\mathrm{p}_l$.
  First, procedure $\mathrm{p}_A$ is called (pos.~1),
  and it installs an exception handler in pos.~2.
  Then, three nested procedures are called,
  and the innermost one ($\mathrm{p}_C$) throws an exception,
  which is caught by the handler.
  Two more functions are called and, finally, $\mathrm{p}_A$ returns.}
\label{fig:potl-example-word}
\end{figure}
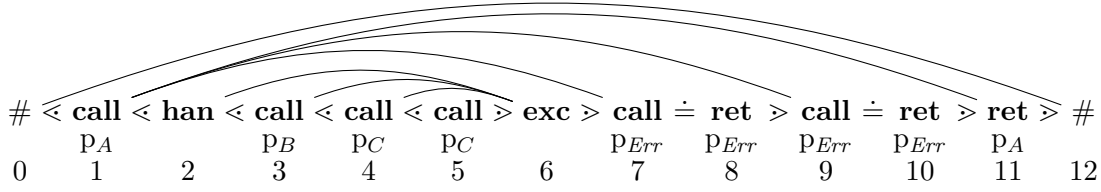

Given a finite set of atomic propositions $AP$, the syntax of POTL follows:
\begin{align*}
\varphi ::= &\; \mathrm{a}
\mid \neg \varphi
\mid \varphi \lor \varphi
\mid \lnext^t \varphi
\mid \lback^t \varphi
\mid \lcnext{t} \varphi
\mid \lcback{t} \varphi
\mid \lguntil{t}{\chi}{\varphi}{\varphi}
\mid \lgsince{t}{\chi}{\varphi}{\varphi} \\
&\mid \lhnext{t} \varphi
\mid \lhback{t} \varphi
\mid \lguntil{t}{H}{\varphi}{\varphi}
\mid \lgsince{t}{H}{\varphi}{\varphi}
\end{align*}
where $\mathrm{a} \in AP$, and $t \in \{d, u\}$.

The semantics of POTL is based on the \emph{word structure}
--also called \emph{OP word} for short--
$\langle U, <, M_{AP}, P \rangle$, where
  $U = \{0, 1, \dots, n, n+1\}$, with $n \in \mathbb{N}$ is a set of word positions;
  $<$ is a linear order on $U$;
  $M_{AP}$ is an operator precedence matrix on $\powset{AP}$;
  $P \colon AP \to \powset{U}$ is a function associating each atomic proposition
  with the set of positions in which it holds, with $0, (n+1) \in P(\#)$.
Given two positions $i, j$ and a PR $\prf$,
we write $i \pr j$ to say $a \pr b$,
where $a = \{\mathrm{p} \mid i \in P(\mathrm{p})\}$, and
$b = \{\mathrm{p} \mid j \in P(\mathrm{p})\}$.

We define the chain relation $\chain \subseteq U \times U$
so that $\chain(i, j)$ holds between two positions $i,j$ iff $i < j-1$,
and $i$ and $j$ are resp.\ the left and right contexts of the same chain.
For composed chains, $\chain$ may not be one-to-one,
but also one-to-many or many-to-one.
Given $i,j \in U$, relation $\chain$ has the following properties:
\begin{enumerate}
\item It never crosses itself: if $\chain(i,j)$ and $\chain(h,k)$, for any $h,k \in U$,
then we have $i < h < j \implies k \leq j$
and $i < k < j \implies i \leq h$.
\item If $\chain(i,j)$, then $i \lessdot i+1$ and $j-1 \gtrdot j$.
\item \label{item:downward-prop}
  There exists at most one single position $h$, called \emph{leftmost context} of $j$,
  s.t.\ $\chain(h,j)$ and $h \lessdot j$ or $h \doteq j$;
  for any $k$ s.t.\ $\chain(k,j)$ and $k \gtrdot j$ we have $k > h$.
\item \label{item:upward-prop}
  There exists at most one single position $h$, called \emph{rightmost context} of $i$,
  s.t.\ $\chain(i,h)$ and $i \gtrdot h$ or $i \doteq h$;
  for any $k$ s.t.\ $\chain(i,k)$ and $i \lessdot k$ we have $k < h$.
\end{enumerate}

\begin{wrapfigure}{R}{0pt}
\includegraphics{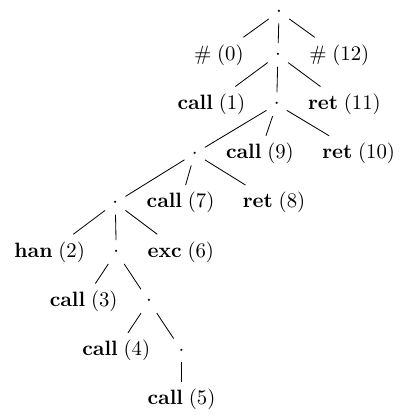}
\caption{The ST corresponding to the word of Fig.~\ref{fig:potl-example-word}.
  Dots represent non-terminals.}
\label{fig:sttree-example}
\end{wrapfigure}
Property \ref{item:upward-prop} says that when the chain relation is one-to-many,
the contexts of the outermost chains are in the $\doteq$ or $\gtrdot$ relation,
while the inner ones are in the $\lessdot$ relation.
Property \ref{item:downward-prop} says that contexts of outermost
many-to-one chains are in the $\doteq$  or $\lessdot$ relation,
the inner ones being in the $\gtrdot$ relation.
In the ST, the right context $j$ of a chain is at the \emph{same level}
as the left one $i$ when $i \doteq j$ (e.g., in Fig.~\ref{fig:sttree-example}, pos.\ 1 and 11),
at a \emph{lower level} when $i \lessdot j$ (e.g., pos.\ 1 with 7, and 9),
at a \emph{higher level} if $i \gtrdot j$ (e.g., pos.\ 3 and 4 with 6).

The truth of POTL formulas is defined w.r.t.\ a single word position.
Let $w$ be an OP word, and $\mathrm{a} \in AP$.
Then, for any position $i \in U$ of $w$,
we have $(w, i) \models \mathrm{a}$ if $i \in P(\mathrm{a})$.
Operators such as $\land$ and $\neg$ have the usual semantics from propositional logic.
Next, while giving the formal semantics of POTL operators, we illustrate it by showing
how it can be used to express properties on program execution traces,
such as the one of Fig.~\ref{fig:potl-example-word}.

\noindent \textbf{Next/back operators.}
The \emph{downward} next and back operators $\ldnext$ and $\ldback$
are like their LTL counterparts, except they are true only if the next
(resp.\ current) position is at a lower or equal ST level than the current (resp.\ preceding) one.
The \emph{upward} next and back, $\lunext$ and $\luback$, are symmetric.
Formally, $(w,i) \models \ldnext \varphi$ iff $(w,i+1) \models \varphi$
and $i \lessdot (i+1)$ or $i \doteq (i+1)$,
and $(w,i) \models \ldback \varphi$ iff $(w,i-1) \models \varphi$,
and $(i-1) \lessdot i$ or $(i-1) \doteq i$.
Substitute $\lessdot$ with $\gtrdot$ to obtain the semantics for $\lunext$ and $\luback$.
E.g., we can write $\ldnext \lcall$ to say that the next position is
an inner call (it holds in pos.\ 2, 3, 4 of
Fig.~\ref{fig:potl-example-word}), $\ldback \lcall$ to say that the
previous position is a $\lcall$, and the
current is the first of the body of a function (pos.\ 2,
4, 5), or the $\lret$ of an empty one (pos.\ 8, 10), and $\luback
\lcall$ to say that the current position terminates an empty function
frame (holds in 6, 8, 10).  In pos.\ 2 $\ldnext \mathrm{p}_B$ holds, but
$\lunext \mathrm{p}_B$ does not.

The \emph{chain} next and back operators $\lcnext{t}$ and $\lcback{t}$
evaluate their argument respectively on future and past positions
in the chain relation with the current one.
The \emph{downward} (resp.\ \emph{upward}) variant only considers chains
whose right context goes down (resp.\ up) or remains at the same level in the ST.
E.g., in pos.\ 1 of Fig.~\ref{fig:potl-example-word}, $\lcdnext \mathrm{p}_{\mathit{Err}}$ holds
because $\chain(1,7)$ and $\chain(1,9)$,
meaning that $\mathrm{p}_A$ calls $\mathrm{p}_{\mathit{Err}}$ at least once.
Formally, $(w,i) \models \lcdnext \varphi$
iff there exists a position $j > i$ such that $\chain(i,j)$,
$i \lessdot j$ or $i \doteq j$, and $(w,j) \models \varphi$.
$(w,i) \models \lcdback \varphi$ iff there exists a position $j < i$
such that $\chain(j,i)$, $j \lessdot i$ or $j \doteq i$, and $(w,j) \models \varphi$.
Replace $\lessdot$ with $\gtrdot$ for the upward versions.
In Fig.~\ref{fig:potl-example-word}, $\lcunext \lthrow$ is true in
$\lcall$ positions whose procedure is terminated by an exception thrown by
an inner procedure (e.g.\ pos.\ 3 and 4).
$\lcuback \lcall$ is true in $\lthrow$ statements that terminate at least one procedure
other than the one raising it, such as the one in pos.\ 6.
$\lcdnext \lret$ and $\lcunext \lret$ hold in $\lcall$s
to non-empty procedures that terminate normally, and not due to an uncaught exception
(e.g., pos.\ 1).

\noindent \textbf{Until/Since operators.}
POTL has two kinds of until and since operators.
They express properties on paths, which are sequences of positions
obtained by iterating the different kinds of next or back operators.
In general, a \emph{path} of length $n \in \mathbb{N}$ between
$i, j \in U$ is a sequence of positions $i = i_1 < i_2 < \dots < i_n = j$.
The \emph{until} operator on a set of paths $\Gamma$ is defined as follows:
for any word $w$ and position $i \in U$,
and for any two POTL formulas $\varphi$ and $\psi$,
$(w, i) \models \lfuntil{\Gamma}{\varphi}{\psi}$
iff there exist a position $j \in U$, $j \geq i$,
and a path $i_1 < i_2 < \dots < i_n$ between $i$ and $j$ in
$\Gamma$
such that $(w, i_k) \models \varphi$ for any $1 \leq k < n$, and $(w, i_n) \models \psi$.
\emph{Since} operators are defined symmetrically.
Note that, depending on $\Gamma$, a path from $i$ to $j$ may not exist.
We define until/since operators by associating them with different sets of paths.

The \emph{summary} until $\lguntil{t}{\chi}{\psi}{\theta}$
(resp.\ since $\lgsince{t}{\chi}{\psi}{\theta}$) operator is obtained by inductively applying
the $\lnext^t$ and $\lcnext{t}$ (resp.\ $\lback^t$ and $\lcback{t}$) operators.
It holds in a position in which either $\theta$ holds,
or $\psi$ holds together with $\lnext^t (\lguntil{t}{\chi}{\psi}{\theta})$
(resp.\ $\lback^t (\lgsince{t}{\chi}{\psi}{\theta})$)
or $\lcnext{t} (\lguntil{t}{\chi}{\psi}{\theta})$
(resp.\ $\lcback{t} (\lgsince{t}{\chi}{\psi}{\theta})$).
It is an until operator on paths that can move not only between consecutive positions,
but also between contexts of a chain, skipping its body.
With the OPM of Fig.~\ref{fig:opm-mcall}, this means skipping function bodies.
The downward variants can move between positions at the same level in the ST
(i.e., in the same simple chain body), or down in the nested chain structure.
The upward ones remain at the same level, or move to higher levels of the ST.

Formula $\lcuuntil{\top}{\lthrow}$ is true in positions contained in the frame
of a function that is terminated by an exception.
It is true in pos.\ 3 of Fig.~\ref{fig:potl-example-word} because of path 3-6,
and false in pos.\ 1, because no path can enter the chain whose contexts are pos.\ 1 and 11.
Formula $\lcduntil{\top}{\lthrow}$ is true in call positions whose function frame
contains $\lthrow$s, but that are not directly terminated by one of them,
such as the one in pos.\ 1 (with path 1-2-6).

We define \emph{Downward Summary Paths} (DSP) as follows.
Given an OP word $w$, and two positions $i \leq j$ in $w$,
the DSP between $i$ and $j$, if it exists,
is a sequence of positions $i = i_1 < i_2 < \dots < i_n = j$ such that, for each $1 \leq p < n$,
\[
i_{p+1} =
\begin{cases}
  k & \text{iff $k = \max\{ h \mid h \leq j \land \chain(i_p,h) \land (i \lessdot k \lor i \doteq k)\}$;} \\
  i_p + 1 & \text{if $i_p \lessdot (i_p + 1)$ or $i_p \doteq (i_p + 1)$.}
\end{cases}
\]
The Downward Summary (DS) until and since operators $\lcduntil{}{}$
and $\lcdsince{}{}$ use as $\Gamma$ the set of DSP starting in the position in which
they are evaluated.
The definition for the upward counterparts is, again,
obtained by substituting $\lessdot$ with $\gtrdot$.
In Fig.~\ref{fig:potl-example-word},
$\lcduntil{\lcall}{(\lret \land \mathrm{p}_{\mathit{Err}})}$ holds in pos.~1 because of path 1-7-8 and 1-9-10,
$\lcusince{(\lcall \lor \lthrow)}{\mathrm{p}_B}$ in pos.~7 because of path 3-6-7,
and $\lcuuntil{(\lcall \lor \lthrow)}{\lret}$ in 3
because of path 3-6-7-8.

\noindent \textbf{Hierarchical operators.}
A single position may be the left or right context of multiple chains.
The operators seen so far cannot keep this fact into account,
since they ``forget'' about a left context when they jump to the right one.
Thus, we introduce the \emph{hierarchical} next and back operators.
The \emph{upward} hierarchical next (resp.\ back),
$\lhunext \psi$ (resp.\ $\lhuback \psi$), is true iff the current position $j$
is the right context of a chain whose left context is $i$,
and $\psi$ holds in the next (resp.\ previous) pos.\ $j'$ that is the right context of $i$,
with $i \lessdot j, j'$.
So, $\lhunext \mathrm{p}_{\mathit{Err}}$ holds in pos.\ 7 of Fig.~\ref{fig:potl-example-word}
because $\mathrm{p}_{\mathit{Err}}$ holds in 9,
and $\lhuback \mathrm{p}_{\mathit{Err}}$ in 9 because $\mathrm{p}_{\mathit{Err}}$ holds in 7.
In the ST, $\lhunext$ goes \emph{u}p between $\lcall$s to $\mathrm{p}_{\mathit{Err}}$,
while $\lhuback$ goes down.
Their \emph{downward} counterparts behave symmetrically,
and consider multiple inner chains sharing their right context.
They are formally defined as:
\begin{itemize}
\item $(w,i) \models \lhunext \varphi$ iff
  there exist a position $h < i$ s.t.\ $\chain(h,i)$ and $h \lessdot i$
  and a position $j = \min\{ k \mid i < k \land \chain(h,k) \land h \lessdot k \}$
  and $(w,j) \models \varphi$;
\item $(w,i) \models \lhuback \varphi$ iff
  there exist a position $h < i$ s.t.\ $\chain(h,i)$ and $h \lessdot i$
  and a position $j = \max\{ k \mid k < i \land \chain(h,k) \land h \lessdot k \}$
  and $(w,j) \models \varphi$;
\item $(w,i) \models \lhdnext \varphi$ iff
  there exist a position $h > i$ s.t.\ $\chain(i,h)$ and $i \gtrdot h$
  and a position $j = \min\{ k \mid i < k \land \chain(k,h) \land k \gtrdot h \}$
  and $(w,j) \models \varphi$;
\item $(w,i) \models \lhdback \varphi$ iff
  there exist a position $h > i$ s.t.\ $\chain(i,h)$ and $i \gtrdot h$
  and a position $j = \max\{ k \mid k < i \land \chain(k,h) \land k \gtrdot h \}$
  and $(w,j) \models \varphi$.
\end{itemize}
In the ST of Fig.~\ref{fig:sttree-example}, $\lhdnext$ and $\lhdback$ go \emph{down} and up among
$\lcall$s terminated by the same $\lthrow$.
For example, in pos.~3 $\lhdnext \mathrm{p}_C$ holds,
because both pos.~3 and 4 are in the chain relation with 6.
Similarly, in pos.~4 $\lhdback \mathrm{p}_B$ holds.
Note that these operators do not consider leftmost/rightmost contexts,
so $\lhunext \lret$ is false in pos.\ 9, as $\lcall \doteq \lret$,
and pos.\ 11 is the rightmost context of pos.\ 1.

The hierarchical until and since operators are defined by iterating these next and back operators.
The upward hierarchical path (UHP) between $i$ and $j$ is a sequence of positions
$i = i_1 < i_2 < \dots < i_n = j$ such that there exists a position $h < i$ such that
for each $1 \leq p \leq n$ we have $\chain(h,i_p)$ and $h \lessdot i_p$,
and for each $i \leq q < n$ there exists no position $k$
such that $i_q < k < i_{q+1}$ and $\chain(h,k)$.
The until and since operators based on the set of UHP
starting in the position in which they are evaluated are denoted
as $\lhuuntil{}{}$ and $\lhusince{}{}$.
E.g., $\lhuuntil{\lcall}{\mathrm{p}_{\mathit{Err}}}$ holds in pos.~7
because of the singleton path 7 and path 7-9,
and $\lhusince{\lcall}{\mathrm{p}_{\mathit{Err}}}$ in pos.~9
because of paths 9 and 7-9.

The downward hierarchical path (DHP) between $i$ and $j$ is a sequence of positions
$i = i_1 < i_2 < \dots < i_n = j$ such that there exists a position $h > j$ such that
for each $1 \leq p \leq n$ we have $\chain(i_p,h)$ and $i_p \gtrdot h$,
and for each $1 \leq q < n$ there exists no position $k$
such that $i_q < k < i_{q+1}$ and $\chain(k,h)$.
The until and since operators based on the set of DHP
starting in the position in which they are evaluated are denoted
as $\lhduntil{}{}$ and $\lhdsince{}{}$.
In Fig.~\ref{fig:potl-example-word}, $\lhduntil{\lcall}{\mathrm{p}_C}$ holds in pos.~3,
and $\lhdsince{\lcall}{\mathrm{p}_B}$ in pos.~4, both because of path 3-4.

The POTL until and since operators enjoy expansion laws similar to those of LTL.
Here we give those for two until operators,
those for their since and downward counterparts being symmetric.
All such laws are proved in Appendix~\ref{subsec:expansion-proofs}.
\begin{align*}
  \lguntil{t}{\chi}{\varphi}{\psi} &\equiv
    \psi \lor \Big(\varphi \land \big(\lnext^t (\lguntil{t}{\chi}{\varphi}{\psi})
      \lor \lcnext{t} (\lguntil{t}{\chi}{\varphi}{\psi})\big)\Big) \\
  \lhuuntil{\varphi}{\psi} &\equiv
    (\psi \land \lcdback \top \land \neg \lcuback \top) \lor
     \big(\varphi \land \lhunext (\lhuuntil{\varphi}{\psi})\big)
\end{align*}

\subsection{Motivating Examples}

In Corollary~\ref{cor:optl-in-potl}, we show that OPTL $\subseteq$ POTL,
and CaRet \cite{AlurEM04} $\subseteq$ NWTL \cite{lmcs/AlurABEIL08} $\subset$ POTL.
More importantly, POTL can express many useful requirements of procedural programs. To emphasize the potential practical applications in automatic verification, we supply a few examples of typical program properties expressed as POTL formulas, not all of them being expressible in the other above languages.

Let $\llglob \psi := \neg (\lcuuntil{\top}{(\lcduntil{\top}{\neg \psi})})$
be the LTL \emph{globally} operator.
POTL can express Hoare-style pre/postconditions
with formulas such as $\llglob (\lcall \land \rho \implies \lcdnext (\lret \land \theta))$,
where $\rho$ is the precondition, and $\theta$ is the postcondition.

Unlike NWTL, POTL can easily express properties related to exception handling
and interrupt management \cite{MP18}.
E.g., the shortcut
$\lthrnext(\psi) := \lunext (\lthrow \land \psi) \lor \lcunext (\lthrow \land \psi)$,
evaluated in a $\lcall$, states that the procedure currently started
is terminated by an $\lthrow$ in which $\psi$ holds.
So, $\llglob (\lcall \land \rho \land \lthrnext(\top) \implies \lthrnext(\theta))$
means that if precondition $\rho$ holds when a procedure is called,
then postcondition $\theta$ must hold if that procedure is terminated by an exception.
In object oriented programming languages,
if $\rho \equiv \theta$ is a class invariant asserting that a class instance's state is valid,
this formula expresses \emph{weak exception safety} \cite{Abrahams00},
and \emph{strong exception safety} if $\rho$ and $\theta$
express particular states of the class instance.
The \emph{no-throw guarantee} can be stated with
$\llglob (\lcall \land \mathrm{p}_A \implies \neg \lthrnext(\top))$,
meaning procedure $\mathrm{p}_A$ is never interrupted by an exception.

\emph{Stack inspection} \cite{EsparzaKS03,JensenLT99},
i.e.\ properties regarding the sequence of procedures
active in the program's stack at a certain point of its execution,
is an important class of requirements that can be expressed with shortcut
$\lcallsince(\varphi, \psi) := \lcdsince{(\lcall \implies \varphi)}{(\lcall \land \psi)}$,
which subsumes the \emph{call since} of CaRet, as it also works with exceptions.
E.g., $\llglob \big((\lcall \land \mathrm{p}_B \land
    \lcallsince(\top, \mathrm{p}_A))
    \implies \lcduntil{\top}
    {\lthrnext(\top)}\big)$
means that whenever $\mathrm{p}_B$ is executed and
at least one instance of $\mathrm{p}_A$ is on the stack,
$\mathrm{p}_B$ or a subfunction thereof throw an exception.
The OPA of Fig.~\ref{fig:example-prog} satisfies this formula,
because $\mathrm{p}_B$ is always called by $\mathrm{p}_A$,
and $\mathrm{p}_C$ always throws.



\subsection{Comparison with the state of the art}

\subsubsection{Logics on Nested Words}
The first temporal logics with explicit context-free aware modalities
were based on Nested Words \cite{jacm/AlurM09}.
A nested word is a tuple $\langle U, P, <, \allowbreak \mu, \allowbreak \mathtt{call}, \allowbreak \mathtt{ret} \rangle$,
where $U$ is a set of word positions, $P \colon AP \to \powset{U}$ is a labeling function,
$<$ is a linear order on $U$, and $\mu$ is a binary relation
and $\mathtt{call}$, $\mathtt{ret}$ are two unary relations on $U$.
$\mu$ is a one-to-one nesting relation which never crosses itself.
For any $i,j \in U$, if $\mu(i,j)$ then $i \in \mathtt{call}$ is a \emph{call}
and $j \in \mathtt{ret}$ is a \emph{return}.
Call and return positions model function calls and returns,
while other positions, called \emph{internal}, model all other program operations.
The main limitation of the $\mu$ with respect to the $\chain$ relation is
its being strictly one-to-one, and the fact that a position
cannot be both a call and a return.
When seen as context-free languages, nested words generate syntax trees
where each right-hand-side (rhs) starts with a call, and ends with a return.

CaRet was the first temporal logic on nested words to be introduced,
and it focuses on expressing properties on procedural programs,
which explains its choice of modalities.
The \emph{abstract} next and until operators are defined on paths of positions
in the frame of the same function, skipping frames of nested calls.
The \emph{caller} next and until are actually past modalities,
and they operate on paths made of the calls of function frames
containing the current position.
LTL Next and Until are also present.
The caller operators enable upward movement in the ST of a nested word,
and abstract operators enable movement in the same rhs.
However, no CaRet operator allows pure downward movement in the ST,
which is needed to express properties limited to a single subtree.
While the LTL until can go downward, it can also go past the rightmost leaf of a subtree,
thus effectively jumping upwards.

This seems to be the main expressive limitation of CaRet,
which is conjectured not to be FO-complete \cite{lmcs/AlurABEIL08}.
In fact, FO-complete temporal logics were introduced in \cite{lmcs/AlurABEIL08}
by adding various kinds of \emph{within} modalities to CaRet.
Such operators limit their operands to span only positions within the same
call-return pair, and hence the same subtree of the ST, at the cost of an exponential
jump in the complexity of model checking.

Another approach to FO-completeness is that of NWTL \cite{lmcs/AlurABEIL08},
which is based on \emph{summary} until and since operators.
Summary paths are made of either consecutive positions, or matched call-return pairs.
Thus, they can skip function bodies, and enter or exit them.
Summary-up and down paths, and the respective operators, can be obtained
from summary paths, enabling exclusive upward or downward movement in the ST.
In particular, summary-down operators may express properties limited to a single subtree.

\subsubsection{Logics on OPL}
The only way to overcome the limitations of nested words is to base a temporal logic
on a more general algebraic structure.
OPTL \cite{ChiariMP20a} was introduced with this aim,
but it shares some of the limitations that CaRet has on nested words.
It features all LTL past and future operators,
plus the \emph{matching} next ($\lanext$) and back ($\laback$) operators,
resp.\ equivalent to POTL $\lcunext$ and $\lcuback$, OP \emph{summary} until and since,
and \emph{hierarchical} until and since.
POTL has several advantages over OPTL, regarding both the ease of expressing certain requirements and, as we conjecture, expressive power.

Given a set of PR $\Pi$, OPTL summary until $\luntil{\Pi}{}{}$ considers paths made of either
consecutive positions in a relation in $\Pi$, or positions $i, j$ s.t.\ $\chain(i,j)$ and $i \doteq j$ or $i \gtrdot j$.
The summary since $\lsince{\Pi}{}{}$ is symmetric, except positions in the $\chain$ relation must be in the $\lessdot$ or $\doteq$ PR.
Thus, none of such operators can go downward in the syntax tree, but only upward (e.g., any OPTL until may go upward if evaluated in pos.\ 4 of Fig.~\ref{fig:potl-example-word}).
This prevents OPTL from expressing function-local properties limited to a single subtree.
E.g., POTL formula $\llglob (\lthrow \implies \lcdback (\lhandle \land \lcduntil{\top}{\mathrm{p}_A}))$ means that if an exception is thrown and caught, procedure $\mathrm{p}_A$ is called at some point inside the $\lhandle$-$\lthrow$ block.
Any OPTL formula containing a summary until, such as $\llglob (\lthrow \implies \laback (\lhandle \land \luntil{\lessdot \doteq}{\top}{\mathrm{p}_A}))$, would fail, because it could go past the $\lthrow$ position by skipping one of the chains that terminated calls form with it.

OPTL has yield-precedence hierarchical until ($\lhyuntil{}{}$) and since ($\lhysince{}{}$) operators that, evaluated on a position $i$, consider paths made of positions $j$ s.t.\ $\chain(i,j)$ and $i \lessdot j$, all starting from the rightmost of such positions.
Their take-precedence counterparts ($\lhtuntil{}{}$ and $\lhtsince{}{}$) are symmetric.
One could try to express the POTL formula above with an OPTL formula such as $\llglob (\lthrow \implies \laback \lhandle \land (\lback (\lsince{\doteq \gtrdot}{\top}{\mathrm{p}_A}) \lor \lhyuntil{\top}{(\lsince{\doteq \gtrdot}{\top}{\mathrm{p}_A})}))$, but this would not work with nested $\lhandle$-$\lthrow$ blocks. 

The fact that its hierarchical until and since are evaluated on the left chain context for the yield-precedence versions, and on the right context for the take-precedence ones, is another limitation of OPTL.
It is not possible to concatenate them to express complex properties on right (resp.\ left) contexts of chains sharing their left (resp.\ right) context, such as several function calls issued by the same function, or multiple function calls terminated by the same exception.
POTL has both hierarchical next/back and until/since pairs, which make it expressively complete on such positions.
For example, we conjecture that formula
\(
\llglob (\lcall \land \mathrm{p}_B \implies
   \lhuuntil{\neg \mathrm{p}_C}{\mathrm{p}_\mathit{Err}})
\)
is not expressible in OPTL, and thus OPTL $\subset$ POTL.
It means that if procedure $\mathrm{p}_B$ is called by a function,
the same function must later call $\mathrm{p}_\mathit{Err}$
after $\mathrm{p}_B$ returns,
without calling $\mathrm{p}_C$ in the meantime.

In Appendix~\ref{sec:potl-translation}, we provide a direct translation of OPTL into POTL.

\section{First-Order Completeness}
\label{sec:fo-completeness}

We give a translation of POTL into FOL, and one of Conditional XPath (CXPath) \cite{Marx2004},
a logic on trees, into POTL on OP words.
From CXPath being equivalent to FOL on trees \cite{Marx2005},
we derive a FO-completeness result for POTL.

\subsection{First Order  Semantics of POTL}
\label{sec:fo-semantics}

We show that POTL can be expressed with FOL
equipped with monadic relations for atomic propositions,
a total order on positions, and the chain relation between pairs of positions.
We define below the translation function $\nu$,
such that for any POTL formula $\varphi$, word $w$ and position $x$,
$(w, x) \models \nu_\varphi(x)$ iff $(w, x) \models \varphi$.
The translation for propositional operators is trivial. \\
For temporal operators, we first need to define a few auxiliary formulas.
We define the successor relation as the FO formula
\[
  \lsucc(x, y) := x < y \land \neg \exists z (x < z \land z < y).
\]
In the following, $\prf \in \{\lessdot, \doteq, \gtrdot\}$
and $\Pi \subseteq \{\lessdot, \doteq, \gtrdot\}$.
The PR between positions can be expressed by means of propositional combinations
of monadic atomic relations only.
Given a set of atomic propositions $a \subseteq AP$, we define formula $\sigma_a(x)$,
stating that all and only propositions in $a$ hold in position $x$, as follows:
\[
  \sigma_a(x) :=
    \bigwedge_{\mathrm{p} \in a} \mathrm{p}(x)
    \land
    \bigwedge_{\mathrm{p} \not\in AP \setminus a} \neg \mathrm{p}(x)
\]
For any pair of FO variables $x, y$ and $\prf \in \{\lessdot, \doteq, \gtrdot\}$,
we can build formula
\[
  x \pr y :=
    \bigvee_{a, b \subseteq AP \mid a \pr b} (\sigma_a(x) \land \sigma_b(y)).
\]

The following translations employ the three FO variables $x, y, z$, only.
This, in addition to the FO-completeness result for POTL,
proves that FO on OP words retains the three-variable property, which holds in regular words.

\subsubsection{Next and Back Operators}
\[
  \nu_{\ldnext \varphi}(x) :=
    \exists y \Big(\lsucc(x, y) \land \bigvee _{\pr \in \{\lessdot, \doteq\}} (x \pr y) \land \exists x \big(x = y \land \nu_\varphi(x)\big)\Big)
\]
$\nu_{\ldback \varphi}(x)$ is defined similarly, and $\nu_{\lunext \varphi}(x)$ and $\nu_{\luback \varphi}(x)$
by replacing $\lessdot$ with $\gtrdot$.
\[
  \nu_{\lcdnext \varphi}(x) :=
    \exists y \big(x < y \land \chain(x,y) \land \bigvee _{\pr \in \{\lessdot, \doteq\}} (x \pr y) \land \exists x (x = y \land \nu_\varphi(x))\big)
\]
$\nu_{\lcdback{\Pi} \varphi}(x)$, $\nu_{\lcunext \varphi}(x)$ and $\nu_{\lcuback \varphi}(x)$
are defined similarly.

\subsubsection{Downward/Upward Summary Until/Since}
The translation for the DS until operator can be obtained by noting that,
given two positions $x$ and $y$,
the DSP between them, if it exists, is the one that skips all chain bodies
entirely contained between them, among those whose contexts
are in a relation in $\Pi = \{\lessdot, \doteq\}$.
The fact that a position $z$ is part of such path can be expressed with
formula $\neg \gamma(x,y,z)$ as follows:
\begin{align*}
  \gamma(x,y,z) &:=
    \gamma_L(x,z) \land \gamma_R(y,z) \\
  \gamma_L(x,z) &:=
    \exists y \Big(x \leq y \land y < z \land \exists x \big(z < x \land \chain(y,x) \land \bigvee_{\pr \in \Pi} (y \pr x)\big)\Big) \\
  \gamma_R(y,z) &:=
    \exists x \Big(z < x \land x \leq y \land \exists y \big(y < z \land \chain(y,x) \land \bigvee_{\pr \in \Pi} (y \pr x)\big)\Big)
\end{align*}
$\gamma(x,y,z)$ is true iff $z$ is not part of the DSP between $x$ and $y$, while $x \leq z \leq y$.
In particular, $\gamma_L(x,z)$ asserts that $z$ is part of the body of a chain whose left context is after $x$,
and $\gamma_R(y,z)$ states that $z$ is part of the body of a chain whose right context is before $y$.
Only chains whose contexts are in a relation in $\Pi$ are considered.
Since chain bodies cannot cross, either the two chain bodies are actually the same one,
or one of them is a sub-chain nested into the other.
In both cases, $z$ is part of a chain body entirely contained between $x$ and $y$,
and is thus not part of the path.

Moreover, for such a path to exist, each one of its positions must be
in one of the admitted PR with the next one.
Formula
\[
\delta(y,z) :=
  \exists x \big(z < x \land x \leq y
  \land \bigvee_{\pr \in \Pi} (z \pr x)
  \land \neg \gamma(z, y, x)
  \land (\lsucc(z, x) \lor \chain(z, x))\big)
\]
asserts this for each position $z$, with the path ending in $y$.
(Note that by exchanging $x$ and $z$ in the definition of $\gamma(x,y,z)$ above,
one can obtain $\gamma(z, y, x)$ without using any additional variable.)
Finally, $\lcduntil{\varphi}{\psi}$ can be translated as follows:
\begin{align*}
  \nu_{\lcduntil{\varphi}{\psi}}(x) :=
    \exists y \Big(&x \leq y \land \exists x (x = y \land \nu_\psi(x)) \\
      &\land \forall z \big(x \leq z \land z < y \land \neg \gamma(x, y, z)
        \implies \exists x (x = z \land \nu_\varphi(x)) \land \delta(y,z) \big)\Big)
\end{align*}

The translation for the DS since operator is similar:
\begin{align*}
  \nu_{\lcdsince{\varphi}{\psi}}(x) :=
    \exists y \Big(&y \leq x \land \exists x (x = y \land \nu_\psi(x)) \\
      &\land \forall z \big(y < z \land z \leq x \land \neg \gamma(y, x, z)
        \implies \exists x (x = z \land \nu_\varphi(x)) \land \delta(x,z) \big)\Big)
\end{align*}
$\nu_{\lcuuntil{\varphi}{\psi}}(x)$ and $\nu_{\lcusince{\varphi}{\psi}}(x)$
are defined as above, but with $\Pi = \{\doteq, \gtrdot\}$.

\subsubsection{Hierarchical Operators}
Finally, below are the translations for two hierarchical operators,
the others being symmetric.
\begin{align*}
  \nu_{\lhunext \varphi}(x) :=
    &\exists y \Bigg(y < x \land \chain(y,x) \land y \lessdot x \land \\
        &\qquad \exists z \Big(x < z \land \chain(y,z) \land y \lessdot z \land \exists x (x = z \land \nu_\varphi(x)) \\
        &\qquad \quad \land \forall y \big(x < y \land y < z \implies \forall z (\chain(z,x) \land z \lessdot x \implies \neg \chain(z,y))\big)\Big)\Bigg)
\end{align*}
\begin{align*}
  \nu_{\lhuuntil{\varphi}{\psi}}(x) :=
    \exists z \bigg(&z < x \land z \lessdot x \land \chain(z,x) \land \\
      &\exists y \Big(x \leq y \land \chain(z,y) \land z \lessdot y \land \exists x (x = y \land \nu_\psi(x)) \land \\
        &\quad \forall z \big(x \leq z \land z < y \land \exists y (y < x \land y \lessdot x \land \chain(y,x) \land \chain(y,z)) \\
          &\qquad\qquad \implies \exists x (x = z \land \nu_\varphi(x)))\big)\Big)\bigg)
\end{align*}

\subsection{Translation of Conditional XPath}

To translate CXPath to POTL, we give an isomorphism between OP words
and (a subset of) unranked ordered trees (UOT),
the algebraic structures on which CXPath is based.
First, we show how to translate OP words into trees, and then the reverse.

A UOT is a tuple $T = \langle S, \rchild, \rsibl, L \rangle$.
Each node is a sequence of child numbers, representing the path from the root to it.
$S$ is a finite set of finite sequences of natural numbers closed under the prefix operation,
and for any sequence $s \in S$,
if $s \cdot k \in S$, $k \in \mathbb{N}$, then either $k = 0$ or $s \cdot (k-1) \in S$
(by $\cdot$ we denote concatenation).
$\rchild$ and $\rsibl$ are two binary relations called the \emph{descendant} and
\emph{following sibling} relation, respectively.
For $s, t \in S$, $s \rchild t$ iff $t$ is any child of $s$
($t = s \cdot k$, $k \in \mathbb{N}$, i.e.\ $t$ is the $k$-th child of $s$),
and $s \rsibl t$ iff $t$ is the immediate sibling to the right of $s$
($s = r \cdot h$ and $t = r \cdot (h+1)$, for $r \in S$ and $h \in \mathbb{N}$).
$L \colon AP \rightarrow \powset{S}$ is a function that maps each
atomic proposition to the set of nodes labeled with it.
We denote as $\uotrees$ the set of all UOT.

\begin{wrapfigure}{R}{0pt}
\includegraphics{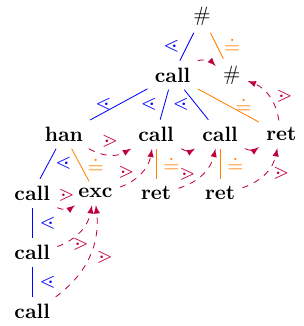}
\caption{The UOT
  corresponding to the word of Fig.~\ref{fig:potl-example-word},
  and to the ST of Fig.~\ref{fig:sttree-example}.
  PR are highlighted,
  and dashed arrows point to the Rcc of each node,
  decorated with $\lessdot$ when they are actual right contexts.}
\label{fig:uotree-example}
\end{wrapfigure}

Given an OP word $w = \langle U, <, M_{\powset{AP}}, P \rangle$,
it is possible to build an UOT
$T_w = \langle S_w, \rchild, \rsibl, L_w \rangle \in \uotrees$
with labels in $\powset{AP}$ isomorphic to $w$.
To do so, we define a function $\tau \colon U \rightarrow S_w$,
which maps positions of $w$ into nodes of $T_w$.
\begin{itemize}
\item
  $\tau(0) = 0$: position 0 is the root node.
\item
  Given any position $i \in U$, if $i \doteq i+1$,
  then $\tau(i+1) = \tau(i) \cdot 0$ is the only child of $i$.
\item
  If $i \gtrdot i+1$, then $i$ has no children.
\item
  If $i \lessdot i+1$, then the leftmost child of $i$ is $i+1$
  ($\tau(i+1) = \tau(i) \cdot 0$).
\item
  If $j_1 < j_2 < \dots < j_n$ is the largest set of positions such that $\chain(i,j_k)$
  and either $i \lessdot j_k$ or $i \doteq j_k$ for $1 \leq k \leq n$,
  then $\tau(j_k) = \tau(i) \cdot k$.
\end{itemize}
In general, $i$ is in the $\lessdot$ relation with all of its children,
except possibly the rightmost one, with which $i$ may be in the $\doteq$ relation
(cf. property~\ref{item:upward-prop} of the $\chain$ relation).
This way, every position $i$ in $w$ appears in the tree exactly once.
Indeed, if the position preceding $i$ is in the $\doteq$ or $\lessdot$ relation
with it, then $i$ is one of its children.
If $(i-1) \gtrdot i$, then at least a chain ends in $i$.
In particular, consider $j$, the leftmost context of $i$, s.t.\ $\chain(j,i)$,
and for no $j' < j$ we have $\chain(j',i)$:
by property~\ref{item:downward-prop} of the $\chain$ relation, either $j \doteq i$ or $j \lessdot i$
(or $i$ would be the right context of another chain containing $j$,
which would not be the leftmost context of $i$).
So, $i$ is a child of $j$.
Finally, $\tau(i) \in L_w(\mathrm{a})$ iff $i \in P(\mathrm{a})$ for all $\mathrm{a} \in AP$,
so each node in $T_w$ is labeled with the set of atomic
propositions that hold in the corresponding word position.
We denote as $T_w = \tau(w)$ the tree obtained by applying $\tau$
to every position of an OP word $w$.
Fig.~\ref{fig:uotree-example} shows the translation of the word of Fig.~\ref{fig:potl-example-word}
into an UOT.

As for the other way of the isomorphism, notice that we are considering
only a subset of UOTs.
In fact, we only consider UOT whose node labels are compatible with a given OPM $M_{\powset{AP}}$.
In order to define the notion of OPM compatibility for trees,
we need to introduce the \emph{right context candidate} (Rcc) of a node.
Given a tree $T$ and a node $s \in T$, the Rcc of $s$ is denoted $\ra(s)$.
If $r$ is the leftmost right sibling of $s$, then $\ra(s) = r$.
If $s$ has no right siblings, $\ra(s) = \ra(p)$, where $p$ is the parent of $s$.

We denote the set of trees compatible with an OPM $M$ as $\uotrees_M$.
A tree $T$ is in $\uotrees_M$ iff the following properties hold.
The root node is labeled with $\#$, and its rightmost child is labeled with $\#$.
No other node is labeled with $\#$.
In the following, for any $s, s' \in S$ and $\prf \in \{ \lessdot, \doteq, \gtrdot \}$,
we write $s \pr s'$ meaning that $a \pr b$,
where $a = \{\mathrm{p} \mid s \in L(\mathrm{p})\}$, and
$b = \{\mathrm{p} \mid s' \in P(\mathrm{p})\}$.
For any node $s \in T$,
let $r \in T$ be the rightmost child of $s$. Then either $s \lessdot r$ or $s \doteq r$.
For any child $s' \in T$ of $s$ s.t.\ $s'$ is a (left) sibling of $r$, we have $s \lessdot s'$.
If $s$ has no child $s'$ such that $s \doteq s'$, then $s \gtrdot \ra(s)$, if the latter exists.
Note that $\ra(s)$ always exists for all nodes not labeled with $\#$,
because it may be the rightmost child of the root.

Given a tree $T \in \uotrees_M$ with labels on $\powset{AP}$,
it is possible to build an OP word $w_T$ isomorphic to $T$.
We define function $\invtau : S \rightarrow \powset{AP}^+$,
which maps a tree node to the subword corresponding to the subtree rooted in it.
For any node $s \in T$, let $a = \{\mathrm{p} \mid s \in L(\mathrm{p})\}$ be its label,
and let $c_0, c_1 \dots c_n$ be its children,
if any. Then $\invtau(s)$ is defined as $\invtau(s) = a$ if $s$ has no children,
and $\invtau(s) = a \cdot \invtau(c_0) \cdot \invtau(c_1) \cdots \invtau(c_n)$ otherwise.

The string obtained in this way is a valid OP word.
To show this, we need to prove by induction on the tree structure
that for any tree node $s$, $\invtau(s)$ is of the form
$a_0 x_0 a_1 x_1 \dots a_n x_n$, with $n \geq 0$,
and such that for $0 \leq k < n$, $a_k \doteq a_{k+1}$
and either $x_k = \varepsilon$ or $\ochain{a_k}{x_k}{a_{k+1}}$.
In the following, we denote as $\first(x)$ the first position of a string $x$,
and as $\last(x)$ the last one.
Indeed, for each $0 \leq i < n$ we have $a \lessdot \first(\invtau(c_i))$, and
the rightmost leaf $f_i$ of the tree rooted in $c_i$ is such that $\ra(f_i) = c_{i+1}$.
Since $f_i = \tau(\last(\invtau(c_i)))$ and $c_{i+1} = \tau(\first(\invtau(c_{i+1})))$,
we have $\last(\invtau(c_i)) \gtrdot \first(\invtau(c_{i+1}))$.
So, $\ochain{a}{\invtau(c_i)}{\first(\invtau(c_{i+1}))}$.
As for $\invtau(c_n)$, if $a \lessdot c_n$ then $\invtau(s) = a_0 x_0$
(and $a_0 \lessdot \first(x_0)$),
with $a_0 = a$ and $x_0 = \invtau(c_0) \cdot \invtau(c_1) \cdots \invtau(c_n)$.
If $a \doteq c_n$, consider that, by hypothesis, $\invtau(c_n)$ is of the form $a_1 x_1 a_2 \dots a_n x_n$.
So $\invtau(s) = a_0 x_0 a_1 x_1 a_2 \dots a_n x_n$,
with $a_0 = a$ and $x_0 = \invtau(c_0) \cdot \invtau(c_1) \cdots \invtau(c_{n-1})$.

The root $0$ of $T$ is labeled with $\#$, and so is its rightmost child $c_\#$,
and let $c_l$ s.t.\ $c_l \rsibl c_\#$.
So, $\invtau(c_\#) = \#$, $\invtau(c_l) = a_1 x_1 \dots a_n x_n$,
and $\invtau(0) = \# x_0 a_1 x_1 \dots a_n x_n \#$.
Let $f$ be the rightmost leaf of the subtree rooted in $c_l$:
we have $\ra(f) = c_\#$, and $\invtau(f) \gtrdot \#$.
So $\# a_0 x_0 a_1 x_1 \dots a_n x_n \#$ is a finite OP word.

$\tau^{-1} : S \rightarrow U$ can be derived from $\invtau$.
From the existence of $\tau^{-1}$ follows
\begin{lem}
Given an OP word $w$ and the
tree $T_w = \tau(w)$, function $\tau$ is an isomorphism
between positions of $w$ and nodes of $T_w$.
\end{lem}
Consequently,
\begin{prop}
Let $M_{AP}$ be an OPM on $\powset{AP}$.
For any FO formula $\varphi(x)$ on OP words compatible with $M_{AP}$,
there exists a FO formula $\varphi'(x)$ on trees in $\uotrees_{M_{AP}}$ such that for any
OP word $w$ and position $i$ in it, $w \models \varphi(i)$ iff $T_w \models \varphi'(\tau(i))$,
with $T_w = \tau(w)$.
\end{prop}

We now give the full translation of the logic \xuntil{} from \cite{Marx2004} into POTL.
The syntax of \xuntil{} formulas is
\(
\varphi ::=
\mathrm{p} \mid
\top \mid
\neg \varphi \mid
\varphi \land \varphi \mid
\rho(\varphi, \varphi),
\)
with $\mathrm{a} \in AP$ and $\rho \in \{\Downarrow, \Uparrow, \Rightarrow, \Leftarrow\}$.
The semantics of propositional operators is the usual one,
while $\rho(\varphi, \varphi)$ is a strict until/since operator on the child and sibling relations.
Let $T \in \uotrees$ be a tree with nodes in $S$.
For any $r, s \in S$, let $\rparen, \rlsibl$ be s.t.\ $r \rparen s$ iff $s \rchild r$,
and $r \rlsibl s$ iff $s \rsibl r$.
We denote as $R_\rho^+$ the transitive (but not reflexive) closure of relation $R_\rho$.
For $s \in S$,
$(T,s) \models \rho(\varphi, \psi)$ iff there exists a node $t \in S$ s.t.\ $s R_\rho^+ t$
and $(T,t) \models \psi$, and for any $r \in S$ s.t.\ $s R_\rho^+ r$ and $r R_\rho^+ t$ we have
$(T,r) \models \varphi$.
\xuntil{} was proved to be equivalent to FOL
on finite UOTs in \cite{Marx2005}.
This result is valid for any labeling of tree nodes,
and so is on OPM-compatible trees.
\begin{thm}
\label{thm:xuntil-fo-completeness}
Let $M_{AP}$ be an OPM on $AP$.
For any FO formula $\varphi(x)$ on trees in $\uotrees_{M_{AP}}$,
there exists a \xuntil{} formula $\varphi'$ such that,
for any $T \in \uotrees_{M_{AP}}$ and node $t \in T$, we have
$T \models \varphi(t)$ iff $(T,t) \models \varphi'$ \cite{Marx2005}.
\end{thm}

We define function $\iota_\mathcal{X}$, which translates any \xuntil{} formula $\varphi$
into a POTL formula s.t.\ $\varphi$ holds on a tree $T$ iff $\iota_\mathcal{X}(\varphi)$
holds on the isomorphic word $w_T$.
$\iota_\mathcal{X}$ is defined as the identity for the propositional operators,
and with the equivalences below for the other \xuntil{} operators.
In the following, for any $a \subseteq AP$,
$\sigma_a := \bigwedge_{\mathrm{p} \in a} \mathrm{p} \land \bigwedge_{\mathrm{q} \not\in a} \neg \mathrm{q}$
holds in a pos.\ $i$ iff $a$ is the set of atomic propositions holding in $i$.
For any POTL formula $\gamma$, let
$\lganext{\lessdot} \gamma := \bigvee_{a,b \subseteq AP, a \lessdot b} (\sigma_a \land \lcdnext (\sigma_b \land \gamma))$
be the restriction of $\lcdnext \gamma$ to chains with contexts in the $\lessdot$ PR,
and $\lganext{\doteq} \gamma$, $\lgaback{\lessdot} \gamma$, $\lgaback{\doteq} \gamma$,
$\lgnext{\lessdot} \gamma$, $\lgback{\lessdot} \gamma$ are defined analogously.
For any \xuntil{} formulas $\varphi$ and $\psi$,
let $\varphi' = \iota_\mathcal{X}(\varphi)$ and $\psi' = \iota_\mathcal{X}(\psi)$.
We define $\iota_\mathcal{X}$ as follows:
\[
\begin{array}{l l}
\multicolumn{2}{l}{
\iota_\mathcal{X}(\Downarrow(\varphi, \psi)) :=
\ldnext (\lcduntil{\varphi'}{\psi'})
\lor \lcdnext (\lcduntil{\varphi'}{\psi'})
} \\[.5em]
\multicolumn{2}{l}{
\iota_\mathcal{X}(\Uparrow(\varphi, \psi)) :=
\ldback (\lcdsince{\varphi'}{\psi'})
\lor \lcdback (\lcdsince{\varphi'}{\psi'})
} \\[.5em]
\begin{aligned}
&\iota_\mathcal{X}(\Rightarrow(\varphi, \psi)) :=
\lhunext (\lhuuntil{\varphi'}{\psi'})  \\
&\qquad \lor \big( \neg \lhunext (\lhuuntil{\top}{\neg \varphi'})
            \land \lgaback{\lessdot} (\lganext{\doteq} \psi')\big) \\
&\qquad \lor \lgback{\lessdot} \Big(
      \lganext{\lessdot} \big(\psi' \land \neg \lhuback (\lhusince{\top}{\neg \varphi'}) \big) \Big) \\
&\qquad \lor \lgback{\lessdot} (\lganext{\doteq} \psi' \land \neg \lganext{\lessdot} \neg \varphi')
\end{aligned}
&
\begin{aligned}
&\iota_\mathcal{X}(\Leftarrow(\varphi, \psi)) :=
\lhuback (\lhusince{\varphi'}{\psi'}) \\
&\qquad \lor \lgaback{\doteq} \big(
      \lganext{\lessdot} (\neg \lhunext \top \land \lhusince{\varphi'}{\psi'}) \big) \\
&\qquad \lor \big( \lgaback{\lessdot} (\lgnext{\lessdot} \psi')
      \land \neg \lhuback (\lhusince{\top}{\neg \varphi'}) \big) \\
&\qquad \lor \lgaback{\doteq} (
      \lgnext{\lessdot} \psi' \land \neg \lganext{\lessdot} \neg \varphi' )
\end{aligned}
\end{array}
\]

We prove the correctness of this translation in the following theorems.
\begin{lem}
\label{lemma:iota-downarrow}
  Given a set of atomic propositions $AP$, and OPM $M_{AP}$,
  for every \xuntil{} formula $\Downarrow(\varphi, \psi)$,
  and for any OP word $w$ based on $M_{AP}$ and position $i$ in $w$, we have
  \[
    (T_w,\tau(i)) \models \Downarrow(\varphi, \psi) \iff
    (w,i) \models \iota_\mathcal{X}(\Downarrow(\varphi, \psi)).
  \]
  $T_w \in \uotrees_{M_{AP}}$ is the tree obtained by applying function $\tau$
  to every position in $w$, such that for any position $j$ in $w$
  $(T_w,\tau(i')) \models \varphi \iff (w,i') \models \iota_\mathcal{X}(\varphi)$,
  and likewise for $\psi$.
\end{lem}
\begin{proof}
  Let $\varphi' = \iota_\mathcal{X}(\varphi)$ and $\psi' = \iota_\mathcal{X}(\psi)$.
  We report the translation for convenience:
  \begin{equation}
    \iota_\mathcal{X}(\Downarrow(\varphi, \psi)) :=
    \ldnext (\lcduntil{\varphi'}{\psi'})
    \lor \lcdnext (\lcduntil{\varphi'}{\psi'}) \label{eq:iota-downarrow}
  \end{equation}

  $[\Rightarrow]$
  Suppose $(T_w,\tau(i)) \models \Downarrow(\varphi, \psi)$.
  Let $s = \tau(j)$, with $\tau(i) \rchild s$,
  be the first tree node of the path witnessing $\Downarrow(\varphi, \psi)$,
  and $r$ s.t.\ $r \rchild s$ be its parent.

  We shall now inductively prove that $\lcduntil{\varphi'}{\psi'}$ holds in $j$.
  If $s$ is the last node of the path, then $\psi'$ holds in $j$ and so does, trivially,
  $\lcduntil{\varphi'}{\psi'}$.
  Otherwise, consider any node $t = \tau(k)$ of the path, except the last one,
  and suppose $\lcduntil{\varphi'}{\psi'}$ holds in $k'$ s.t.\ $t' = \tau(k')$
  is the next node in the path.
  If $t'$ is the leftmost child of $t$, then $k' = k+1$ and either $k \lessdot k'$ or $k \doteq k'$.
  In both cases, $\ldnext (\lcduntil{\varphi'}{\psi'})$ holds in $k$.
  If $t'$ is not the leftmost child, then $\chain(k,k')$ and $k \lessdot k'$ or $k \doteq k'$.
  In both cases, $\lcdnext (\lcduntil{\varphi'}{\psi'})$ holds in $k$.
  So, by expansion law
  $\lcduntil{\varphi'}{\psi'} \equiv \psi'
  \lor \Big(\varphi' \land \big(\ldnext (\lcduntil{\varphi'}{\psi'})
         \lor \lcdnext (\lcduntil{\varphi'}{\psi'})\big)\Big)$,
  $\lcduntil{\varphi'}{\psi'}$ holds in $k$ and, by induction, also in $j$.

  Suppose $s$ is the leftmost child of $r$: $j = i+1$, and either $i \lessdot j$ or $i \doteq j$,
  so $\ldnext (\lcduntil{\varphi'}{\psi'})$ holds in $i$.
  Otherwise, $\chain(i,j)$ and either $i \lessdot j$ or $i \doteq j$.
  In both cases, $\lcdnext (\lcduntil{\varphi'}{\psi'})$ holds in $i$.

  $[\Leftarrow]$
  Suppose \eqref{eq:iota-downarrow} holds in $i$.
  If $\ldnext (\lcduntil{\varphi'}{\psi'})$ holds in $i$,
  then $\lcduntil{\varphi'}{\psi'}$ holds in $j = i+1$,
  and either $i \lessdot j$ or $i \doteq j$: then $s = \tau(j)$ is the leftmost child of $\tau(i)$.
  If $\lcdnext (\lcduntil{\varphi'}{\psi'})$ holds in $i$,
  then $\lcduntil{\varphi'}{\psi'}$ holds in $j$ s.t.\ $\chain(i,j)$
  and $i \lessdot j$ or $i \doteq j$: $s = \tau(j)$ is a child of $\tau(i)$ in this case as well.

  We shall now prove that if $\lcduntil{\varphi'}{\psi'}$ holds in a position
  $j$ s.t.\ $\tau(i) \rchild \tau(j)$, then $\Downarrow(\varphi, \psi)$ holds in $\tau(i)$.
  If $\lcduntil{\varphi'}{\psi'}$ holds in $j$, then there exists a DSP of minimal length
  from $j$ to $h > j$ s.t.\ $(w,h) \models \psi'$ and $\varphi'$ holds
  in all positions $j \leq k < h$ of the path, and $(T_w,\tau(k)) \models \varphi$.
  In any such $k$,
  $\lcduntil{\varphi'}{\psi'} \equiv \psi'
  \lor \Big(\varphi' \land \big(\ldnext (\lcduntil{\varphi'}{\psi'})
         \lor \lcdnext (\lcduntil{\varphi'}{\psi'})\big)\Big)$
  holds.
  Since this DSP is the minimal one, $\psi'$ does not hold in $k$.
  Either $\ldnext (\lcduntil{\varphi'}{\psi'})$
  or $\lcdnext (\lcduntil{\varphi'}{\psi'})$ hold in it.
  Therefore, the next position in the path is $k'$ s.t.\ either $k' = k+1$ or $\chain(k,k')$,
  and either $k \lessdot k'$ or $k \doteq k'$, and
  $(w,k') \models \lcduntil{\varphi'}{\psi'}$.
  Therefore, $\tau(k')$ is a child of $\tau(k)$.
  So, there is a sequence of nodes $s_0, s_1, \dots, s_n$ in $T_w$ s.t.\ $\tau(i) \rchild s_0$,
  and $s_i \rchild s_{i+1}$ and $(T_w,s_i) \models \varphi$ for $0 \leq i < n$,
  and $(T_w,s_n) \models \psi$.
  This is a path making $\Downarrow(\varphi, \psi)$ true in $\tau(i)$.
\end{proof}

\begin{lem}
\label{lemma:iota-uparrow}
  Given a set of atomic propositions $AP$, and OPM $M_{AP}$,
  for every \xuntil{} formula $\Uparrow(\varphi, \psi)$,
  and for any OP word $w$ based on $M_{AP}$ and position $i$ in $w$, we have
  \[
    (T_w,\tau(i)) \models \Uparrow(\varphi, \psi) \iff
    (w,i) \models \iota_\mathcal{X}(\Uparrow(\varphi, \psi)).
  \]
  $T_w \in \uotrees_{M_{AP}}$ is the tree obtained by applying function $\tau$
  to every position in $w$, such that for any position $j$ in $w$
  $(T_w,\tau(i')) \models \varphi \iff (w,i') \models \iota_\mathcal{X}(\varphi)$,
  and likewise for $\psi$.
\end{lem}
\begin{proof}
  The proof is analogous to the one of Lemma~\ref{lemma:iota-downarrow},
  and is therefore omitted.
\end{proof}

\begin{lem}
\label{lemma:iota-rightarrow}
  Given a set of atomic propositions $AP$, and OPM $M_{AP}$,
  for every \xuntil{} formula $\Rightarrow(\varphi, \psi)$,
  and for any OP word $w$ based on $M_{AP}$ and position $i$ in $w$, we have
  \[
    (T_w,\tau(i)) \models \Rightarrow(\varphi, \psi) \iff
    (w,i) \models \iota_\mathcal{X}(\Rightarrow(\varphi, \psi)).
  \]
  $T_w \in \uotrees_{M_{AP}}$ is the tree obtained by applying function $\tau$
  to every position in $w$, such that for any position $j$ in $w$
  $(T_w,\tau(i')) \models \varphi \iff (w,i') \models \iota_\mathcal{X}(\varphi)$,
  and likewise for $\psi$.
\end{lem}
\begin{proof}
  For any $a \subseteq AP$, recall
  $\sigma_a := \bigwedge_{\mathrm{p} \in a} \mathrm{p} \land \bigwedge_{\mathrm{q} \not\in a} \neg \mathrm{q}$,
  and for any POTL formula $\gamma$ and a PR $\pr \in \{\lessdot, \doteq\}$,
  $\lganext{\pr} \gamma := \bigvee_{a,b \subseteq AP, a \pr b} (\sigma_a \land \lcdnext (\sigma_b \land \gamma))$;
  $\lgaback{\pr} \gamma := \bigvee_{a,b \subseteq AP, a \pr b} (\sigma_a \land \lcdback (\sigma_b \land \gamma))$;
  $\lgback{\pr} \gamma := \bigvee_{a,b \subseteq AP, a \pr b} (\sigma_a \land \ldback (\sigma_b \land \gamma))$.
  Let $\varphi' = \iota_\mathcal{X}(\varphi)$ and $\psi' = \iota_\mathcal{X}(\psi)$:
  \begin{align}
    &\iota_\mathcal{X}(\Rightarrow(\varphi, \psi)) :=
    \lhunext (\lhuuntil{\varphi'}{\psi'}) \label{eq:iota-rightarrow-1} \\
    &\qquad \lor \big( \neg \lhunext (\lhuuntil{\top}{\neg \varphi'})
      \land \lgaback{\lessdot} (\lganext{\doteq} \psi')\big) \label{eq:iota-rightarrow-2} \\
    &\qquad \lor \lgback{\lessdot} \Big(
      \lganext{\lessdot} \big(\psi' \land \neg \lhuback (\lhusince{\top}{\neg \varphi'}) \big) \Big) \label{eq:iota-rightarrow-3} \\
    &\qquad \lor \lgback{\lessdot} (\lganext{\doteq} \psi' \land \neg \lganext{\lessdot} \neg \varphi') \label{eq:iota-rightarrow-4}
  \end{align}

  $[\Rightarrow]$
  Suppose $\Rightarrow(\varphi, \psi)$ holds in $s = \tau(i)$.
  Then, node $r = \tau(h)$ s.t.\ $r \rchild s$ has at least two children,
  and $\Rightarrow(\varphi, \psi)$ is witnessed by a path starting in $t = \tau(j)$
  s.t.\ $s \rsibl t$, and ending in $v = \tau(k)$.
  We have the following cases:
  \begin{enumerate}
  \item \label{item:proof-iota-ra-1} \emph{$s$ is not the leftmost child of $r$.}
    \begin{enumerate}
    \item \label{item:proof-iota-ra-1a} \emph{$h \lessdot k$.}
      By the construction of $T_w$, for any node $t'$ in the path, there exists a position
      $j' \in w$ s.t.\ $t' = \tau(j')$, $\chain(h,j')$ and $h \lessdot j'$.
      The path made by such positions is a UHP, and $\lhuuntil{\varphi'}{\psi'}$ is true in $j$.
      Since $s$ is not the leftmost child of $r$, we have $\chain(h,i)$, and $h \lessdot i$,
      so \eqref{eq:iota-rightarrow-1} ($\lhunext (\lhuuntil{\varphi'}{\psi'})$) holds in $i$.
    \item \label{item:proof-iota-ra-1b} \emph{$h \doteq k$, so $v$ is the rightmost child of $r$.}
      $\varphi$ holds in all siblings between $s$ and $v$ (excluded),
      and $\varphi'$ holds in the corresponding positions of $w$.
      All such positions $j$, if any, are s.t.\ $\chain(h,j)$ and $h \lessdot j$,
      and they form a UHP, so $\lhunext (\lhuuntil{\top}{\neg \varphi'})$
      never holds in $i$.
      Moreover, since $\psi$ holds in $v$, $\psi'$ holds in $k$.
      Note that $\lgaback{\lessdot}$ in $i$ uniquely identifies position $h$,
      and $\lganext{\doteq}$ evaluated in $h$ identifies $k$.
      So, \eqref{eq:iota-rightarrow-2} holds in $i$.
    \end{enumerate}
  \item \label{item:proof-iota-ra-2} \emph{$s$ is the leftmost child of $r$.}
    In this case, we have $i = h+1$ and $h \lessdot i$
    (if $h \doteq i$, then $r$ would have only one child).
    \begin{enumerate}
    \item \label{item:proof-iota-ra-2a} \emph{$h \lessdot k$.}
      $\lgback{\lessdot}$ evaluated in $i$ identifies position $h$.
      $\psi'$ holds in $k$, and $\lhuback (\lhusince{\top}{\neg \varphi'})$ does not,
      because in all positions between $i$ and $k$ (excluded) corresponding to children of $r$,
      $\varphi'$ holds. Note that all such positions form a UHP, but $i$ is not part of it
      ($i = h+1$, so $\neg \chain(h,i)$), and is not considered by $\lhusince{\top}{\neg \varphi'}$.
      So, \eqref{eq:iota-rightarrow-3} holds in $i$.
      \item \label{item:proof-iota-ra-2b} \emph{$h \doteq k$, so $v$ is the rightmost child of $r$.}
        $\psi$ holds in $v$, and $\varphi$ holds in all children of $r$,
        except possibly the first ($s$) and the last one ($v$).
        These are exactly all positions s.t.\ $\chain(h,j)$ and $h \lessdot j$.
        Since $\varphi'$ holds in all of them by hypothesis,
        $\neg \lganext{\lessdot} \neg \varphi'$ holds in $h$.
        Since $\psi$ holds in $v$, $\psi'$ holds in $k$, and $\lganext{\doteq} \psi'$ in $h$.
        So, \eqref{eq:iota-rightarrow-4} holds in $i$.
    \end{enumerate}
  \end{enumerate}

  $[\Leftarrow]$
  Suppose \eqref{eq:iota-rightarrow-1} ($\lhunext (\lhuuntil{\varphi'}{\psi'})$)
  holds in a position $i$ in $w$.
  Then, there exists a position $h$ s.t.\ $\chain(h,i)$ and $h \lessdot i$,
  and a position $j$ s.t.\ $\chain(h,j)$ and $h \lessdot j$
  that is the hierarchical successor of $i$, and $\lhuuntil{\varphi'}{\psi'}$ holds in $j$.
  So, $i$ and $j$ are consecutive children of $r = \tau(h)$.
  Moreover, there exists a UHP between $j$ and a position $k \geq j$.
  The tree nodes corresponding to all positions in the path are consecutive children of $r$,
  so we fall in case \ref{item:proof-iota-ra-1a}
  of the proof of the other side of the implication.
  In $T_w$, a path between $t = \tau(j)$ and $v = \tau(k)$ witnesses the truth of
  $\Rightarrow(\varphi, \psi)$ in $s$.

  Suppose \eqref{eq:iota-rightarrow-2}
  ($\neg \lhunext (\lhuuntil{\top}{\neg \varphi'})
   \land \lgaback{\lessdot} (\lganext{\doteq} \psi')$)
  holds in position $i \in w$ (this corresponds to case \ref{item:proof-iota-ra-1b}).
  If $\lgaback{\lessdot} (\lganext{\doteq} \psi')$  holds in $i$,
  then there exists a position $h$ s.t.\ $\chain(h,i)$ and $h \lessdot i$,
  and a position $k$ s.t.\ $\chain(h,k)$ and $h \doteq k$, and $\psi'$ holds in $k$.
  $v = \tau(k)$ is the rightmost child of $r = \tau(h)$, parent of $s = \tau(i)$.
  Moreover, if $\neg \lhunext (\lhuuntil{\top}{\neg \varphi'})$ holds in $i$, then either:
  \begin{itemize}
  \item $\neg \lhunext \top$ holds, i.e.\ there is no position $j > i$
    s.t.\ $\chain(h,j)$ and $h \lessdot j$,
    so $v$ is the immediate right sibling of $s$.
    In this case $\Rightarrow(\varphi, \psi)$ holds in $s$ because $\psi$ holds in $v$.
  \item $\neg (\lhuuntil{\top}{\neg \varphi'})$ holds in $j > i$,
    the first position after $i$ s.t.\ $\chain(h,j)$ and $h \lessdot j$.
    This means $\varphi'$ holds in all positions $j' \geq j$
    s.t.\ $\chain(h,j')$ and $h \lessdot j'$.
    Consequently, the tree nodes corresponding to these positions plus $v = \tau(k)$ form
    a path witnessing $\Rightarrow(\varphi, \psi)$, which holds in $s = \tau(i)$.
  \end{itemize}

  Suppose \eqref{eq:iota-rightarrow-3}
  ($\lgback{\lessdot} \Big(
      \lganext{\lessdot} \big(\psi' \land
      \neg \lhuback (\lhusince{\top}{\neg \varphi'}) \big) \Big)$)
  holds in $i$.
  Let $h = i-1$, with $h \lessdot i$ (it exists because $\lgback{\lessdot}$ is true).
  There exists a position $k$, $\chain(h,k)$ and $h \lessdot k$,
  in which $\psi'$ holds, so $\psi$ does in $v = \tau(k)$,
  and $\lhuback (\lhusince{\top}{\neg \varphi'})$ is false in it.
  If it is false because $\neg \lhuback \top$ holds,
  there is no position $j < k$ s.t.\ $\chain(h,j)$ and $h \lessdot j$,
  so $v$ is the second child of $r = \tau(h)$, $s = \tau(i)$ being the first one.
  So, $\Rightarrow(\varphi, \psi)$ trivially holds in $s$ because $\psi$ holds in the next sibling.
  Otherwise, let $j < k$ be the rightmost position lower than $k$
  s.t.\ $\chain(h,j)$ and $h \lessdot j$.
  $\neg (\lhusince{\top}{\neg \varphi'})$ holds in it, so $\varphi'$ holds
  in all positions $j'$ between $i$ and $k$ that are part of the hierarchical path,
  i.e.\ s.t.\ $\chain(h,j')$ and $h \lessdot j'$.
  The corresponding tree nodes form a path ending in $v = \tau(k)$
  that witnesses the truth of $\Rightarrow(\varphi, \psi)$ in $s$
  (case \ref{item:proof-iota-ra-2a}).

  If \eqref{eq:iota-rightarrow-4}
  ($\lgback{\lessdot} (\lganext{\doteq} \psi' \land \neg \lganext{\lessdot} \neg \varphi')$)
  holds in $i$, then let $h = i-1$, $h \lessdot i$, and $S = \tau(i)$ is the leftmost
  child of $r = \tau(h)$.
  Since $\lganext{\doteq} \psi'$ holds in $h$, there exists a position $k$,
  s.t.\ $\chain(h,k)$ and $h \doteq k$, in which $\psi'$ holds.
  So, $\psi$ holds in $v = \tau(k)$, which is the rightmost child of $r$, by construction.
  Moreover, in all positions s.t.\ $\chain(h,j)$ and $h \lessdot j$, $\psi'$ holds.
  Hence, $\varphi$ holds in all corresponding nodes $t = \tau(j)$,
  which are all nodes between $s$ and $v$, excluded.
  This, together with $\psi$ holding in $v$, makes a path that verifies
  $\Rightarrow(\varphi, \psi)$ in $s$
  (case \ref{item:proof-iota-ra-2b}).
\end{proof}

\begin{lem}
\label{lemma:iota-leftarrow}
  Given a set of atomic propositions $AP$, and OPM $M_{AP}$,
  for every \xuntil{} formula $\Leftarrow(\varphi, \psi)$,
  and for any OP word $w$ based on $M_{AP}$ and position $i$ in $w$, we have
  \[
    (T_w,\tau(i)) \models \Leftarrow(\varphi, \psi) \iff
    (w,i) \models \iota_\mathcal{X}(\Leftarrow(\varphi, \psi)).
  \]
  $T_w \in \uotrees_{M_{AP}}$ is the tree obtained by applying function $\tau$
  to every position in $w$, such that for any position $j$ in $w$
  $(T_w,\tau(i')) \models \varphi \iff (w,i') \models \iota_\mathcal{X}(\varphi)$,
  and likewise for $\psi$.
\end{lem}
\begin{proof}
  For any $a \subseteq AP$, recall
  $\sigma_a := \bigwedge_{\mathrm{p} \in a} \mathrm{p} \land \bigwedge_{\mathrm{q} \not\in a} \neg \mathrm{q}$,
  and for any POTL formula $\gamma$ and a PR $\pr \in \{\lessdot, \doteq\}$,
  $\lganext{\pr} \gamma := \bigvee_{a,b \subseteq AP, a \pr b} (\sigma_a \land \lcdnext (\sigma_b \land \gamma))$;
  $\lgaback{\pr} \gamma := \bigvee_{a,b \subseteq AP, a \pr b} (\sigma_a \land \lcdback (\sigma_b \land \gamma))$;
  $\lgnext{\pr} \gamma := \bigvee_{a,b \subseteq AP, a \pr b} (\sigma_a \land \ldnext (\sigma_b \land \gamma))$.
  Let $\varphi' = \iota_\mathcal{X}(\varphi)$ and $\psi' = \iota_\mathcal{X}(\psi)$:
  \begin{align}
    &\iota_\mathcal{X}(\Leftarrow(\varphi, \psi)) :=
      \lhuback (\lhusince{\varphi'}{\psi'}) \label{eq:iota-leftarrow-1} \\
    &\qquad \lor \lgaback{\doteq} \big(
      \lganext{\lessdot} (\neg \lhunext \top \land \lhusince{\varphi'}{\psi'}) \big)
      \label{eq:iota-leftarrow-2} \\
    &\qquad \lor \big( \lgaback{\lessdot} (\lgnext{\lessdot} \psi')
      \land \neg \lhuback (\lhusince{\top}{\neg \varphi'}) \big)
      \label{eq:iota-leftarrow-3} \\
    &\qquad \lor \lgaback{\doteq} (
      \lgnext{\lessdot} \psi' \land \neg \lganext{\lessdot} \neg \varphi' )
      \label{eq:iota-leftarrow-4}
  \end{align}

  $[\Rightarrow]$
  Suppose $\Leftarrow(\varphi, \psi)$ holds in $s = \tau(i)$.
  Then node $r = \tau(h)$ s.t.\ $r \rchild s$ has at least two children,
  and $\Leftarrow(\varphi, \psi)$ is true because of a path starting in $v = \tau(k)$,
  s.t.\ $r \rchild v$ and $(T_w,v) \models \psi$
  and ending in $t = \tau(j)$ s.t.\ $t \rsibl s$.
  We distinguish between the following cases:
  \begin{enumerate}
  \item \emph{$v$ is not the leftmost child of $r$.}
    \begin{enumerate}
    \item \label{item:proof-iota-la-1a} \emph{$h \lessdot i$.}
      By construction, all nodes in the path correspond to positions $j' \in w$
      s.t.\ $\chain(h,j')$ and $h \lessdot j'$, so they form a UHP.
      Hence, $\lhusince{\varphi'}{\psi'}$ holds in $j$, and \eqref{eq:iota-leftarrow-1}
      ($\lhuback (\lhusince{\varphi'}{\psi'})$) holds in $i$.
    \item \label{item:proof-iota-la-1b} \emph{$h \doteq i$.}
      In this case, $s$ is the rightmost child of $r$, and $\chain(h,i)$.
      The path made of positions between $k$ and $j$ corresponding to nodes between $v$ and $t$
      (included) form a UHP.
      So $\lhusince{\varphi'}{\psi'}$ holds in $j$, which is the rightmost position of any
      possible such UHP: so $\neg \lhunext \top$ also holds in $j$.
      Hence, \eqref{eq:iota-leftarrow-2}
      ($\lgaback{\doteq} \big(
        \lganext{\lessdot} (\neg \lhunext \top \land \lhusince{\varphi'}{\psi'}) \big)$)
      holds in $i$.
    \end{enumerate}
  \item \emph{$v$ is the leftmost child of $r$.}
    \begin{enumerate}
    \item \label{item:proof-iota-la-2a} \emph{$h \lessdot i$.}
      In this case, $k = h+1$ and $\psi'$ holds in $k$.
      So, $\lgnext{\lessdot} \psi'$ holds in $h$,
      and $\lgaback{\lessdot} (\lgnext{\lessdot} \psi')$ holds in $i$.
      Moreover, in all positions $j' \in w$, $k < j' < j$, corresponding to tree nodes,
      $\varphi'$ holds.
      Such positions form a UHP.
      So $\neg \lhuback (\lhusince{\top}{\neg \varphi'})$ holds in $i$.
      Note that this is also true if $s$ is the first right sibling of $v$.
      In conclusion, \eqref{eq:iota-leftarrow-3} holds in $i$.
    \item \label{item:proof-iota-la-2b} \emph{$h \doteq i$.}
      $\psi'$ holds in $k = h+1$, so $\lgnext{\lessdot} \psi'$ holds in $h$.
      Since $\chain(h,i)$ and $h \doteq i$,
      $\lgaback{\doteq} (\lgnext{\lessdot} \psi')$ holds in $i$.
      Moreover, $\varphi$ holds in all children of $r$ except the first and last one,
      i.e.\ $\varphi'$ holds in all positions $j'$ s.t.\ $\chain(h,j')$ and $h \lessdot j'$.
      So $\neg \lganext{\lessdot} \neg \varphi'$ holds in $h$, and \eqref{eq:iota-leftarrow-4}
      ($\lgaback{\doteq} (\lgnext{\lessdot} \psi' \land \neg \lganext{\lessdot} \neg \varphi')$)
      in $i$.
    \end{enumerate}
  \end{enumerate}

  $[\Leftarrow]$
  Suppose \eqref{eq:iota-leftarrow-1} ($\lhuback (\lhusince{\varphi'}{\psi'})$)
  holds in $i$.
  Then, there exists a position $h$ s.t.\ $\chain(h,i)$ and $h \lessdot i$,
  and a position $j < i$ s.t.\ $\chain(h,j)$ and $h \lessdot j$.
  Since $j \neq h+1$ and $h \lessdot i$, the corresponding tree nodes are not the leftmost
  nor the rightmost one.
  So, this corresponds to case \ref{item:proof-iota-la-1a},
  and $\Leftarrow(\varphi, \psi)$ holds in $s = \tau(i)$.

  Suppose \eqref{eq:iota-leftarrow-2}
  ($\lgaback{\doteq} \big(\lganext{\lessdot} (\neg \lhunext \top \land \lhusince{\varphi'}{\psi'}) \big)$)
  holds in $i$.
  Then, there exists a position $h$ s.t.\ $\chain(h,i)$ and $h \doteq i$.
  Moreover, at least a position $j'$ s.t.\ $\chain(h,j')$ and $h \lessdot j'$ exists.
  Let $j$ be the rightmost one, i.e.\ the only one in which $\neg \lhunext \top$ holds.
  The corresponding tree node $t = \tau(j)$ is s.t.\ $t \rsibl s$, with $s = \tau(i)$.
  Since $\lhusince{\varphi'}{\psi'}$ holds in $j$, a UHP starts from it,
  and $\psi$ and $\varphi$ hold in the tree nodes corresponding to, respectively,
  the first and all other positions in the path.
  This is case \ref{item:proof-iota-la-1b}, and $\Leftarrow(\varphi, \psi)$ holds in $s$.

  Suppose \eqref{eq:iota-leftarrow-3}
  ($\lgaback{\lessdot} (\lgnext{\lessdot} \psi')
    \land \neg \lhuback (\lhusince{\top}{\neg \varphi'})$)
  holds in $i$.
  Then, there exists a position $h$ s.t.\ $\chain(h,i)$ and $h \lessdot i$.
  $\psi'$ holds in $k = h+1$, so $\psi$ holds in the leftmost child of $r = \tau(h)$.
  Moreover, $\varphi'$ holds in all positions $j' < i$ s.t.\ $\chain(h,j')$ and $h \lessdot j'$,
  so $\varphi$ holds in all children of $r$ between $v = \tau(k)$ and $s = \tau(i)$, excluded.
  This is case \ref{item:proof-iota-la-2a}, and $\Leftarrow(\varphi, \psi)$ holds in $s$.

  Finally, suppose \eqref{eq:iota-leftarrow-4}
  ($\lgaback{\doteq} (\lgnext{\lessdot} \psi' \land \neg \lganext{\lessdot} \neg \varphi' )$)
  holds in $i$.
  Then, there exists a position $h$ s.t.\ $\chain(h,i)$ and $h \doteq i$.
  $\lgnext{\lessdot} \psi'$ holds in $h$, so $\psi$ holds in node $v = \tau(h+1)$,
  which is the leftmost child of $r = \tau(h)$.
  Since $\neg \lganext{\lessdot} \neg \varphi'$ holds in $h$,
  $\psi'$ holds in all positions $j'$ s.t.\ $\chain(h,j')$ and $h \lessdot j$.
  So, $\psi$ holds in all children of $r$ except (possibly) the leftmost ($v$)
  and the rightmost ($s = \tau(i)$) ones.
  This is case \ref{item:proof-iota-la-2b}, and $\Leftarrow(\varphi, \psi)$ holds in $s$.
\end{proof}

It is possible to express all POTL operators in FOL,
by following the semantics described in Section~\ref{sec:potl-syntax-semantics}.
The translation of DS/US until/since operators is similar to the one employed for NWTL
in \cite{lmcs/AlurABEIL08}. The full translation can be found in Appendix \ref{sec:fo-semantics}.
From this, and Lemmas~\ref{lemma:iota-downarrow}, \ref{lemma:iota-uparrow},
\ref{lemma:iota-rightarrow}, and \ref{lemma:iota-leftarrow}.
together with Theorem~\ref{thm:xuntil-fo-completeness}, we derive
\begin{thm}
\label{thm:potl-completeness}
POTL = FO with one free variable on finite OP words.
\end{thm}
\begin{cor}
\label{cor:complete-subset}
The propositional operators plus
\(
  \ldnext,
  \ldback,
  \lcdnext,
  \lcdback,
  \lcduntil{}{},
  \lcdsince{}{},
  \lhunext, \allowbreak
  \lhuback, \allowbreak
  \lhuuntil{}{}, \allowbreak
  \lhusince{}{}
\)
are expressively complete on OP words.
\end{cor}
\begin{cor}
\label{cor:optl-in-potl}
NWTL $\subset$ OPTL $\subseteq$ POTL over finite OP words.
\end{cor}

Corollary~\ref{cor:complete-subset} follows from the definition of $\iota_\mathcal{X}$
and Theorem~\ref{thm:potl-completeness}.
In Corollary~\ref{cor:optl-in-potl}, NWTL $\subset$ OPTL was proved in \cite{ChiariMP18},
and OPTL $\subseteq$ POTL comes from Theorem~\ref{thm:potl-completeness} and
the semantics of OPTL being expressible in FOL similarly to POTL.

\section{Model Checking}
\label{sec:mc}

We present an automata-theoretic model checking procedure for POTL based on OPA.
Given an OP alphabet $(\powset{AP}, M_{AP})$, where $AP$ is a finite set of atomic propositions,
and a formula $\varphi$, let
\(
  \mathcal{A}_\varphi =
  \langle
     \powset{AP}, \allowbreak
     M_{AP}, \allowbreak
     Q, \allowbreak
     I, \allowbreak
     F, \allowbreak
     \delta
  \rangle
\)
be an OPA.
The construction of $\mathcal{A}_\varphi$ resembles the classical one for LTL and the ones for NWTL and OPTL,
diverging from them significantly when dealing with temporal obligations involving positions in the $\chain$ relation.

We first introduce $\clos{\varphi}$, the \emph{closure} of $\varphi$, containing all subformulas of $\varphi$,
plus a few auxiliary operators.
Initially, $\clos{\varphi}$ is the smallest set such that
\begin{enumerate}
\item
  $\varphi \in \clos{\varphi}$,
\item
  $AP \subseteq \clos{\varphi}$,
\item
  if $\psi \in \clos{\varphi}$ and $\psi \neq \neg \theta$, then $\neg \psi \in \clos{\varphi}$
  (we identify $\neg \neg \psi$ with $\psi$);
\item
  if $\neg \psi \in \clos{\varphi}$, then $\psi \in \clos{\varphi}$;
\item
  if any of $\psi \land \theta$ or $\psi \lor \theta$ is in $\clos{\varphi}$,
  then $\psi \in \clos{\varphi}$ and $\theta \in \clos{\varphi}$;
\item
  if any of the unary temporal operators (such as $\ldnext$, $\lcdnext$, ...)
  is in $\clos{\varphi}$, and $\psi$ is its argument, then $\psi \in \clos{\varphi}$;
\item
  if any of the until- and since-like operators is in $\clos{\varphi}$,
  and $\psi$ and $\theta$ are its operands, then $\psi, \theta \in \clos{\varphi}$.
\end{enumerate}
The set $\atoms{\varphi}$ contains all consistent subsets of $\clos{\varphi}$,
i.e.\ all $\Phi \subseteq \clos{\varphi}$ s.t.
\begin{enumerate}
\item
  for every $\psi \in \clos{\varphi}$, $\psi \in \Phi$ iff $\neg \psi \notin \Phi$;
\item
  $\psi \land \theta \in \Phi$, iff $\psi \in \Phi$ and $\theta \in \Phi$;
\item
  $\psi \lor \theta \in \Phi$, iff $\psi \in \Phi$ or $\theta \in \Phi$, or both.
\end{enumerate}
The consistency constraints on $\atoms{\varphi}$ will be augmented incrementally in the following, for each operator.

The set of states of $\mathcal{A}_\varphi$ is $Q = \atoms{\varphi}^2$,
and its elements, which we denote with Greek capital letters,
are of the form $\Phi = (\Phi_c, \Phi_p)$,
where $\Phi_c$, called the \emph{current} part of $\Phi$,
is the set of formulas that hold in the current position,
and $\Phi_p$, or the \emph{pending} part of $\Phi$, is the set of temporal obligations.
The latter keep track of arguments of temporal operators
that must be satisfied after a chain body, skipping it.
The way they do so depends on the transition relation $\delta$,
which we also define incrementally.
Each automaton state is associated to word positions.
So, for $(\Phi, a, \Psi) \in \delta_\mathit{push/shift}$, with $\Phi \in \atoms{\varphi}^2$
and $a \in \powset{AP}$, we have $\Phi_c \cap AP = a$
(by $\Phi_c \cap AP$ we mean the set of atomic propositions in $\Phi_c$).
\emph{Pop} moves do not read input symbols, and the automaton remains stuck at the same position
when performing them: for any $(\Phi, \Theta, \Psi) \in \delta_\mathit{pop}$
we impose $\Phi_c = \Psi_c$.
The initial set $I$ contains states of the form $(\Phi_c, \Phi_p)$, with $\varphi \in \Phi_c$,
and the final set $F$ states of the form $(\Psi_c, \Psi_p)$,
s.t.\ $\Psi_c \cap AP = \{\#\}$ and $\Psi_c$ contains no future operators.
$\Phi_p$ and $\Psi_c$ may contain only operators according to rules explicitly
stated in the following.

\subsection{Next/Back Operators}
Let $\ldnext \psi \in \clos{\varphi}$: then $\psi \in \clos{\varphi}$.
Let $(\Phi, a, \Psi) \in \delta_\mathit{shift} \cup \delta_\mathit{push}$,
with $\Phi, \Psi \in \atoms{\varphi}^2$, $a \in \powset{AP}$, and let $b = \Psi_c \cap AP$:
we have $\ldnext \psi \in \Phi_c$ iff $\psi \in \Psi_c$
and either $a \lessdot b$ or $a \doteq b$.
The constraints introduced for the $\ldback$ operator are symmetric,
and for their upward counterparts it suffices to replace $\lessdot$ with $\gtrdot$.

\subsection{Chain Next Operators}

In Section~\ref{sec:fo-completeness}, we introduced operators
$\lganext{\prf}, \lgaback{\prf}$, with $\prf \in \{\lessdot, \doteq\}$,
which restrict their downward counterparts to a single PR.
Their semantics can be defined directly:
given an OP word $w$ and a position $i$, we have $(w,i) \models \lganext{\prf} \psi$
iff there exists a position $j > i$ such that $\chain(i,j)$ and $i \pr j$,
and $(w,j) \models \psi$.
Since they are needed for model-checking hierarchical operators,
we include them in the construction.
We also use them to model check downward/upward chain next and back operators.

If $\lcdnext \psi \in \clos{\varphi}$, we add
$\lganext{\lessdot} \psi, \lganext{\doteq} \psi \in \clos{\varphi}$,
and for each $\Phi \in \atoms{\varphi}^2$ we impose that $\lcdnext \psi \in \Phi_c$,
iff $\lganext{\lessdot} \psi \in \Phi_c$ or $\lganext{\doteq} \psi \in \Phi_c$.
To model check $\lcunext \psi$, we add the consistency constraint that,
for any $\Phi \in \atoms{\varphi}$,
$\lcunext \psi \in \Phi_c$ iff either $\lganext{\doteq} \psi \in \Phi_c$,
$\lganext{\gtrdot} \psi \in \Phi_c$, or both.

Moreover, we add into $\clos{\varphi}$ the auxiliary symbol $\chain_L$,
which forces the current position to be the first one of a chain body.
Let the current state of the OPA be $\Phi \in \atoms{\varphi}^2$: $\chain_L \in \Phi_p$
iff the next transition (i.e.\ the one reading the current position) is a push.
Formally, if $(\Phi, a, \Psi) \in \delta_\mathit{shift}$
or $(\Phi, \Theta, \Psi) \in \delta_\mathit{pop}$, for any $\Phi, \Theta, \Psi$ and $a$,
then $\chain_L \not\in \Phi_p$.
If $(\Phi, a, \Psi) \in \delta_\mathit{push}$, then $\chain_L \in \Phi_p$.
For any initial state $(\Phi_c, \Phi_p) \in I$,
we have $\chain_L \in \Phi_p$ iff $\# \not\in \Phi_c$.

If $\lganext{\doteq} \psi \in \clos{\varphi}$,
its satisfaction is ensured by the following constraints on $\delta$.
\begin{enumerate}
\item \label{rule:lganext-doteq-start}
  Let $(\Phi, a, \Psi) \in \delta_\mathit{push/shift}$:
  then $\lganext{\doteq} \psi \in \Phi_c$ iff $\lganext{\doteq} \psi, \chain_L \in \Psi_p$;
\item \label{rule:lganext-doteq-pop}
  let $(\Phi, \Theta, \Psi) \in \delta_\mathit{pop}$:
  then $\lganext{\doteq} \psi \not\in \Phi_p$,
  and $\lganext{\doteq} \psi \in \Theta_p$ iff $\lganext{\doteq} \psi \in \Psi_p$;
\item \label{rule:lganext-doteq-shift}
  let $(\Phi, a, \Psi) \in \delta_\mathit{shift}$:
  then $\lganext{\doteq} \psi \in \Phi_p$ iff $\psi \in \Phi_c$.
\end{enumerate}
\noindent If $\lganext{\lessdot} \psi \in \clos{\varphi}$,
$\lganext{\lessdot} \psi$ is allowed in the pending part of initial states,
and we add the following constraints.
\begin{enumerate}[resume]
\item \label{rule:lganext-lessdot-push-shift}
  Let $(\Phi, a, \Psi) \in \delta_\mathit{push/shift}$:
  then $\lganext{\lessdot} \psi \in \Phi_c$ iff $\lganext{\lessdot} \psi, \chain_L \in \Psi_p$;
\item \label{rule:lganext-lessdot-pop}
  let $(\Phi, \Theta, \Psi) \in \delta_\mathit{pop}$:
  then $\lganext{\lessdot} \psi \in \Theta_p$ iff $\chain_L \in \Psi_p$, and either
  \begin{enumerate*}
  \item
    $\lganext{\lessdot} \psi \in \Psi_p$ or
  \item
    $\psi \in \Phi_c$.
  \end{enumerate*}
\end{enumerate}
\noindent The rules for $\lganext{\gtrdot} \psi$ only differ in $\psi$
being enforced by a pop transition, triggered by the $\gtrdot$ relation between the
left and right contexts of the chain on which $\lganext{\gtrdot} \psi$ holds.
Thus, if $\lganext{\gtrdot} \psi \in \clos{\varphi}$ we have:
\begin{enumerate}[resume]
\item \label{rule:lganext-gtrdot-start}
  Let $(\Phi, a, \Psi) \in \delta_\mathit{push/shift}$:
  then $\lganext{\gtrdot} \psi \in \Phi_c$ iff $\lganext{\gtrdot} \psi, \chain_L \in \Psi_p$;
\item
  let $(\Phi, \Theta, \Psi) \in \delta_{pop}$:
  $\lganext{\gtrdot} \psi \in \Theta_p$ iff $\lganext{\gtrdot} \psi \in \Psi_p$,
  and $\lganext{\gtrdot} \psi \in \Phi_p$ iff $\psi \in  \Phi_c$;
\item \label{rule:lganext-gtrdot-shift}
  let $(\Phi, a, \Psi) \in \delta_\mathit{shift}$: then $\lganext{\gtrdot} \psi \not\in \Phi_p$.
\end{enumerate}

\begin{figure}[bt]
\centering
\begin{tabular}{| c | r | l | r | c | c |}
  \hline
  & input & state & stack & PR & move \\
  \hline
  1 & $\lcall \, \lhandle \, \lthrow \, \lret \, \#$
  & \(
    \begin{aligned}
      &\Phi^0_c = \{\lcall, \lcdnext \lret, \lganext{\doteq} \lret \}, \\
      &\Phi^0_p = \{\chain_L\}
    \end{aligned}
    \)
  & $\bot$
  & $\# \lessdot \lcall$
  & push \\
\hline
  2 & $\lhandle \, \lthrow \, \lret \, \#$
  & $\Phi^1 = (\{\lhandle\}, \{\lganext{\doteq} \lret, \chain_L\})$
  & $[\lcall, \Phi^0] \bot$
  & $\lcall \lessdot \lhandle$
  & push \\
\hline
  3 & $\lthrow \, \lret \, \#$
  & $\Phi^2 = (\{\lthrow\}, \emptyset)$
  & $[\lhandle, \Phi^1] [\lcall, \Phi^0] \bot$
  & $\lhandle \doteq \lthrow$
  & shift \\
\hline
  4 & $\lret \, \#$
  & $\Phi^3 = (\{\lret\}, \emptyset)$
  & $[\lthrow, \Phi^1] [\lcall, \Phi^0] \bot$
  & $\lthrow \gtrdot \lret$
  & pop \\
\hline
  5 & $\lret \, \#$
  & $\Phi^4 = (\{\lret\}, \{\lganext{\doteq} \lret\})$
  & $[\lcall, \Phi^0] \bot$
  & $\lcall \doteq \lret$
  & shift \\
\hline
  6 & $\#$
  & $\Phi^5 = (\{\#\}, \emptyset)$
  & $[\lret, \Phi^0] \bot$
  & $\lret \gtrdot \#$
  & pop \\
\hline
  7 & $\#$
  & $\Phi^5 = (\{\#\}, \emptyset)$
  & $\bot$
  & --
  & -- \\
\hline
\end{tabular}
\caption{Example accepting run of the automaton for $\lcdnext \lret$.}
\label{fig:mc-example-run}
\end{figure}

We illustrate how the construction works for $\lganext{\doteq}$
with the example of Fig.~\ref{fig:mc-example-run}.
The OPA starts in state $\Phi^0$, with $\lcdnext \lret \in \Phi^0_c$,
and guesses that $\lcdnext$ will be fulfilled by $\lganext{\doteq}$, so
$\lganext{\doteq} \lret \in \Phi^0_c$.
$\lcall$ is read by a push move, resulting in state $\Phi^1$.
The OPA guesses the next move will be a push, so $\chain_L \in \Phi^1_p$.
By rule \ref{rule:lganext-doteq-start}, we have $\lganext{\doteq} \lret \in \Phi^1_p$.
The last guess is immediately verified by the next push (step 2-3).
Thus, the pending obligation for $\lganext{\doteq} \lret$ is stored onto the stack in $\Phi^1$.
The OPA, then, reads $\lthrow$ with a shift,
and pops the stack symbol containing $\Phi^1$ (step 4-5).
By rule \ref{rule:lganext-doteq-pop}, the temporal obligation
is resumed in the next state $\Phi^4$,
so $\lganext{\doteq} \lret \in \Phi^4_p$.
Finally, $\lret$ is read by a shift which, by rule \ref{rule:lganext-doteq-shift},
may occur only if $\lret \in \Phi^4_c$.
Rule \ref{rule:lganext-doteq-shift} verifies the guess that
$\lganext{\doteq} \lret$ holds in $\Phi_0$,
and fulfills the temporal obligation contained in $\Phi^4_p$,
by preventing computations in which $\lret \not\in \Phi^4_c$ from continuing.
Had the next transition been a pop
(e.g.\ because there was no $\lret$ and $\lcall \gtrdot \#$),
the run would have been blocked by rule \ref{rule:lganext-doteq-pop},
preventing the OPA from reaching an accepting state, and from emptying the stack.

\begin{figure}
\centering
\begin{tabular}{m{0.4\textwidth} m{0.5\textwidth}}
\centering
\includegraphics[width=.5\textwidth]{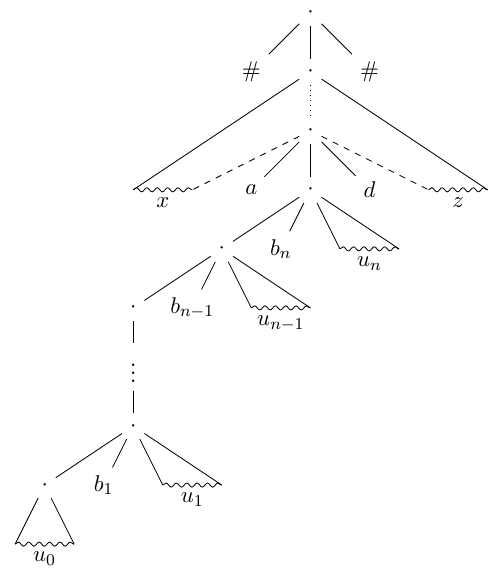}
&
\centering
\includegraphics[width=.5\textwidth]{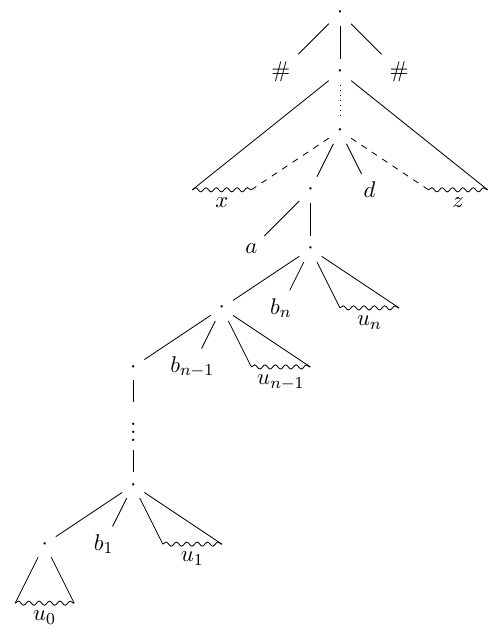}
\end{tabular}
\begin{tikzpicture}
\matrix (m) [matrix of math nodes, column sep=-4, row sep=-4]
{
  \#
  & \lessdot & x
  & \pi_x & a
  & \lessdot & u_0
  & \gtrdot & b_1
  & \pi_1 & u_1
  & \gtrdot & \dots
  & \gtrdot & b_{n-1}
  & \pi_{n-1} & u_{n-1}
  & \gtrdot & b_n
  & \pi_n & u_n
  & \gtrdot & d
  & \pi_z & z
  & \gtrdot & \# \\
  0
  & &
  & & i
  & &
  & & i_{b_1}
  & &
  & &
  & & i_{b_{n-1}}
  & &
  & & i_{b_n}
  & &
  & & j
  & &
  & & \\
};
\draw (m-1-1) to [out=30, in=150] (m-1-27);
\draw (m-1-5) to [out=30, in=150] node[xshift=8em, yshift=-0.9em] {\footnotesize $\doteq$/$\gtrdot$} (m-1-23);
\draw (m-1-5) to [out=30, in=150] node[xshift=7em, yshift=-1em] {\footnotesize $\lessdot$} (m-1-19);
\draw (m-1-5) to [out=30, in=150] node[xshift=4em, yshift=-0.3em] {\footnotesize $\lessdot$} (m-1-15);
\draw (m-1-5) to [out=30, in=150] node[xshift=1em, yshift=0.2em] {\footnotesize $\lessdot$} (m-1-9);
\end{tikzpicture}
\caption{The two possible STs of a generic OP word $w = xyz$ (top),
  and its flat representation with chains (bottom).
  Wavy lines are placeholders for subtree frontiers.
  We have either $a \doteq d$ (top left), or $a \gtrdot d$ (top right).
  In both trees, $a \lessdot b_k$ for $1 \leq k \leq n$,
  and the corresponding word positions are in the chain relation.
  For $1 \leq k \leq n$, $u_k$ is the word generated by the right part
  of the rhs whose first terminal is $b_k$.
  So, either $^{b_k}[u_k]^{b_{k+1}}$, or $u_k$ is of the form
  $v^k_0 c^k_0 v^k_1 c^k_1 \dots c^k_{m_k} v^k_{m_k+1}$,
  where $c^k_p \doteq c^k_{p+1}$ for $0 \leq p < m_k$,
  $b_k \doteq c^k_0$, and resp. $c^k_{m_k} \gtrdot b_{k+1}$
  and $c^n_{m_n} \gtrdot d$ (cf.\ Figure~\ref{fig:uk-structure}).
  Moreover, for each $0 \leq p < m_k$, either $v^k_{p+1} = \varepsilon$ or
  $^{c^k_p}[v^k_{p+1}]^{c^k_{p+1}}$;
  either $v^k_0 = \varepsilon$ or $^{b_k}[v^k_0]^{c^k_0}$,
  and either $v^k_{m_k+1} = \varepsilon$ or $^{c^k_{m_k}}[v^k_{m_k+1}]^{b_{k+1}}$
  (resp. $^{c^n_{m_n}}[v^n_{m_n+1}]^{d}$).
  $u_0$ has this latter form, except $v^0_0 = \varepsilon$ and $a \lessdot c^0_0$.
  In the bottom representation, the $\pi_k$s are placeholders for precedence relations,
  that depend on the surrounding characters.}
\label{fig:proof-general-word}
\end{figure}
\begin{figure}
\centering
\includegraphics{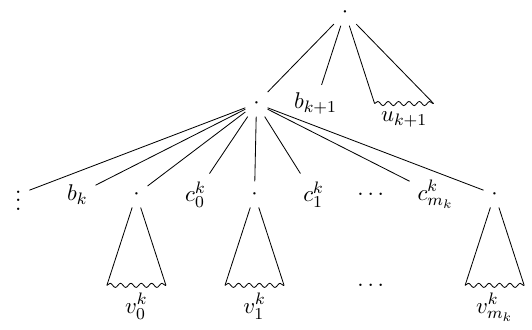}
\begin{tikzpicture}
\matrix (m) [matrix of math nodes, column sep=-4, row sep=-4]
{
  \dots
  & \gtrdot & b_k
  & \lessdot & v^k_0
  & \gtrdot & c^k_0
  & \lessdot & v^k_1
  & \gtrdot & c^k_1
  & \dots & c^k_{m_k}
  & \lessdot & v^k_{m_k}
  & \gtrdot & b_{k+1}
  & \odot_{k+1} & u_{k+1} \\
  & & i_{b_k}
  & &
  & &
  & &
  & &
  & &
  & &
  & & i_{b_{k+1}}
  & & \\
};
\draw (m-1-3) to [out=30, in=150] node[yshift=0.5em] {\footnotesize $\doteq$} (m-1-7);
\draw (m-1-7) to [out=30, in=150] node[yshift=0.5em] {\footnotesize $\doteq$} (m-1-11);
\draw (m-1-13) to [out=30, in=150] node[xshift=-1em, yshift=0.2em] {\footnotesize $\gtrdot$} (m-1-17);
\draw (-4.5,1) to [out=0, in=120] node[below] {\footnotesize $\lessdot$} (m-1-3);
\draw (-4.5,1.2) to [out=0, in=150] node[below] {\footnotesize $\lessdot$} (m-1-17);
\end{tikzpicture}
\caption{The structure of $u_k$ in the word of Fig.~\ref{fig:proof-general-word}.}
\label{fig:uk-structure}
\end{figure}

We now prove the correctness of this construction.
For each operator, we show that in all accepting computations it appears
in an OPA state iff it holds in the corresponding word position.
While doing so, we assume that the construction is correct for the operands of each operator.
This allows us to prove the correctness of the whole construction
inductively on the formula's structure, in Section~\ref{subsec:mc-proof}.
In the following, we we denote as $\first(w)$ the first position of a word $w$.
We also use Figure~\ref{fig:uk-structure},
which represents the generic structure of any composed chain.

\begin{lem}
\label{thm:mc-lganext-doteq}
Given a finite set of atomic propositions $AP$, an OP alphabet $(\powset{AP}, M_{AP})$,
a word $w = \# x y z \#$ on it, and a position $i = |x| + 1$ in $w$, we have
\[
(w,i) \models \lganext{\doteq} \psi
\]
if and only if all accepting computations of an OPA satisfying rules
\ref{rule:lganext-doteq-start}-\ref{rule:lganext-doteq-shift}
bring it from a configuration
$\config{y z}{\Phi}{\alpha \gamma}$ with $\lganext{\doteq} \psi \in \Phi_c$
to a configuration $\config{z}{\Phi'}{\alpha' \gamma}$
such that $\lganext{\doteq} \psi \not\in \Phi'_p$, $|\alpha| = 1$ and
$|\alpha'| = 1$ if $\first(y)$ is read by a shift move,
$|\alpha'| = 2$ if it is read by a push move.
If $\lganext{\doteq} \psi \not\in \Phi_p$ and it is not in the pending part
of the state in $\alpha$, then it is not in the pending parts of states in $\alpha'$.
If no other rules constrain the transition relation, at least one computation is accepting.
\end{lem}
\begin{proof}
In the following, we denote by $\Phi^a$ the state of the automaton before reading symbol $a$,
so $\Phi^a \cap AP = a$, for any $a \subseteq AP$.

$[\Rightarrow]$
Suppose $\lganext{\doteq} \psi$ holds in position $i$,
corresponding to terminal symbol $a$.
In all accepting computations, the OPA reaches configuration
$\config{a \dots z}{\Phi^a}{[f, \Phi^f] \gamma}$, where $\alpha = [f, \Phi^f]$,
and guesses that $\lganext{\doteq} \psi$ holds in $i$,
so $\lganext{\doteq} \psi \in \Phi^a_c$.
We show later in the proof that all accepting computations must make this guess.
$a$ is read by either a push or a shift transition,
leading the OPA to configuration $\config{c^0_0 \dots z}{\Phi^{c^0_0}}{\delta}$,
with either $\delta = [a, \Phi^a] [f, \Phi^f] \gamma$ or $\delta = [a, \Phi^f] \gamma$, respectively.
Moreover, $\lganext{\doteq} \psi \in \Phi^{c^0_0}_p$ and $\chain_L \in \Phi^{c^0_0}_p$
due to rule~\eqref{rule:lganext-doteq-start}.
Since $\lganext{\doteq} \psi$ holds in $i$, $a$ is the left context of a chain,
so the next transition is a push, satisfying the requirement for $\chain_L$.
This also means $w$ has the form of Fig.~\ref{fig:proof-general-word},
possibly with $n = 0$ (cf.\ the caption for notation).
Any accepting computation must go through the support for this chain.
The next configuration is $\config{v^0_1 \dots z}{\Phi^{v^0_1}}{[c^0_0, \Phi^{c^0_0}] \delta}$,
with $\lganext{\doteq} \psi \in \Phi^{c^0_0}_p$.
Then, the computation goes on normally.
Note that, when reading an inner chain body such as $v^0_1$,
the automaton does not touch the stack symbol containing $\Phi^{c^0_0}$,
and other symbols in the body of the same simple chain, i.e.\ $c^0_1, c^0_2 \dots$,
are read with shift moves that update the topmost stack symbol with the new terminal,
leaving state $\Phi^{c^0_0}$ untouched.

If $a$ is the left context of more than one chain (i.e.\ $n > 0$ in the figure),
the OPA then reaches configuration $\config{b_1 \dots z}{\Phi^{b_1}}{[c^0_{m_0}, \Phi^{c^0_0}] \delta}$.
Since $c^0_{m_0} \gtrdot b_1$, the next transition is a pop.
$\lganext{\doteq} \psi \in \Phi^{c^0_0}$, so by rule \eqref{rule:lganext-doteq-pop},
the automaton reaches configuration $\config{b_1 \dots z}{\Phi^{\prime b_1}}{\delta}$
with $\lganext{\doteq} \psi \in \Phi^{\prime b_1}_p$.
Then, since $a$ is contained in the topmost stack symbol and $a \lessdot b_1$,
the next move is a push, leading to
$\config{v^1_0 \dots z}{\Phi^{v^1_0}}{[b_1, \Phi^{\prime b_1}] \delta}$.
Notice how $\lganext{\doteq} \psi$ is again stored as a pending obligation
in the topmost stack symbol.
The OPA run goes on in the same way for each terminal $b_p$, $1 \leq p \leq n$,
until the automaton reaches configuration $\config{d \dots z}{\Phi^d}{[c^n_{m_n}, \Phi^{b_n}] \delta}$
with $\lganext{\doteq} \psi \in \Phi^{b_n}_p$.
If $a$ was the left context of one chain only,
this is the configuration reached after reading the body of such chain, with $n = 0$.
Since $c^n_{m_n} \gtrdot d$, a pop transition leads to
$\config{d \dots z}{\Phi^{\prime d}}{\delta}$, with $\lganext{\doteq} \psi \in \Phi^{\prime d}_p$,
by rule~\eqref{rule:lganext-doteq-pop}.
Note that there exists a computation in which $\lganext{\doteq} \psi \not\in \Phi^d_p$,
so rule \eqref{rule:lganext-doteq-pop} applies.
The fact that a computation with $\lganext{\doteq} \psi \not\in \Phi^d_p$
is blocked by rule \eqref{rule:lganext-doteq-pop} is correct,
because this implies $\lganext{\doteq} \psi$ holds in the position preceding $d$.
This would be wrong, because such a position is in the $\gtrdot$ relation with $d$,
and it cannot be the left context of a chain, so $\lganext{\doteq} \psi$ must be false in it.
Then, if $\lganext{\doteq} \psi$ holds in $i$, since $a$ is the terminal in the topmost stack symbol,
we must have $a \doteq d$. So $d$ is read by a shift move, leading to
$\config{z}{\Phi^z}{\alpha' \gamma}$
with $\alpha' = [d, \Phi^a][f, \Phi^f]$ or $\alpha' = [d, \Phi^f]$,
depending on which kind of move previously read $a$.
Note that if $\lganext{\doteq} \psi \not\in \Phi^a_p, \Phi^f_p$,
the claim about states in $\alpha'$ is satisfied.
Since $\lganext{\doteq} \psi$ holds in $i$, $\psi$ holds in $j$ (the position corresponding to $d$),
and $\psi \in \Phi^{\prime d}_c$, because we assume the correctness of the construction
for all other operators.
This satisfies rule \eqref{rule:lganext-doteq-shift},
and verifies the initial guess that $\lganext{\doteq} \psi$ holds in $i$.
By rule \eqref{rule:lganext-doteq-shift}, any computation in which $\psi$ holds in $j$
must have $\lganext{\doteq} \psi \in \Phi^{\prime d}_p$,
which is only the case if the OPA makes such initial guess.
Finally, there exists a computation in which $\lganext{\doteq} \psi \not\in \Phi^z$,
satisfying the thesis statement.
Note that all computations of this form may then proceed normally until acceptance,
if they are not blocked by rules other than
\ref{rule:lganext-doteq-start}-\ref{rule:lganext-doteq-shift}.

$[\Leftarrow]$
Suppose an accepting computation starts from configuration
$\config{a \dots z}{\Phi^a}{[f, \Phi^f] \gamma}$,
with $\lganext{\doteq} \psi \in \Phi^a_c$, $\alpha = [f, \Phi^f]$, and $f \lessdot a$
(the case $f \doteq a$ is analogous).
$a$ is read by a push move in this case, which leads the OPA to configuration
$\config{c^0_0 \dots z}{\Phi^{c^0_0}}{[a, \Phi^a][f, \Phi^f] \gamma}$,
with $\lganext{\doteq} \psi, \chain_L \in \Phi^{c^0_0}_p$.
Since $\chain_L \in \Phi^{c^0_0}_p$, the next transition must be a push,
so $a \lessdot c^0_0$, $a$ is the left context of a chain and $w$ is of the form
of Fig.~\ref{fig:proof-general-word}.
The push move brings the OPA to configuration
$\config{v^0_0 \dots z}{\Phi^{v^0_0}}{[c^0_0, \Phi^{c^0_0}][a, \Phi^a][f, \Phi^f] \gamma}$.
Notice that the stack size is now $|\gamma| + 3$.
By the thesis, the automaton eventually reaches a configuration
in which the stack size is $|\gamma| + 2$.
This can be achieved if $[c^0_0, \Phi^{c^0_0}]$ is popped,
so $\alpha' = [a, \Phi^a][f, \Phi^f]$.
In a generic word such as the one of Fig.~\ref{fig:proof-general-word},
this happens only before reading $b_p$, $1 \leq i \leq n$, or $d$.

In both cases, let $[c^k_{m_k}, \Phi^{b_k}]$ be the popped stack symbol.
We have $\lganext{\doteq} \psi \in \Phi^{b_k}_p$.
By rule \eqref{rule:lganext-doteq-pop}, if $\Phi'$ is the destination state of the pop move,
$\lganext{\doteq} \psi \in \Phi'_p$, which does not satisfy the thesis statement.
If the next move is a push (such as when reading any $b_p$, $1 \leq p \leq n$),
the stack length increases again, which also does not satisfy the thesis.
If the next move is a pop, rule \eqref{rule:lganext-doteq-pop} blocks the computation.
So, the next move must be a shift, updating symbol $[a, \Phi^a]$ to $[d, \Phi^a]$,
where $d$ is the just-read terminal symbol.
This means the OPA reached the right context of the chain whose left context is $i$
(i.e.\ $a$), and the two positions are in the $\doteq$ relation.
By rule \eqref{rule:lganext-doteq-shift}, $\psi$ is part of the starting state of this move,
so $\psi$ holds in this position, satisfying $\lganext{\doteq} \psi$ in $i$.
The state resulting from the shift move may not contain $\lganext{\doteq} \psi$
as a pending obligation, thus satisfying the thesis.
\end{proof}

The proof for $\lganext{\gtrdot}$ is very similar to Lemma~\ref{thm:mc-lganext-doteq},
and is therefore omitted.

\begin{lem}
\label{thm:mc-lganext-lessdot}
Given a finite set of atomic propositions $AP$, an OP alphabet $(\powset{AP}, M_{AP})$,
a word $w = \# x y z \#$ on it, and a position $i = |x| + 1$ in $w$, we have
\[
(w,i) \models \lganext{\lessdot} \psi
\]
if and only if all accepting computations of an OPA satisfying rules
\ref{rule:lganext-lessdot-push-shift}-\ref{rule:lganext-lessdot-pop}
bring it from configuration
$\config{y z}{\Phi}{\alpha \gamma}$ with $\lganext{\lessdot} \psi \in \Phi_c$
to a configuration $\config{z}{\Phi'}{\alpha' \gamma}$
such that $\lganext{\lessdot} \psi \not\in \Phi'_p$, $|\alpha| = 1$ and
$|\alpha'| = 1$ if $\first(y)$ is read by a shift move,
$|\alpha'| = 2$ if it is read by a push move.
If $\lganext{\lessdot} \psi \not\in \Phi_p$ and it is not in the pending part
of the state in $\alpha$, then it is not in the pending parts of states in $\alpha'$.
If no other rules constrain the transition relation, at least one computation is accepting.
\end{lem}
\begin{proof}
  $[\Rightarrow]$
  Suppose $\lganext{\lessdot} \psi$ holds in position $i$,
  corresponding to terminal $a$.
  Then, $a$ must be the left context of more than one chain
  (by property \ref{item:upward-prop} of the $\chain$ relation),
  and the word being read is of the form of Fig.~\ref{fig:proof-general-word}, with $n \geq 1$.
  Let us call $b_p$, $1 \leq p \leq n$, the right contexts of those of these chains
  that are s.t.\ $a \lessdot b_p$ (i.e., all except the rightmost context of $i$).
  There exists an index $q$, $1 \leq q \leq n$,
  such that $\psi$ holds in $i_{b_q}$, the word position labeled with $b_q$.
  All accepting computations reach a configuration
  $\config{a \dots z}{\Phi^a}{[f, \Phi^f] \gamma}$, where $\alpha = [f, \Phi^f]$,
  and $\lganext{\lessdot} \psi \in \Phi^a_c$, because the OPA guesses that
  $\lganext{\lessdot} \psi$ holds in $i$.
  $a$ is read by a shift or a push transition, which leads the OPA to configuration
  $\config{c^0_0 \dots z}{\Phi^{c^0_0}}{\delta}$, with $\delta = \alpha' \gamma$,
  and either $\alpha' = [a, \Phi^a] [f, \Phi^f]$ or $\alpha' = [a, \Phi^f]$, respectively.
  The claim on the pending part of states in $\alpha'$ is trivially satisfied.
  Due to rule~\eqref{rule:lganext-lessdot-push-shift},
  we have $\lganext{\lessdot} \psi \in \Phi^{c^0_0}_p$ and $\chain_L \in \Phi^{c^0_0}_p$.
  As a result, the next move must be a push, consistently with the hypothesis implying
  $a$ is the left context of a chain.
  Then, starting with $c^0_0$, the OPA reads the body of the innermost chain whose left context is $a$,
  until it reaches its right context $b_1$.
  In this process, the topmost stack symbol $[c^0_0, \Phi^{c^0_0}]$ may be updated by shift transitions
  reading other terminals $c^0_p$, $1 \leq p \leq m_0$,
  that are part of the same simple chain as $c^0_0$.
  However, it is never popped until $b_1$ is reached,
  since subchains cause the OPA to only push, pop and update new stack symbols,
  but not existing ones.
  So, the OPA reaches configuration
  $\config{b_1 \dots z}{\Phi^{b_1}}{[c^0_{m_0}, \Phi^{c^0_0}] \delta}$,
  with $\lganext{\lessdot} \psi \in \Phi^{c^0_0}_p$.

  Suppose $q \neq 1$, so $\psi$ does not hold in $b_1$.
  Since $c^0_{m_0} \gtrdot b_1$, the next transition is a pop.
  Due to rule~\eqref{rule:lganext-lessdot-pop}, it leads the OPA to configuration
  $\config{b_1 \dots z}{\Phi^{\prime b_1}}{\delta}$
  with $\lganext{\lessdot} \psi \in \Phi^{\prime b_1}_p$ and $\chain_L \in \Phi^{\prime b_1}_p$.
  The presence of $\chain_L$ implies the next move is a push,
  a requirement that is satisfied because $a \lessdot b_1$.
  So, the OPA transitions to configuration
  $\config{v^1_0 \dots z}{\Phi^{v^1_0}}{[b_1, \Phi^{\prime b_1}] \delta}$.
  The computation, then, goes on in the same way for each $b_p$, $1 \leq p < q$.
  Before $b_q$ is read, (and possibly $q = 1$), the OPA is in configuration
  $\config{b_q \dots z}{\Phi^{b_q}}{[c^{q-1}_{m_q - 1}, \Phi^{b_{q-1}}] \delta}$,
  with $\lganext{\lessdot} \psi \in \Phi^{b_{q-1}}_p$.
  Since $c^{q-1}_{m_q - 1} \gtrdot b_q$, a pop transition brings the OPA to
  $\config{b_q \dots z}{\Phi^{\prime b_q}}{\delta}$.
  Since by hypothesis $\psi \in \Phi^{b_q}_c$, by rule~\eqref{rule:lganext-lessdot-pop}
  we just have $\chain_L \in \Phi^{\prime b_q}_p$, and the initial guess is verified.
  Since the topmost stack symbol contains $a$, and $a \lessdot b_q$,
  the next transition is a push, which satisfies the requirement of $\chain_L$.
  Note that $\lganext{\lessdot} \psi \not\in \Phi^{\prime b_q}_p$,
  and the current stack is $\delta$, which satisfies the thesis statement,
  also ensuring that a computation of this form may be finally accepting.

  $[\Leftarrow]$
  Suppose that during an accepting computation the OPA reaches configuration
  $\config{a \dots z}{\Phi^a}{[f, \Phi^f] \gamma}$,
  with $\lganext{\lessdot} \psi \in \Phi^a_c$.
  Again, $a$ must be read by either a push or a shift move.
  Since $\chain_L$ is inserted as a pending requirement into the state resulting from this move,
  the next transition must be a push, so $a$ is the left context of at least a chain.
  This chain has the form of Fig.~\ref{fig:proof-general-word}.
  By rule~\eqref{rule:lganext-lessdot-push-shift},
  the OPA reaches configuration $\config{c^0_0 \dots z}{\Phi^{c^0_0}}{\delta}$,
  with $\lganext{\lessdot} \psi, \chain_L \in \Phi^{c^0_0}_p$,
  and $\delta$ as in the $[\Rightarrow]$ part after reading $a$.
  Let $[c^0_0, \Phi^{c^0_0}]$ be the stack symbol pushed with $c^0_0$.
  The stack size at this time is greater by one w.r.t.\ what is required by the thesis statement,
  so $[c^0_0, \Phi^{c^0_0}]$ must be popped.

  This happens when the OPA reaches a symbol $e$ s.t.\ the terminal symbol in the topmost stack
  symbol takes precedence from $e$.
  $e$ must be s.t.\ $a \lessdot e$ (and $e = b_1$ in Fig.~\ref{fig:proof-general-word}).
  Otherwise, suppose by contradiction that $a \gtrdot e$ or $a \doteq e$
  (so $e = d$ in Fig.~\ref{fig:proof-general-word}, in which $n = 0$ and $c^0_{m_0}$ precedes $d$).
  In this case, after popping $[c^0_{m_0}, \Phi^{c^0_0}]$, the automaton reaches configuration
  $\config{d z}{\Phi^{\prime d}}{\delta'}$.
  Since $\lganext{\lessdot} \psi \in \Phi^{c^0_0}_p$, by rule~\eqref{rule:lganext-lessdot-pop}
  we have $\lganext{\lessdot} \psi \in \Phi^{\prime d}_p$,
  so this configuration does not satisfy the thesis statement.
  Moreover, $\chain_L \in \Phi^{\prime d}_p$, which requires the next transition to be a push.
  But $a \doteq d$ or $a \gtrdot d$, and $a$ is the topmost stack symbol,
  so such a computation is blocked by $\chain_L$, never reaching a configuration
  complying with the thesis statement.

  So, $e = b_1$, and the OPA reaches configuration
  $\config{b_1 \dots z}{\Phi^{b_1}}{[c^0_{m_0}, \Phi^{c^0_0}] \delta}$.
  The subsequent pop move leads to $\config{b_1 \dots z}{\Phi^{\prime b_1}}{\delta}$.
  Suppose $\psi \in \Phi^{b_1}_c$.
  Then, by rule~\eqref{rule:lganext-lessdot-pop} we only have $\chain_L \in \Phi^{\prime b_1}_p$,
  and $\lganext{\lessdot} \psi \not\in \Phi^{\prime b_1}_p$.
  This configuration satisfies the thesis statement,
  and since $a \lessdot b_1$, $a$ and $b_1$ are the context of a chain, and $\psi$ holds in $b_1$,
  we can conclude that $\lganext{\lessdot} \psi$ holds in $a$.

  Otherwise, if $\psi \not\in \Phi^{b_1}_c$, by rule~\eqref{rule:lganext-lessdot-pop}
  we have $\lganext{\lessdot} \psi, \chain_L \in \Phi^{\prime b_1}_p$.
  The next transition will therefore push the symbol $[b_1, \Phi^{\prime b_1}]$ onto the stack,
  again with $\lganext{\lessdot} \psi$ as a pending obligation in it.
  Then, the same reasoning done with $[c^0_0, \Phi^{c^0_0}]$ (and its subsequent updates)
  can be repeated.
  The only way the thesis statement can be satisfied is by reading a position $b_q$,
  s.t.\ $a \lessdot b_q$, the terminal in the topmost stack symbol takes precedence from $b_q$
  (so $a$ and $b_q$ are the context of a chain), and $\psi \in \Phi^{b_q}_c$,
  so $\psi$ holds in $b_q$.
  This implies $\lganext{\lessdot} \psi$ holds in $a$.
\end{proof}

\subsection{Chain Back Operators}

We now give the construction for the chain back operators, and their proofs.

To model check the $\lcdback \psi$ and $\lcuback \psi$ operators,
we employ the auxiliary operator $\lgaback{\prf} \psi$,
with $\prf \in \{\lessdot, \doteq, \gtrdot\}$.
Given an OP word $w$ and a position $i$ in it,
we have $(w,i) \models \lgaback{\prf} \psi$ iff there exists a position $j < i$
such that $\chain(j,i)$ and $j \pr i$, and $(w,j) \models \psi$.
For any $\Phi \in \atoms{\varphi}$, we have
$\lcdback \psi \in \Phi_c$ iff either $\lgaback{\doteq} \psi \in \Phi_c$,
$\lgaback{\lessdot} \psi \in \Phi_c$, or both;
$\lcuback \psi \in \Phi_c$ iff either $\lgaback{\doteq} \psi \in \Phi_c$,
$\lgaback{\gtrdot} \psi \in \Phi_c$, or both.

We add symbol $\chain_R$,
which lets the computation go on only if the previous transition was a pop,
and the position associated with the current state is the right context of a chain.
So, for any $(\Phi, a, \Psi) \in \delta_\mathit{push/shift}$,
we have $\chain_R \not\in \Psi_p$;
for any $(\Phi, \Theta, \Psi) \in \delta_\mathit{pop}$,
we have $\chain_R \in \Psi_p$.
$\chain_R$ is allowed in the pending part of final states.

If $\lgaback{\doteq} \psi \in \clos{\varphi}$,
we add the following constraints on the transition relation.
\begin{enumerate}[resume]
\item \label{rule:lgaback-doteq-shift}
  Let $(\Phi, a, \Psi) \in \delta_\mathit{shift}$:
  then $\lgaback{\doteq} \psi \in \Phi_c$ iff $\lgaback{\doteq} \psi, \chain_R \in \Phi_p$;
\item \label{rule:lgaback-doteq-push}
  let $(\Phi, a, \Psi) \in \delta_\mathit{push}$:
  then $\lgaback{\doteq} \psi \not\in \Phi_c$;
\item \label{rule:lgaback-doteq-pop}
  let $(\Phi, \Theta, \Psi) \in \delta_\mathit{pop}$:
  then $\lgaback{\doteq} \psi \in \Psi_p$ iff $\lgaback{\doteq} \psi \in \Theta_p$;
\item \label{rule:lgaback-doteq-push-shift}
  let $(\Phi, a, \Psi) \in \delta_\mathit{push/shift}$:
  then $\lgaback{\doteq} \psi \in \Psi_p$ iff $\psi \in \Phi_c$.
\end{enumerate}

\noindent The constraints added if $\lgaback{\lessdot} \psi \in \clos{\varphi}$ now follow.
\begin{enumerate}[resume]
\item
  Let $(\Phi, a, \Psi) \in \delta_\mathit{push}$:
  then $\lgaback{\lessdot} \psi \in \Phi_c$ iff $\lgaback{\lessdot} \psi, \chain_R \in \Phi_p$;
\item
  let $(\Phi, a, \Psi) \in \delta_\mathit{shift}$:
  then $\lgaback{\lessdot} \psi \not\in \Phi_c$;
\item
  let $(\Phi, \Theta, \Psi) \in \delta_\mathit{pop}$:
  then $\lgaback{\lessdot} \psi \in \Psi_p$ iff $\lgaback{\lessdot} \psi \in \Theta_p$;
\item
  let $(\Phi, a, \Psi) \in \delta_\mathit{push/shift}$:
  then $\lgaback{\lessdot} \psi \in \Psi_p$ iff $\psi \in \Phi_c$.
\end{enumerate}

\noindent Finally, when $\lgaback{\gtrdot} \psi \in \clos{\varphi}$,
we add symbol $\chain_{\doteq}$, which appears in a state iff the next transition
will be a shift: for any $(\Phi, a, \Psi) \in \delta_\mathit{push}$
and $(\Phi, \Theta, \Psi) \in \delta_\mathit{pop}$, $\chain_{\doteq} \not\in \Phi_p$,
and for any $(\Phi, a, \Psi) \in \delta_\mathit{shift}$, $\chain_{\doteq} \in \Phi_p$.
$\lgaback{\gtrdot} \psi$ and $\chain_{\doteq}$ are allowed in the pending part of final states.
We also add the constraints below. \\
Let $(\Phi, a, \Psi) \in \delta_\mathit{push/shift}$:
\begin{enumerate}[resume]
\item \label{rule:lgaback-gtrdot-push-shift-stop}
  $\lgaback{\gtrdot} \psi \not\in \Psi_p$;
\item \label{rule:lgaback-gtrdot-push-shift-end}
  $\lgaback{\gtrdot} \psi \in \Phi_c$ iff $\lgaback{\gtrdot} \psi, \chain_R \in \Phi_p$;
\end{enumerate}
let $(\Phi, \Theta, \Psi) \in \delta_\mathit{pop}$:
\begin{enumerate}[resume]
\item \label{rule:lgaback-gtrdot-pop-end}
  if ($\chain_L \in \Psi_p$ or $\chain_{\doteq} \in \Psi_p$),
  then $\lgaback{\gtrdot} \psi \in \Psi_p$ iff $\lgaback{\gtrdot} \psi \in \Phi_p$;
\item \label{rule:lgaback-gtrdot-pop-prop}
  if $\chain_L, \chain_{\doteq} \not\in \Psi_p$, then
  $\lgaback{\gtrdot} \psi \in \Psi_p$ iff either
  $\lgaback{\lessdot} \psi \lor \ldback \psi \in \Theta_c$
  or $\lgaback{\gtrdot} \psi \in \Phi_p$.
\end{enumerate}

We proceed by proving the correctness of the construction for each operator,
as we did for their future counterparts.

\begin{lem}
\label{thm:mc-lgaback-doteq}
Given a finite set of atomic propositions $AP$, an OP alphabet $(\powset{AP}, M_{AP})$,
a word $w = \# x y z \#$ on it, and a position $j = |x y|$ in $w$, we have
\[
(w,j) \models \lgaback{\doteq} \psi
\]
if and only if all accepting computations of an OPA satisfying rules
\ref{rule:lgaback-doteq-shift}-\ref{rule:lgaback-doteq-push-shift}
bring it from configuration
$\config{y z}{\Phi}{\alpha \gamma}$ to a configuration $\config{z}{\Phi'}{\alpha' \gamma}$
such that $|\alpha| = 1$,
$|\alpha'| = 1$ if $\first(y)$ is read by a shift move,
$|\alpha'| = 2$ if it is read by a push move,
and $\lgaback{\doteq} \psi \in \Phi^j_c$, where $\Phi^j$ is the state of the OPA before reading $j$,
the last position of $y$.
If no other rules constrain the transition relation, at least one computation is accepting.
\end{lem}
\begin{proof}
$[\Rightarrow]$
Suppose $\lgaback{\doteq} \psi$ holds in position $i$, corresponding to terminal symbol $a$.
Then, there exists a position $j$, labeled with terminal $d$,
s.t.\ $\chain(i,j)$, $a \doteq d$, and $\psi$ holds in $i$.
Since $a$ and $d$ are the context of a chain,
the input word must have the form of Fig.~\ref{fig:proof-general-word}.
All accepting computations of the OPA reach configuration
$\config{a \dots z}{\Phi^a}{[f, \Phi^f] \gamma}$ before reading $a$.
By the inductive assumption, we have $\psi \in \Phi^a$.
$a$ is read by a shift or a push move, bringing the OPA to
$\config{c^0_0 \dots z}{\Phi^{c^0_0}}{\delta}$, with $\delta = \alpha' \gamma$,
and either $\alpha' = [a, \Phi^a] [f, \Phi^f]$ or $\alpha' = [a, \Phi^f]$, respectively.
Due to rule~\eqref{rule:lgaback-doteq-push-shift},
we have $\lgaback{\doteq} \psi \in \Phi^{c^0_0}_p$.
After reading $c^0_0$, the OPA reaches configuration
$\config{v^0_0 \dots z}{\Phi^{v^0_0}}{[c^0_0, \Phi^{c^0_0}] \delta}$.
Then, the automaton proceeds to read the rest of the body of chain $\chain(i,j)$.
If $i$ is the left context of multiple chains, the stack symbol $[c^0_0, \Phi^{c^0_0}]$,
containing $\lgaback{\doteq} \psi$ as a pending obligation, is popped before reaching $d$.
Let $b_p$, $1 \leq p \leq n$, be all labels of positions $i_{b_p}$
s.t.\ $\chain(i, i_{b_p})$ and $a \lessdot b_p$.
It can be proved inductively that, before reading any of such positions, the OPA is in a configuration
$\config{b_p \dots z}{\Phi^{b_p}}{[c^{p-1}_{m_{p-1}}, \Phi^{b_{p-1}}] \delta}$,
with $\lgaback{\doteq} \psi \in \Phi^{b_{p-1}}$.
Since $c^{p-1}_{m_{p-1}} \gtrdot b_p$, the next move is a pop, leading to a configuration
$\config{b_p \dots z}{\Phi^{\prime b_p}}{\delta}$, with $\lgaback{\doteq} \psi \in \Phi^{\prime b_p}_p$,
due to rule~\eqref{rule:lgaback-doteq-pop}.
Then, $b_p$ is read by a push move because $a \lessdot b_p$, so $\lgaback{\doteq} \psi$ is again
stored in the topmost stack symbol as a pending obligation,
in a configuration $\config{v^p_0 \dots z}{\Phi^{v^p_1}}{[b_p, \Phi^{\prime b_p}] \delta}$.
The stack symbol containing $\lgaback{\doteq} \psi$ is only popped in positions $b_p$,
or when reaching $d$, since subchains only cause the OPA to push and pop new symbols.

So, configuration $\config{d z}{\Phi^d}{[c^n_{m_n}, \Phi^{\prime b_n}] \delta}$ is reached,
with $\lgaback{\doteq} \psi \in \Phi^{\prime b_n}$ (note that $d$ labels the last position of $y$).
Due to rule~\eqref{rule:lgaback-doteq-pop}, a pop move leads the OPA to
$\config{d z}{\Phi^{\prime d}}{\delta}$, with $\lgaback{\doteq} \psi \in \Phi^{\prime d}_p$.
Then, since by hypothesis $a \doteq d$, and $a$ is contained in the topmost stack symbol,
$d$ is read by a shift move.
Since this transition is preceded by a pop,
we have a computation in which $\chain_R \in \Phi^{\prime d}_p$.
So, by rule~\eqref{rule:lgaback-doteq-shift},
since $\lgaback{\doteq} \psi, \chain_R \in \Phi^{\prime d}_p$,
we have $\lgaback{\doteq} \psi \in \Phi^{\prime d}_c$,
with the stack equal to $\delta$, satisfying the thesis statement.
Computations of this form can proceed until acceptance,
if not blocked by rules other than \ref{rule:lganext-doteq-start}-\ref{rule:lganext-doteq-shift}.

$[\Leftarrow]$
Suppose that, while reading $w$, an accepting computation of the OPA arrives at a configuration
$\config{d z}{\Phi^{\prime d}}{\delta}$, where $d$ is the last character of $y$,
and $\lgaback{\doteq} \psi \in \Phi^{\prime d}_c$.
By rule~\eqref{rule:lgaback-doteq-shift},
we have $\lgaback{\doteq} \psi, \chain_R \in \Phi^{\prime d}_p$.
$\chain_R \in \Phi^d_p$ requires the previous transition to be a pop,
so $d$ is the right context of a chain. Let $a$ be its left context.
By hypothesis, the computation proceeds reading $d$, and by rule~\eqref{rule:lgaback-doteq-push}
it must be read by a shift transition. So, we have $a \doteq d$,
and $w$ must be of the form of Fig.~\ref{fig:proof-general-word}.
Going back to $\config{d z}{\Phi^{\prime d}}{\delta}$,
consider the pop move leading to this configuration.
It starts from configuration $\config{d z}{\Phi^d}{[c^n_{m_n}, \Phi^{b_n}] \delta}$,
and by rule~\eqref{rule:lgaback-doteq-pop} we have $\lgaback{\doteq} \psi \in \Phi^{b_n}_p$.

Consider the move that pushed $\Phi^{b_n}$ onto the stack.
Suppose it was preceded by a pop move.
Since $\Phi^{b_n}$ is the target state of this transition,
and $\lgaback{\doteq} \psi \in \Phi^{b_n}_p$,
by rule~\eqref{rule:lgaback-doteq-pop} $\lgaback{\doteq} \psi$ must be contained
as a pending obligation in the popped state as well.
So, this obligation is propagated backwards every time the automaton encounters a position that
is the left context of a chain, i.e.\ positions $b_p$, $1 \leq p \leq n$,
in Fig.~\ref{fig:proof-general-word}.
In order to stop the propagation, a push of a state with $\lgaback{\doteq} \psi$
as a pending obligation, preceded by another push or shift move must be encountered.
Such a transition pushes or updates the stack symbol under the one containing $\lgaback{\doteq} \psi$,
which means the left context $a$ s.t.\ $a \doteq d$ of a chain whose right context is $d$
has been reached. In both cases, the target state of the push/shift transitions contains
$\lgaback{\doteq} \psi$ as a pending obligation, so by rule~\eqref{rule:lgaback-doteq-push-shift}
we have $\psi \in \Phi^a_c$.
Hence, by the inductive assumption, $\psi$ holds in position $i$ (corresponding to $a$),
we have $i \doteq j$ and $\chain(i,j)$, which implies $\lgaback{\doteq} \psi$ holds in $j$.
\end{proof}

The proof of the model checking rules of $\lgaback{\lessdot} \psi$
is similar to the one of Lemma~\ref{thm:mc-lgaback-doteq}, and is therefore omitted.

\begin{figure}
\centering
\begin{tabular}{m{0.4\textwidth} m{0.5\textwidth}}
\centering
\includegraphics[width=.5\textwidth]{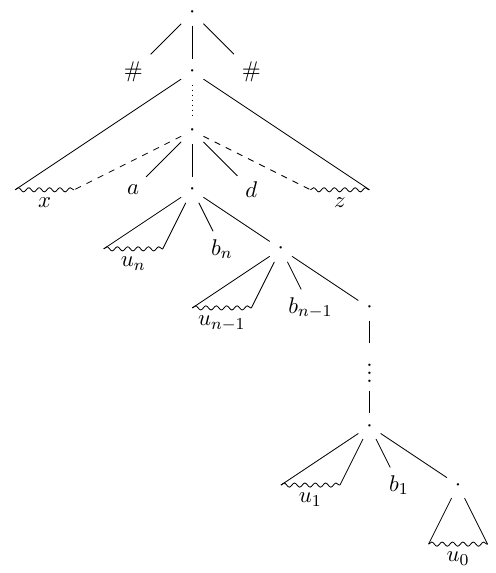}
&
\centering
\includegraphics[width=.5\textwidth]{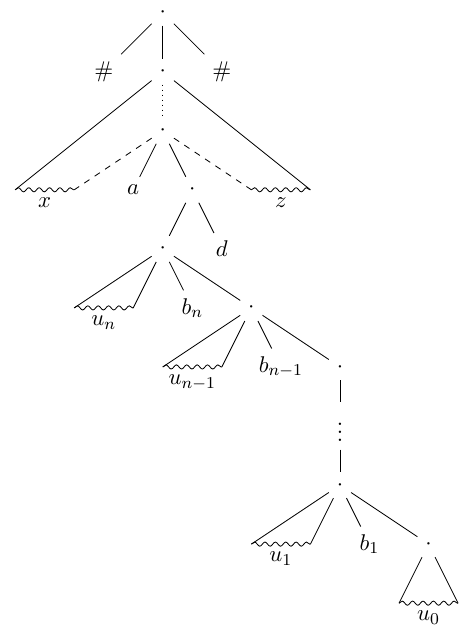}
\end{tabular}
\begin{tikzpicture}
\matrix (m) [matrix of math nodes, column sep=-4, row sep=-4]
{
  \#
  & \lessdot & x
  & \pi_x & a
  & \lessdot & u_n
  & \pi_n & b_n
  & \lessdot & u_{n-1}
  & \pi_{n-1} & b_{n-1}
  & \lessdot & \dots
  & \lessdot & u_1
  & \pi_1 & b_1
  & \lessdot & u_0
  & \gtrdot & d
  & \pi_z & z
  & \gtrdot & \# \\
  0
  & &
  & & i
  & &
  & & i_{b_n}
  & &
  & & i_{b_{n-1}}
  & &
  & &
  & & i_{b_1}
  & &
  & & j
  & &
  & & \\
};
\draw (m-1-1) to [out=30, in=150] (m-1-27);
\draw (m-1-5) to [out=30, in=150] node[xshift=-8em, yshift=-1em] {\footnotesize $\lessdot$/$\doteq$} (m-1-23);
\draw (m-1-9) to [out=30, in=150] node[xshift=-7em, yshift=-1em] {\footnotesize $\gtrdot$} (m-1-23);
\draw (m-1-13) to [out=30, in=150] node[xshift=-4em, yshift=-0.3em] {\footnotesize $\gtrdot$} (m-1-23);
\draw (m-1-19) to [out=30, in=150] node[xshift=-1em, yshift=0.2em] {\footnotesize $\gtrdot$} (m-1-23);
\end{tikzpicture}
\caption{The two possible STs of a generic OP word $w = xyz$ (top)
  expanded on the rightmost non-terminal, and its flat representation with chains (bottom).
  Wavy lines are placeholders for subtree frontiers.
  We have either $a \doteq d$ (top left) or $a \lessdot d$ (top right),
  and $b_k \gtrdot d$ for $1 \leq k \leq n$.
  For $1 \leq k \leq n$, we either have $^{b_{k+1}}[u_k]^{b_k}$,
  or $u_k$ is of the form
  $v^k_0 c^k_0 v^k_1 c^k_1 \dots c^k_{m_k} v^k_{m_k+1}$,
  where $c^k_p \doteq c^k_{p+1}$ for $0 \leq p < m_k$,
  $c^k_{m_k} \doteq b_k$, and resp. $a \lessdot c^n_0$
  and $b_{k+1} \lessdot c^k_0$.
  Moreover, for each $0 \leq p < m_k$, either $v^k_{p+1} = \varepsilon$ or
  $^{c^k_p}[v^k_{p+1}]^{c^k_{p+1}}$;
  either $v^k_{m_k+1} = \varepsilon$ or $^{c^k_{m_k}}[v^k_{m_k+1}]^{b_k}$,
  and either $v^k_0 = \varepsilon$ or $^{b_{k+1}}[v^k_0]^{c^k_0}$
  (resp. $^{a}[v^n_0]^{c^n_0}$).
  $u_0$ has the same form, except $v^0_{m_0} = \varepsilon$ and
  $c^0_{m_0} \gtrdot d$.
  The $\pi_i$s are placeholders for precedence relations,
  and they vary depending on the surrounding terminal characters.
}
\label{fig:proof-general-word-right}
\end{figure}

\begin{lem}
\label{thm:mc-lgaback-gtrdot}
Given a finite set of atomic propositions $AP$, an OP alphabet $(\powset{AP}, M_{AP})$,
a word $w = \# x y z \#$ on it, and a position $j = |x y|$ in $w$, we have
\[
(w,j) \models \lgaback{\gtrdot} \psi
\]
if and only if all accepting computations of an OPA satisfying rules
\ref{rule:lgaback-gtrdot-push-shift-stop}-\ref{rule:lgaback-gtrdot-pop-prop}
bring it from configuration
$\config{y z}{\Phi}{\alpha \gamma}$ to a configuration $\config{z}{\Phi'}{\alpha' \gamma}$
such that $|\alpha| = 1$,
$|\alpha'| = 1$ if $\first(y)$ is read by a shift move,
$|\alpha'| = 2$ if it is read by a push move,
and $\lgaback{\gtrdot} \psi \in \Phi^j_c$, where $\Phi^j$ is the state of the OPA before reading $j$,
the last position of $y$.
If no other rules constrain the transition relation, at least one computation is accepting.
\end{lem}
\begin{proof}
$[\Rightarrow]$
Suppose $\lgaback{\gtrdot} \psi$ holds in position $j$.
Then, $j$ is the right context of at least two chains,
and the word $w$ has the form of Fig.~\ref{fig:proof-general-word-right},
with $i$ being the left context of the outermost chain whose right context is $j$.
Let positions $i_{b_p}$, labeled with $b_p$, $1 \leq p \leq n$,
be all other left contexts of chains sharing $j$ as their right context.
There exists a value $q$, $ \leq q \leq n$, s.t.\ $\psi$ holds in $i_{b_q}$.

During an accepting run, the OPA reads $w$ normally, until it reaches $b_q$, with configuration
\[
\config{b_q \dots z}{\Phi^{b_q}}{[c^q_{m_q}, \Phi^{c^q_0}] [b_{q+1}, \Phi^{c^{q+1}_0}] \dots \delta},
\]
with $\psi \in \Phi^{b_q}_c$,  $\delta = \alpha' \gamma$,
and either $\alpha' = [a, \Phi^a] [f, \Phi^f]$, if $a$ (the label of $i$) was read by a push move,
or $\alpha' = [a, \Phi^f]$ if it was read by a shift.
Note that if $b_q$ is the only character in its simple chain body
($u_q = \varepsilon$ in Fig.~\ref{fig:proof-general-word-right})
$[c^q_{m_q}, \Phi^{c^q_0}]$ is not present on the stack.
In this case, $b_q$ is read by a push move instead of a shift.
Suppose $b_q$ is the left context of one or more chains,
besides the one whose right context is $j$.
In Fig.~\ref{fig:proof-general-word-right}, this means $v^{q-1}_0 \neq \varepsilon$.
Consider the right context of the outermost of such chains:
w.l.o.g.\ we call it $c^{q-1}_0$ (it may as well be $b_{q-1}$).
Since $\psi$ holds in $i_{b_q}$, $\lgaback{\gtrdot} \psi$ holds in $c^{q-1}_0$.
If, instead, $v^{q-1}_0 = \varepsilon$, then $c^{q-1}_0$ is the successor of $b_q$,
and $\lgback{\lessdot} \psi$ holds in it.
In both cases, $\lgaback{\lessdot} \psi \lor \lgback{\lessdot} \psi$ holds in $c^{q-1}_0$.
Since $b_q \lessdot c^{n-1}_0$, the latter is read by a push transition,
pushing stack symbol $[c^{q-1}_0, \Phi^{c^{q-1}_0}]$,
with $\lgaback{\lessdot} \psi \lor \lgback{\lessdot} \psi \in \Phi^{c^{q-1}_0}_c$.
This symbol remains on stack until $d$ is reached,
although its terminal symbol may be updated.
The computation then proceeds normally, until configuration
$\config{d z}{\Phi^{(q-2) d}}{[b_{q-1}, \Phi^{c^{q-1}_0}] \dots \delta}$ is reached.

Since $\lgaback{\lessdot} \psi \lor \lgback{\lessdot} \psi \in \Phi^{c^{q-1}_0}_c$,
by rule~\eqref{rule:lgaback-gtrdot-pop-prop},
the OPA transitions to configuration $\config{d z}{\Phi^{(q) d}}{[b_q, \Phi^{c^q_0}] \dots \delta}$
with $\lgaback{\gtrdot} \psi \in \Phi^{(q) d}_p$ and $\chain_L, \chain_{\doteq} \not\in \Phi^{(q) d}_p$.
(Note that the next transition must be a pop, since the topmost stack symbol is
$b_q$, and $b_q \gtrdot d$.)
Then, by rule~\eqref{rule:lgaback-gtrdot-pop-prop}, all subsequent pop transitions propagate
$\lgaback{\gtrdot} \psi$ as a pending obligation in the OPA state, until configuration
$\config{d z}{\Phi^{(n-1) d}}{\delta}$, with $\lgaback{\gtrdot} \psi \in \Phi^{(n-1) d}_p$.
Now, the automaton guesses that this is the last pop move, and the next one will be a push or a shift.
So, it transitions to $\config{d z}{\Phi^{(n) d}}{\delta}$,
with and $\chain_L \in \Phi^{(n) d}_p$ or $\chain_{\doteq} \in \Phi^{(n) d}_p$,
and $\lgaback{\gtrdot} \psi \in \Phi^{(n) d}_p$,
according to rule~\eqref{rule:lgaback-gtrdot-pop-end}.
Also, $\chain_R \in \Phi^{(n) d}_p$, because the previous move was a pop.
At this point, $d$ is read with either a shift or a push transition.
According to rule~\eqref{rule:lgaback-gtrdot-push-shift-end},
$\lgaback{\gtrdot} \psi \in \Phi^{(n) d}_c$,
which satisfies the thesis statement.

$[\Leftarrow]$
Suppose the automaton reaches a state $\Phi^j = \Phi^{(n) d}$
s.t.\ $\lgaback{\gtrdot} \psi \in \Phi^j_c$ during an accepting computation.
$j$ has to be read by either a push or a shift move,
so either $\chain_L \in \Phi^j_p$ or $\chain_{\doteq} \in \Phi^j_p$.
By rule~\eqref{rule:lgaback-gtrdot-push-shift-end}, for the computation to continue,
we have $\chain_R \in \Phi^j_p$.
So, the transition leading to state $\Phi^j_p$ must be a pop,
and the related word position $d$ is the right context of a chain.
Let $\Phi^{\prime j}$ be the starting state of this transition.
Since $\lgaback{\gtrdot} \psi \in \Phi^j_p$,
by rule~\eqref{rule:lgaback-gtrdot-pop-end} we have $\lgaback{\gtrdot} \psi \in \Phi^{\prime j}_p$.
By rule~\eqref{rule:lgaback-gtrdot-push-shift-stop}, this transition must be preceded by another pop,
so $d$ is the right context of at least two chains, and the word being read is of the form
of Fig.~\ref{fig:proof-general-word-right}, with $n \geq 1$.

So, before reading $d$, the OPA performs a pop transition for each inner chain having $d$
as a right context, i.e.\ those having $b_p$, $1 \leq p \leq n$, as left contexts
in Fig.~\ref{fig:proof-general-word-right}, plus one for the outermost chain
(whose left context is $a$).
By rule~\eqref{rule:lgaback-gtrdot-pop-prop}, $\lgaback{\gtrdot} \psi$ is propagated backwards
through such transitions from the one before $d$ is read, to one in which
$\lgaback{\lessdot} \psi \lor \lgback{\lessdot} \psi$ is contained into the popped state.

By rule~\eqref{rule:lgaback-gtrdot-push-shift-stop}, for the computation to reach such pop
transitions, the propagation of $\lgaback{\gtrdot} \psi$ as a pending obligation must stop.
So, the OPA must reach a configuration $\config{d z}{\Phi^{(q) d}}{[b_q, \Phi^{c^q_0}] \dots \delta}$
with $\lgaback{\lessdot} \psi \lor \lgback{\lessdot} \psi \in \Phi^{c^q_0}_c$.
Note that the following reasoning also applies to the case in which,
in Fig.~\ref{fig:proof-general-word-right}, $u_q = \varepsilon$,
by substituting $b_q$ to $c^q_0$.
The topmost stack symbol was pushed after configuration
$\config{c^q_0 \dots z}{\Phi^{c^q_0}}{[b_{q-1}, \Phi^{c^{q-1}_0}] \dots \delta}$.
We have $b_{q-1} \lessdot c^q_0$.
If $v^q_0 = \varepsilon$, and $c^q_0$ is in the position next to $b_{q-1}$,
$\lgback{\lessdot} \psi$ holds, while if $v^q_0 \neq \varepsilon$,
since $\ochain{b_{q-1}}{v^q_0}{c^q_0}$ is a chain, $\lgaback{\lessdot} \psi$ holds.
Therefore, $\psi$ holds in $b_{q-1}$.
Since $b_{q-1} \gtrdot d$ and $\chain(i_{b_{q-1}}, j)$, $\lgaback{\gtrdot} \psi$ holds in $j$.
\end{proof}

\subsection{Summary Until and Since}
The construction for these operators is based on their expansion laws.
The rules for until follow, those of since being symmetric.
For any $\Phi \in \atoms{\varphi}^2$, we have $\luntil{t}{\psi}{\theta} \in \Phi_c$,
with $t \in \{d, u\}$ being a direction, iff either:
\begin{enumerate*}
\item $\theta \in \Phi_c$,
\item $\lgnext{t}(\luntil{t}{\psi}{\theta}), \psi \in \Phi_c$, or
\item $\lganext{t}(\luntil{t}{\psi}{\theta}), \psi \in \Phi_c$.
\end{enumerate*}

\subsection{Hierarchical Operators}
\label{subsec:mc-hierarchical}

For the hierarchical operators, we do not give an explicit OPA construction,
but we rely on a translation into other POTL operands.
For each hierarchical operator $\eta$ in $\varphi$, we add a propositional symbol $\fprop{\eta}$.
The upward hierarchical operators consider the right contexts of chains
sharing the same left context.
To distinguish such positions, we define formula
\(
  \gamma_{L,\eta} :=
    \lgaback{\lessdot} \big(\fprop{\eta}
      \land \lnext (\llglob \neg \fprop{\eta})
      \land \lback (\llpglob \neg \fprop{\eta})\big),
\)
where $\llglob \psi := \neg (\lcuuntil{\top}{(\lcduntil{\top}{\neg \psi})})$,
and $\llpglob$ is symmetric.
$\lnext$ and $\lback$ are the LTL next and back operators,
for which model checking can be done as for $\ldnext$ and $\ldback$,
but removing the restrictions on PR.
They could be replaced with $\lnext \psi := \ldnext \psi \lor \lunext \psi$,
but this would cause an exponential blowup in the following equivalences,
which can be used for model checking upwards hierarchical operators.
$\gamma_{L,\eta}$, evaluated in one of the right contexts,
asserts that $\fprop{\eta}$ holds in the unique left context of the same chain, only.
\begin{align}
  \lhunext \psi &:=
    \gamma_{L, \lhunext \psi} \land
    \lnext \big(
      \lcuuntil
        {(\neg \lgaback{\lessdot} \fprop{\lhunext \psi})}
        {(\lgaback{\lessdot} \fprop{\lhunext \psi} \land \psi)} \big) \\
  \lhuback \psi &:=
    \gamma_{L, \lhuback \psi} \land
    \lback \big(
      \lcusince
        {(\neg \lgaback{\lessdot} \fprop{\lhuback \psi})}
        {(\lgaback{\lessdot} \fprop{\lhuback \psi} \land \psi)} \big) \\
  \lhuuntil{\psi}{\theta} &:=
    \gamma_{L, \lhuuntil{\psi}{\theta}} \land
      \lcuuntil
        {(\lgaback{\lessdot} \fprop{\lhuuntil{\psi}{\theta}} \implies \psi)}
        {(\lgaback{\lessdot} \fprop{\lhuuntil{\psi}{\theta}} \land \theta)} \label{eq:lhyuntil-mc} \\
  \lhusince{\psi}{\theta} &:=
    \gamma_{L, \lhusince{\psi}{\theta}} \land
      \lcusince
        {(\lgaback{\lessdot} \fprop{\lhusince{\psi}{\theta}} \implies \psi)}
        {(\lgaback{\lessdot} \fprop{\lhusince{\psi}{\theta}} \land \theta)}
\end{align}

We only prove equivalence~\eqref{eq:lhyuntil-mc}, as the others are essentially analogous.
\begin{lem}[Equivalence \eqref{eq:lhyuntil-mc}]
\label{lemma:equivalence-lhyuntil-mc}
Let $w$ be an OP word based on an alphabet of atomic propositions $\powset{AP}$,
and $i$ a position in $w$, and let $\fprop{\lhuuntil{\psi}{\theta}} \not\in AP$,
$\psi$ and $\theta$ being two POTL formulas on $AP$.
Let $w'$ be a word on alphabet $\powset{AP \cup \{\fprop{\lhuuntil{\psi}{\theta}}\}}$
identical to $w$, except $\fprop{\lhuuntil{\psi}{\theta}}$ holds
in position $h < i$ s.t.\ $\chain(h,i)$ and $h \lessdot i$.

Then, $(w,i) \models \lhuuntil{\psi}{\theta}$ iff
$(w',i) \models \Upsilon(\psi,\theta)$, with
\begin{align*}
  \Upsilon(\psi,\theta) &:=
  \gamma_{L, \lhuuntil{\psi}{\theta}} \land
      \lcuuntil
        {(\lgaback{\lessdot} \fprop{\lhuuntil{\psi}{\theta}} \implies \psi)}
        {(\lgaback{\lessdot} \fprop{\lhuuntil{\psi}{\theta}} \land \theta)}, \\
  \gamma_{L,\eta} &:=
    \lgaback{\lessdot} \big(\fprop{\eta}
      \land \lnext(\llglob \neg \fprop{\eta})
      \land \lback(\llpglob \neg \fprop{\eta})\big).
\end{align*}
\end{lem}
\begin{proof}
$[\Rightarrow]$
Suppose $\lhuuntil{\psi}{\theta}$ holds in position $i$ in word $w$.
Then, by its semantics, there exists a UHP $i = i_0 < i_1 < \dots < i_n$, with $n \geq 0$,
and a position $h < i$ s.t.\ for each $i_p$, $0 \leq p \leq n$,
we have $\chain(h, i_p)$ and $h \lessdot i_p$, and for $0 \leq q < n$, $\psi$ holds in $i_q$,
and $\theta$ holds in $i_n$.
We show that in $\Upsilon(\psi,\theta)$ holds in $i$ in $w'$.
By construction, in $w'$, $\fprop{\lhuuntil{\psi}{\theta}}$ holds in $h$ only.
So, $\fprop{\lhuuntil{\psi}{\theta}}
      \land \lnext(\llglob \neg \fprop{\lhuuntil{\psi}{\theta}})
      \land \lback(\llpglob \neg \fprop{\lhuuntil{\psi}{\theta}})$
holds in $h$, and $\gamma_{L, \lhuuntil{\psi}{\theta}}$ holds in $i$.

For $\lcuuntil
        {(\lgaback{\lessdot} \fprop{\lhuuntil{\psi}{\theta}} \implies \psi)}
        {(\lgaback{\lessdot} \fprop{\lhuuntil{\psi}{\theta}} \land \theta)}$
to hold in $i$, there must exist a USP between $i = i_0$ and $i_n$.
Suppose, by contradiction, that no such path exists.
This implies there exist two positions $r, s$, with $i \leq r < s \leq i_n$,
$r \lessdot s$, either $s = r+1$ or $\chain(r,s)$, s.t.\ no USP can skip them.
So, there exist no positions $r', s'$ s.t.\
$i \leq r' < s < s' \leq i_n$ s.t.\ $\chain(r',s')$ and either $r' \doteq s'$ or $r' \gtrdot s'$.
Since $r \lessdot s$, $r$ is the left context of a chain.
Let $k$ be the maximal (i.e.\ rightmost) position s.t.\ $\chain(r,k)$.
There are three cases:
\begin{itemize}
\item $k > i_n$.
  In this case, $i_n$ is part of the body of the chain $\chain(r,k)$.
  However, by hypothesis, $\chain(h, i_n)$, and $h < i \leq r < i_n < k$.
  These two chains cross each other, which is impossible by the definition of chain.
\item $k = i_n$.
  If $r \doteq i_n$ or $r \gtrdot i_n$, then $i_n$ is reachable by the USP.
  Otherwise, we would have $\chain(h, i_n)$ and $\chain(r, i_n)$,
  $h \lessdot i_n$ and $r \lessdot i_n$ with $h \neq r$,
  which is impossible because of property \eqref{item:chain-prop-3-proof}
  of Lemma~\ref{lemma:chain-prop}.
\item $k < i_n$.
  If $r \doteq k$ or $r \gtrdot k$, then $r$ and $k$ can be part of an USP reaching $i_n$.
  If $r \lessdot k$, then $k$ is the first position of the body of another chain
  having $r$ as its left context, which contradicts the assumption that $k$ is maximal.
\end{itemize}

By hypothesis, $\fprop{\lhuuntil{\psi}{\theta}}$ holds in $h$,
so $\lgaback{\lessdot} \fprop{\lhuuntil{\psi}{\theta}}$ holds in all positions $i_p$,
$0 \leq p \leq n$, in the UHP.
Since $\theta$ holds in $i_n$, $\lgaback{\lessdot} \fprop{\lhuuntil{\psi}{\theta}} \land \theta$
holds in it.
Moreover, since $\psi$ holds in all $i_q$, $0 \leq q \leq n$,
$\lgaback{\lessdot} \fprop{\lhuuntil{\psi}{\theta}} \implies \psi$ holds in all positions
in the USP between $i_0$ and $i_n$.

$[\Leftarrow]$
Suppose $(w',i) \models \Upsilon(\psi,\theta)$.
Then, $\gamma_{L, \lhuuntil{\psi}{\theta}}$ holds in $i$.
This implies there exists a position $h$ s.t.\ $\chain(h,i)$ and $h \lessdot i$,
which is unique by Lemma~\ref{lemma:chain-prop}.
By $\gamma_{L, \lhuuntil{\psi}{\theta}}$, $\fprop{\lhuuntil{\psi}{\theta}}$ holds in $h$
and in no other position.
Moreover, $\lcuuntil
        {(\lgaback{\lessdot} \fprop{\lhuuntil{\psi}{\theta}} \implies \psi)}
        {(\lgaback{\lessdot} \fprop{\lhuuntil{\psi}{\theta}} \land \theta)}$
holds in $i$, so there exists an USP $i = j_0 < j_1 < \dots < j_m$.
We show that there exists a sequence of indices $0 = p_0 < p_1 < \dots < p_n = m$
s.t.\ $j_{p_0}, j_{p_1}, \dots, j_{p_n}$ is a UHP satisfying $\lhuuntil{\psi}{\theta}$ in $i$ in $w$.

First, note that $\theta$ holds in $j_{p_n}$, and since $h$ is the only position in which
$\fprop{\lhuuntil{\psi}{\theta}}$ holds, we have $\chain(h, j_{p_n})$ and $h \lessdot j_{p_n}$.
So, $j_{p_n}$ is the last position of a UHP starting in $i$.
For each position $j$ s.t.\ $i < j < j_{p_n}$, $\chain(h, j)$ and $h \lessdot j$,
there exists an index $1 \leq q \leq n-1$ s.t.\ $j_{p_q} = j$.
Since all such positions $j$ are between $j_0$ and $j_m$,
the USP could skip them only if they were part of the body of a chain,
i.e.\ if there exist two positions $j_0 \leq r < s \leq j_m$ s.t.\ $\chain(j_0, j_m)$
and either $r \doteq s$ or $r \gtrdot s$.
Such a chain would, however, cross with $\chain(h, j)$, which contradicts the definition of chain.

Because $\fprop{\lhuuntil{\psi}{\theta}}$ only holds in $h$,
the fact that $\lgaback{\lessdot} \fprop{\lhuuntil{\psi}{\theta}} \implies \psi$
holds in all positions $j_0, j_1, \dots, j_{m-1}$ implies $\psi$ holds in all of
$j_{p_0}, j_{p_1}, \dots, j_{p_{n-1}}$.
So, by construction of $w'$, $j_{p_0}, j_{p_1}, \dots, j_{p_n}$ is a UHP satisfying
$\lhuuntil{\psi}{\theta}$ in position $i$ in $w$.
\end{proof}

We now give the equivalences for downward hierarchical operators.
The following formula, when evaluated in the left context of a chain,
forces symbol $\mathrm{p}_\eta$ in the right context.
Note that if the left context is in the $\gtrdot$ relation with the right one,
the latter is uniquely identified.
\[
  \gamma_{R,\eta} :=
    \lganext{\gtrdot} \big(\fprop{\eta}
      \land \lnext(\llglob \neg \fprop{\eta})
      \land \lback(\llpglob \neg \fprop{\eta})\big)
\]
We give the following equivalences for the take precedence hierarchical operators.
\begin{align}
  \lhdnext \psi &:=
    \gamma_{R, \lhdnext \psi} \land
    \lnext \big(
      \lcduntil
        {(\neg \lganext{\gtrdot} \fprop{\lhdnext \psi})}
        {(\lganext{\gtrdot} \fprop{\lhdnext \psi} \land \psi)} \big) \\
  \lhdback \psi &:=
    \gamma_{R, \lhdback \psi} \land
    \lback \big(
      \lcdsince
        {(\neg \lganext{\gtrdot} \fprop{\lhdback \psi})}
        {(\lganext{\gtrdot} \fprop{\lhdback \psi} \land \psi)} \big) \\
  \lhduntil{\psi}{\theta} &:=
    \gamma_{R, \lhduntil{\psi}{\theta}} \land
      \lcduntil
        {(\lganext{\gtrdot} \fprop{\lhduntil{\psi}{\theta}} \implies \psi)}
        {(\lganext{\gtrdot} \fprop{\lhduntil{\psi}{\theta}} \land \theta)} \\
  \lhdsince{\psi}{\theta} &:=
    \gamma_{R, \lhdsince{\psi}{\theta}} \land
      \lcdsince
        {(\lganext{\gtrdot} \fprop{\lhdsince{\psi}{\theta}} \implies \psi)}
        {(\lganext{\gtrdot} \fprop{\lhdsince{\psi}{\theta}} \land \theta)}
\end{align}

\subsection{Concluding Proof}
\label{subsec:mc-proof}

\begin{thm}[Correctness of Finite Model Checking.]
\label{thm:finite-mc}
Given a finite set of atomic propositions $AP$,
an OP alphabet $(\powset{AP}, \allowbreak M_{AP})$,
a word $w$ on it, and an POTL formula $\varphi$,
the automaton built according to the procedure in this section is such that we have
\[
  (w,1) \models \varphi
\]
if and only if it performs at least one accepting computation on a word $w'$ equal to $w$,
except for the presence of one more propositional symbol for each hierarchical operator in $\varphi$.
\end{thm}
\begin{proof}
We proved that all chain next/back operators hold in a position in $w$
iff in all accepting computations, after reading a subword of $w$,
the OPA is left in a state not containing any pending obligation
related to that instance of the operator
(cf.\ Lemmas~\ref{thm:mc-lganext-doteq}, \ref{thm:mc-lganext-lessdot},
\ref{thm:mc-lgaback-doteq}, \ref{thm:mc-lgaback-gtrdot}).
While the correctness of the upward/downward next/back operators is trivial,
that of summary until/since operators is due to the correctness of the respective expansion laws,
proved in Lemma~\ref{lemma:expansion-law-su}.
Moreover, in Lemma~\ref{lemma:equivalence-lhyuntil-mc} we proved the correctness of the equivalences
for the hierarchical operators.

The results above allow us to prove that, by structural induction on the syntax of $\varphi$,
if $\varphi$ holds in position 1 of $w$,
there exists a word $w'$ identical to $w$, except for the propositional symbols
needed for the hierarchical operators, such that the OPA performs at least a computation
reaching the end of $w$ in a state containing no future operators and no temporal obligations.
By the definition of the set of final states $F$, such a computation is accepting.

Conversely, suppose there exists a word $w'$ with the described features on which
the OPA performs at least one accepting computation starting from a state containing $\varphi$.
Then $\varphi$ holds in the first position of a word $w$ built by removing
the propositional symbols introduced by equivalence formulas for hierarchical operators.
Indeed, such a computation ends with an empty stack, and a state containing no future operators
or temporal obligations which, by the lemmas listed above,
implies all temporal obligations have been satisfied, and $w$ is a model for $\varphi$.
\end{proof}

\noindent \textbf{Complexity.}
The set $\clos{\varphi}$ is linear in $|\varphi|$, the length of $\varphi$.
$\atoms{\varphi}$ has size at most $2^{|\clos{\varphi}|} = 2^{O(|\varphi|)}$,
and the size of the set of states is the square of that.
Moreover, the use of the equivalences for the hierarchical operators causes
only a linear increase in the length of $\varphi$.
Therefore,
\begin{thm}
Given a POTL formula $\varphi$,
it is possible to build an OPA $\mathcal{A}_\varphi$ accepting the language denoted by $\varphi$
with at most $2^{O(|\varphi|)}$ states.
\end{thm}
$\mathcal{A}_\varphi$ can then be intersected~\cite{LonatiEtAl2015}
with an OPA modeling a program (e.g.\ Fig.~\ref{fig:example-prog}),
and emptiness can be decided with \emph{summarization}
techniques~\cite{AlurBE18}.

\section{Conclusions}
\label{sec:conclusion}

We introduced the temporal logic POTL, proved its FO-completeness, and
gave an automata-theoretic model checking procedure.
We argue that the
strong gain in expressive power w.r.t.\ previous approaches to model
checking CFL brought by POTL is worth the technicalities needed to
achieve the present --and future-- results.
The next natural research step is the extension of such results to $\omega$-words,
which, for model checking, may follow the approach sketched in \cite{ChiariMP20a} for OPTL,
and may be done with composition arguments for FO-completeness.
Whether POTL is strictly
more expressive than OPTL remains an open problem, although we
conjecture OPTL is not FO-complete.  A direct explanation of the
completeness result of Corollary~\ref{cor:complete-subset} also
remains to be given.

We already implemented the OPA construction of Section~\ref{sec:mc} in
a prototype model checking tool, which is showing promising results.
We also plan to develop user-friendly domain-specific languages,
to prove that OP languages and logics are suitable in practice to program verification.

\bibliographystyle{abbrv}
\bibliography{optl}

\clearpage
\appendix

\section{Omitted Proofs: Semantics of POTL}
\label{subsec:expansion-proofs}

In the following Lemma, we prove a few properties of the chain relation.

\begin{lem}[Properties of the $\chain$ relation.]
\label{lemma:chain-prop}
Given an OP word $w$ and positions $i,j,h,k$ in it, the following properties hold.
\begin{enumerate}
\item \label{item:chain-prop-1-proof}
  If $\chain(i,j)$ and $\chain(h,k)$,
  then we have $i < h < j \implies k \leq j$
  and $i < k < j \implies i \leq h$.
\item If $\chain(i,j)$, then $i \lessdot i+1$ and $j-1 \gtrdot j$.
\item \label{item:chain-prop-3-proof}
  Given $j$, there exists at most one single position $i$ s.t.\ $\chain(i,j)$
  and $i \lessdot j$ or $i \doteq j$;
  for any $i'$ s.t.\ $\chain(i',j)$ and $i' \gtrdot j$ we have $i' > i$.
\item Given $j$, there exists at most one single position $j$ s.t.\ $\chain(i,j)$
  and $i \gtrdot j$ or $i \doteq j$;
  for any $j'$ s.t.\ $\chain(i,j')$ and $i \lessdot j'$ we have $j' < j$.
\end{enumerate}
\end{lem}
\begin{proof}
In the following, we denote by $c_p$ the character labeling word position $p$,
and by writing $\ochain{c_{-1}}{x_0 c_0 x_1 \dots x_n c_n x_{n+1}}{c_{n+1}}$ we imply
$c_{-1}$ and $c_{n+1}$ are the context of a simple or composed chain,
in which either $x_p = \varepsilon$, or $\ochain{c_{p-1}}{x_{p}}{c_p}$ is a chain, for each $p$.
\begin{enumerate}
\item Suppose, by contradiction, that $\chain(i,j)$, $\chain(h,k)$, and $i < h < j$, but $k > j$.
  Consider the case in which $\chain(i,j)$ is the innermost chain whose body contains $h$,
  so it is of the form $\ochain{c_i}{x_0 c_0 \dots c_h x_p c_p \dots c_n x_{n+1}}{c_j}$
  or $\ochain{c_i}{x_0 c_0 \dots c_h x_{n+1}}{c_j}$.
  By the definition of chain, we have either $c_h \doteq c_p$ or $c_h \gtrdot c_j$, respectively.

  Since $\chain(h,k)$, this chain must be of the form
  $\ochain{c_h}{x_p c_p \dots }{c_k}$ or $\ochain{c_h}{x_{n+1} c_j \dots }{c_k}$,
  implying $c_h \lessdot c_p$ or $c_h \lessdot c_j$, respectively.
  This means there is a conflict in the OPM, contradicting the hypothesis that $w$ is an OP word.

  In case $\chain(i,j)$ is not the innermost chain whose body contains $h$,
  we can reach the same contradiction by inductively considering the chain between $i$ and $j$
  containing $h$ in its body.
  Moreover, it is possible to reach a symmetric contradiction with the hypothesis
  $\chain(i,j)$, $\chain(h,k)$, and $i < k < j$, but $i > h$.

\item Trivially follows from the definition of chain.

\item Suppose, by contradiction, there exists a position $h \neq i$, and w.l.o.g., $h < i$,
s.t.\ $\chain(h,j)$ and $h \lessdot j$.
Since $i \lessdot j$, by the definition of chain, $j$ must be part of the body
of another composed chain whose left context is $i$.
So, $w$ contains a structure of the form $\ochain{c_i}{x_0 c_j \dots}{c_k}$
where $|x_0| \geq 1$, $\ochain{c_i}{x_0}{c_j}$, and $k > j$ is such that $\chain(i, k)$.
This contradicts the hypothesis that $\chain(h, j)$ and $h < i$,
because such a chain would cross $\chain(i,k)$,
contradicting property~\eqref{item:chain-prop-1-proof}.

Similarly, if $\chain(h,j)$, $\chain(i,j)$, $h \doteq j$, and $h < i$,
then $w$ contains a structure $\ochain{c_h}{\dots c_i x_i}{c_j}$,
with $|x_i| \geq 1$ and $\ochain{c_i}{x_i}{c_j}$.
By the definition of chain, we have $i \gtrdot j$, which contradicts the hypothesis
that either $i \lessdot j$ or $i \doteq j$.
This proves that $i$ is unique.

For the second part of the property, suppose there exists a position $i'$
s.t.\ $\chain(i',j)$ and $i' \gtrdot j$, but $i' < i$ (the case $i' = i$ is trivial).
The only way of having both $\chain(i',j)$ and $\chain(i,j)$ in this case is
$\ochain{c_{i'}}{\dots c_i x_i}{c_j}$, with $|x_i| \geq 1$ and $\ochain{c_i}{x_i}{c_j}$.
From the definition of chain follows that $i \gtrdot j$,
which contradicts the hypothesis that $i \lessdot j$ or $i \doteq j$.

\item The proof is symmetric to the previous one.
\end{enumerate}
\end{proof}

In the rest of this section, we prove the following expansion laws of the until and since operators.
\begin{align}
  \lguntil{t}{\chi}{\varphi}{\psi} &\equiv
    \psi \lor \Big(\varphi \land \big(\lnext^t (\lguntil{t}{\chi}{\varphi}{\psi})
      \lor \lcnext{t} (\lguntil{t}{\chi}{\varphi}{\psi})\big)\Big)
  \label{eq:expansion-opsuntil} \\
  \lgsince{t}{\chi}{\varphi}{\psi} &\equiv
    \psi \lor \Big(\varphi \land \big(\lback^t (\lgsince{t}{\chi}{\varphi}{\psi})
      \lor \lcback{t} (\lgsince{t}{\chi}{\varphi}{\psi})\big)\Big)
  \label{eq:expansion-opssince} \\
  \lhuuntil{\varphi}{\psi} &\equiv
    (\psi \land \lcdback \top \land \neg \lcuback \top) \lor
     \big(\varphi \land \lhunext (\lhuuntil{\varphi}{\psi})\big) \\
  \lhusince{\varphi}{\psi} &\equiv
    (\psi \land \lcdback \top \land \neg \lcuback \top) \lor
     \big(\varphi \land \lhuback (\lhusince{\varphi}{\psi})\big) \\
  \lhduntil{\varphi}{\psi} &\equiv
    (\psi \land \lcunext \top \land \neg \lcdnext \top) \lor
     \big(\varphi \land \lhdnext (\lhduntil{\varphi}{\psi})\big) \\
  \lhdsince{\varphi}{\psi} &\equiv
    (\psi \land \lcunext \top \land \neg \lcdnext \top) \lor
     \big(\varphi \land \lhdback (\lhdsince{\varphi}{\psi})\big)
\end{align}

\begin{lem}
\label{lemma:expansion-law-su}
  Given a word $w$ on an OP alphabet $(\powset{AP}, M_{AP})$,
  two POTL formulas $\varphi$ and $\psi$,
  and a non-empty set $\Pi \subseteq \{\lessdot, \doteq, \gtrdot\}$,
  for any position $i \in w$ the following equivalence holds:
  \[
    \luntil{\Pi}{\varphi}{\psi} \equiv
    \psi \lor \Big(\varphi \land \big(\lgnext{\Pi} (\luntil{\Pi}{\varphi}{\psi})
      \lor \lganext{\Pi} (\luntil{\Pi}{\varphi}{\psi})\big)\Big).
  \]
\end{lem}
\begin{proof}
  $[\Rightarrow]$
  Suppose $\luntil{\Pi}{\varphi}{\psi}$ holds in $i$.
  If $\psi$ holds in $i$, the equivalence is trivially verified.
  Otherwise, $\luntil{\Pi}{\varphi}{\psi}$ is verified by an OPSP
  $i = i_0 < i_1 < \dots < i_n = j$ with $n \geq 1$,
  s.t.\ $(w,i_p) \models \varphi$ for $0 \leq p < n$ and $(w,i_n) \models \psi$.
  Note that, by the definition of OPSP, any suffix of such a path is also an OPSP ending in $j$.
  Consider position $i_1$: $\varphi$ holds in it, and it can be either
  \begin{itemize}
  \item $i_1 = i+1$.
    Then there exists $\prf \in \Pi$ s.t.\ $i \pr (i+1)$,
    and path $i_1 < i_2 < \dots < i_n = j$ is the OPSP between $i_1$ and $j$,
    and $\varphi$ holds in all $i_p$ with $1 \leq p < n$, and $\psi$ in $j_n$.
    So, $\luntil{\Pi}{\varphi}{\psi}$ holds in $i_1$,
    and $\lgnext{\Pi} (\luntil{\Pi}{\varphi}{\psi})$ holds in $i$.
  \item $i_1 > i+1$.
    Then, $\chain(i,i_1)$, and there exists $\prf \in \Pi$ s.t.\ $i \pr i_1$.
    Since $i_1 < i_2 < \dots < i_n = j$ is the OPSP from $i_1$ to $j$,
    $\luntil{\Pi}{\varphi}{\psi}$ holds in $i_1$,
    and so does $\lganext{\Pi} (\luntil{\Pi}{\varphi}{\psi})$ in $i$.
  \end{itemize}

  $[\Leftarrow]$
  Suppose $\psi \lor \Big(\varphi \land \big(\lgnext{\Pi} (\luntil{\Pi}{\varphi}{\psi})
  \lor \lganext{\Pi} (\luntil{\Pi}{\varphi}{\psi})\big)\Big)$ holds in $i$.
  The case $(w,i) \models \psi$ is trivial.
  Suppose $\psi$ does not hold in $i$.
  Then $\varphi$ holds in $i$, and either:
  \begin{itemize}
  \item $\lgnext{\Pi} (\luntil{\Pi}{\varphi}{\psi})$ holds in $i$.
    Then, we have $i \pr (i+1)$, $\prf \in \Pi$,
    and there is an OPSP $i+1 = i_1 < i_2 < \dots < i_n = j$,
    with $\varphi$ holding in all $i_p$ with $1 \leq i_p < n$, and $\psi$ in $i_n$.
    If $i$ is not the left context of any chain, then $i = i_0 < i_1 < i_2 < \dots < i_n$
    is an OPSP satisfying $\luntil{\Pi}{\varphi}{\psi}$ in $i$.
    Otherwise, let $k = \min\{h \mid \chain(i,h)\}$.
    Since $i$ is the left context of a chain,
    $\mathord{\lessdot} \in \Pi$, or $\lgnext{\Pi} (\luntil{\Pi}{\varphi}{\psi})$
    would not be true in $i$.

    Suppose $k > j$. This is always the case if $\mathord{\gtrdot} \not\in \Pi$,
    because then there is no position $h \leq i$
    s.t.\ $\chain(h, i_p)$ for any $1 \leq p \leq n$.
    So, adding $i$ to the OPSP generates another OPSP,
    because there is no position $h$ s.t.\ $\chain(i,h)$ with $h \leq j$,
    and the successor of $i$ in the path can only be $i_1 = i+1$.

    Suppose $k \leq j$.
    Let $k' = \max\{h \mid h \leq j \land \chain(i,h) \land \bigvee_{\prf \in \Pi} i \pr k\}$.
    Since $i_1 > i$, and chains cannot cross each other,
    there exists a value $q$, $1 \leq q \leq n$, s.t.\ $i_q = k'$.
    The path $i = i_0 < i_q < \dots < i_n = j$ is an OPSP by definition,
    and $\varphi$ holds both in $i$ and $i_q$.
    So, this path makes $\luntil{\Pi}{\varphi}{\psi}$ true in $i$.

  \item $\lganext{\Pi} (\luntil{\Pi}{\varphi}{\psi})$ holds in $i$.
    Then, there exists a position $k$ s.t.\ $\chain(i,k)$ and $i \pr k$
    with $\prf \in \Pi$, and $\luntil{\Pi}{\varphi}{\psi}$ holds in $k$,
    because of an OPSP $k = i_1 < i_2 < \dots < i_n = j$.
    If $k = \max\{h \mid h \leq j \land \chain(i,h) \land \bigvee_{\prf \in \Pi} i \pr k\}$,
    then $i = i_0 < i_1 < i_2 < \dots < i_n$ is an OPSP by definition,
    and since $\varphi$ holds in $i$, $\luntil{\Pi}{\varphi}{\psi}$ is satisfied in it.
    Otherwise, let
    $k' = \max\{h \mid h \leq j \land \chain(i,h) \land \bigvee_{\prf \in \Pi} i \pr k\}$.
    Since $i_1 > i$ and chains cannot cross,
    there exists a value $q$, $1 < q \leq n$, s.t.\ $i_q = k'$.
    $i_q < i_{q+1} < \dots < i_n = j$ is an OPSP,
    so $\luntil{\Pi}{\varphi}{\psi}$ holds in $i_q$ as well.
    The path $i < i_q < \dots < i_n$ is an OPSP, and $\luntil{\Pi}{\varphi}{\psi}$ holds in $i$.
  \end{itemize}
\end{proof}

The proof for the OPS since operator is analogous.

\begin{lem}
\label{lemma:expansion-law-hu}
  Given a word $w$ on an OP alphabet $(\powset{AP}, M_{AP})$,
  and two POTL formulas $\varphi$ and $\psi$,
  for any position $i \in w$ the following equivalence holds:
  \[
    \lhyuntil{\varphi}{\psi} \equiv
    (\psi \land \lgaback{\lessdot} \top) \lor
     \big(\varphi \land \lhynext (\lhyuntil{\varphi}{\psi})\big).
  \]
\end{lem}
\begin{proof}
  $[\Rightarrow]$
  Suppose $\lhyuntil{\varphi}{\psi}$ holds in $i$.
  Then, there exists a path $i = i_0 < i_1 < \dots < i_n$, $n \geq 0$,
  and a position $h < i$ s.t.\ $\chain(h,i_p)$ and $h \lessdot i_p$ for each $0 \leq p \leq n$,
  $\varphi$ holds in all $i_q$ for $0 \leq q < n$, and $\psi$ holds in $i_n$.
  If $n = 0$, $\psi$ holds in $i = i_0$, and so does $\lgaback{\lessdot} \top$.
  Otherwise, the path $i_1 < \dots < i_n$ is also a YPHP,
  so $\lhyuntil{\varphi}{\psi}$ is true in $i_1$.
  Therefore, $\lhynext (\lhyuntil{\varphi}{\psi})$ holds in $i$, and so does $\varphi$.

  $[\Leftarrow]$
  If $\lgaback{\lessdot} \top$ holds in $i$,
  then there exists a position $h < i$ s.t.\ $\chain(h,i)$ and $h \lessdot i$.
  If $\psi$ also holds in $i$, then $\lhyuntil{\varphi}{\psi}$ is trivially satisfied
  in $i$ by the path made of only $i$ itself.
  Otherwise, if $\lhynext (\lhyuntil{\varphi}{\psi})$ holds in $i$,
  then there exist a position $h < i$ s.t.\ $\chain(h,i)$ and $h \lessdot i$,
  and a position $i_1$ which is the minimum one s.t.\ $i_1 > i$, $\chain(h,i_1)$ and $h \lessdot i_1$.
  In $i_1$, $\lhyuntil{\varphi}{\psi}$ holds,
  so it is the first position of a YPHP $i_1 < i_2 < \dots < i_n$.
  Since $\varphi$ also holds in $i$, the path $i = i_0 < i_1 < \dots < i_n$ is also a YPHP,
  satisfying $\lhyuntil{\varphi}{\psi}$ in $i$.
\end{proof}
The proofs for the other hierarchical operators are analogous.

\section{Omitted Proofs: Conditional XPath Translation}
\label{sec:xpath-proofs}

\subsection{Completeness of CXPath on OPM-compatible trees}

First, we give an argument for Theorem~\ref{thm:xuntil-fo-completeness},
by proving a more general result.
\begin{lem}
Let $\mathcal{M}$ be the set of algebraic structures with common signature $\sigma$,
let $\mathcal{L}$ be a logic formalism that is FO-complete on $\mathcal{M}$,
and let $\mathcal{N}$ be FO-definable subset of $\mathcal{M}$.
Then, $\mathcal{L}$ is also FO-complete on $\mathcal{N}$.
\end{lem}
\begin{proof}
Since $\mathcal{N}$ is FO-definable, there exists a FO formula $\varphi_\mathcal{N}$
such that, for any $M \in \mathcal{M}$, we have $M \models \varphi_\mathcal{N}$
iff $M \in \mathcal{N}$.
Thus, any FO formula $\psi$ on $\mathcal{N}$ is equivalent to $\psi \land \varphi_\mathcal{N}$
on $\mathcal{M}$.

Since $\mathcal{L}$ is FO-complete, there exists an $\mathcal{L}$-formula $\Phi$ such that,
for any $M \in \mathcal{M}$, $M \models \Phi$ iff $M \models \psi \land \varphi_\mathcal{N}$.
Therefore, since $\mathcal{N} \subseteq \mathcal{M}$,
we also have $N \models \Phi$ iff $N \models \psi \land \varphi_\mathcal{N}$
for any $N \in \mathcal{N}$.
By construction, $\varphi_\mathcal{N} \equiv \top$ on any $N \in \mathcal{N}$,
and thus $N \models \Phi$ iff $N \models \psi$.
\end{proof}

In our case, $\mathcal{M}$ is the set of all unranked ordered trees $\uotrees$,
while $\mathcal{N}$ is $\uotrees_{M_{AP}}$, for a given OPM $M_{AP}$.
$\mathcal{L}$ is the logic CXPath, proved to be FO-complete in \cite{Marx2005}.
We only need to show that the set $\uotrees_{M_{AP}}$ is FO-definable.

Note that the actual signature of $\uotrees$ differs from the one reported
in Section~\ref{sec:fo-completeness} in the fact that the transitive and reflexive closures
of the $\rchild$ and $\rsibl$ relations are used (denoted resp. $\rchild^*$ and $\rsibl^*$).
We also use $\rchild^+$ to denote the transitive closure of $\rchild$.
Moreover, the signature contains monadic predicates for propositional symbols,
instead of the labeling function $L$.
First, we define the following FO formula on $\uotrees$,
which is true iff a node $y$ is the \emph{right context candidate} of another node $x$:
\begin{align*}
  \ra(x,y) :=
  x \rsibl y
  \lor \big(&\neg \exists z (x \rsibl z) \\
            & \land \exists z (z \rchild^* x \land z \rsibl y \\
            & \qquad \land \forall y (z \rchild^+ y \land y \rchild^* x \implies \neg \exists x (y \rsibl x))))\big)
\end{align*}
We also express sets of atomic propositions and PR
as detailed at the beginning of Section~\ref{sec:fo-semantics},
and we define the following shortcuts:
\newcommand*{\llmost}{\operatorname{leftmost}}
\newcommand*{\lrmost}{\operatorname{rightmost}}
\begin{align*}
  \llmost(x) &:= \neg \exists y (y \rsibl x) \\
  \lrmost(x) &:= \neg \exists y (x \rsibl y)
\end{align*}
The following formula $\varphi_{\uotrees_{M_{AP}}}$ defines the set $\uotrees_{M_{AP}}$.
\begin{align*}
\varphi_{\uotrees_{M_{AP}}} :=
  \forall x \Big[
    &\Big(\neg \exists y (y \rchild x) \implies
    \big(\sigma_{\#}(x)
    \land \exists y (x \rchild y \land \lrmost(y) \land \sigma_{\#}(y) \\
      &\qquad \qquad \land (\llmost(y) \lor \exists z (z \rsibl y \land \llmost(z) \land \neg \#(z))))\big)\Big) \\
  &\land (\exists y (y \rchild x) \implies \neg \#(x)) \\
  &\land \big(
    \forall y (x \rchild y \land \lrmost(y) \implies x \lessdot y \lor x \doteq y) \\
    &\qquad \land \forall y (x \rchild y \land \neg \lrmost(y) \implies x \lessdot y) \\
    &\qquad \land \neg \exists y (x \rchild y \land x \doteq y) \implies
      \forall y (\ra(x,y) \implies x \gtrdot y)\big)\Big]
\end{align*}
The first two lines say that the root is labeled with $\#$ and it has at most two children,
the rightmost one labeled with $\#$.
The third line states that no other position is labeled with $\#$.
The remaining lines describe the PR among sets of labels of each node,
as described in Section~\ref{sec:fo-completeness}.

\subsection{POTL Translation of OPTL}
\label{sec:potl-translation}

As an alternative proof of Corollary~\ref{cor:optl-in-potl},
we provide a direct translation of OPTL into POTL.
We define function $\kappa$, which given an OPTL formula $\varphi$,
yields a POTL formula $\kappa(\varphi)$ such that, for any OP word $w$ and position $i$,
we have $(w,i) \models \varphi$ iff $(w,i) \models \kappa(\varphi)$.
$\kappa$ is defined as the identity for propositional operators.
In the following, we use the abbreviations $\varphi' := \kappa(\varphi)$
and $\psi' := \kappa(\psi)$.
All operators $\lgnext{\pr}, \lgback{\pr}, \lganext{\pr}, \lgaback{\pr}$,
with $\pi \in \{\lessdot, \doteq, \gtrdot\}$,
are defined as in Section~\ref{sec:fo-completeness}.
\begin{align*}
\kappa(\lnext \varphi) &:= \ldnext \varphi' \lor \lunext \varphi'
&
\kappa(\lback \varphi) &:= \ldback \varphi' \lor \luback \varphi' \\
\kappa(\lanext \varphi) &:= \lcunext \varphi'
&
\kappa(\laback \varphi) &:= \lcdback \varphi'
\end{align*}

The translation for LTL until and since is much more involved:
\[
\kappa(\lluntil{\varphi}{\psi}) :=
  \psi' \lor
  \lcuuntil
    {\big(\varphi' \land \alpha(\varphi')\big)}
    {\big(\psi' \lor
       \lcduntil
         {(\varphi' \land \beta(\varphi'))}
         {(\psi' \land \beta(\varphi'))}\big)}
\]
where
\begin{align*}
\alpha(\varphi') &:=
  \lcunext \top \implies
  \neg \big(\lgnext{\lessdot} (\lcduntil{\top}{\neg \varphi'})
        \lor \lganext{\lessdot} (\lcduntil{\top}{\neg \varphi'})\big) \\
\beta(\varphi') &:=
  \lcdback \top \implies
    \neg \big(\lgback{\gtrdot} (\lcusince{\top}{\neg \varphi'})
          \lor \lgaback{\gtrdot} (\lcusince{\top}{\neg \varphi'})\big)
\end{align*}
The main formula is the concatenation of a US until and a DS until.
Whenever a USP contains the left context of a chain,
either the path ends there or it continues with the right context of that chain.
Instead, whenever a DSP contains a right chain context, it must contain the left context too.

When evaluated in the left context of a chain, subformula $\alpha(\varphi)$
makes sure $\varphi'$ holds in all positions of the body of the outermost
chain with that left context (i.e.\ the one whose right context
is the rightmost one).
Therefore, including it in the left side of the US until makes sure
$\varphi'$ holds in all chain bodies skipped by its paths.

Symmetrically, when evaluated in the right context of a chain,
$\beta(\varphi)$ makes sure $\varphi'$ holds in all positions
in the body of the outermost chain with that right context
(i.e., whose left context is the leftmost).
It is included in both sides of the DS until, so that $\varphi'$ holds
in all chain bodies skipped by its paths.

The translation for the since operator is symmetric.

The translations of the summary operators changes depending on the allowed PR.
The main difference between the semantics of summary until in OPTL and POTL
is that, in the former, PR are checked only on consecutive positions,
and the path can follow all ``maximal'' chains,
whose contexts are in the $\doteq$ or $\gtrdot$ relations.
In POTL, the allowed PR must holds between all positions consecutive in the path,
including contexts of the same chain, and also ``non-maximal chains'' are considered.
Since maximal chains have their contexts in the $\doteq$ or $\gtrdot$ relations, we have
\[
\kappa(\luntil{\doteq \gtrdot}{\varphi}{\psi}) :=
  \lcuuntil{\varphi'}{\psi'}.
\]
When only one of such relation is allowed, we must prevent the path from spanning
consecutive positions in the wrong relation.
\begin{align*}
\kappa(\luntil{\doteq}{\varphi}{\psi}) &:=
  \lcuuntil{(\varphi' \land \neg \lgnext{\gtrdot} \top)}{\psi'} \\
\kappa(\luntil{\gtrdot}{\varphi}{\psi}) &:=
  \lcuuntil{(\varphi' \land \neg \lgnext{\doteq} \top)}{\psi'}
\end{align*}

Things become more complicated when the $\lessdot$ relation is also allowed.
\[
\kappa(\luntil{\lessdot \doteq \gtrdot}{\varphi}{\psi}) :=
  \lcuuntil
         {\varphi'}
         {[\psi' \lor \lcduntil
                             {(\varphi' \land \gamma(\varphi'))}
                             {(\psi \land \gamma(\varphi'))}]}
\]
\begin{align*}
\gamma(\varphi') := \lgaback{\lessdot} \top \implies
  [&\lhuback \lhupglob (\varphi' \land (\lgnext{\doteq} \top \lor \lganext{\doteq} \top \implies \lgnext{\doteq} \delta(\varphi') \lor \lganext{\doteq} \delta(\varphi'))) \\
   &\land \lgaback{\lessdot} \lgnext{\lessdot} \delta(\varphi')]
\end{align*}
where $\lhupglob \theta := \neg [\lhusince{\top}{(\neg \theta)}]$, and
\(
\delta(\varphi') :=
   \neg [\lcuuntil
                {(\neg \lgaback{\lessdot} \top)}
                {(\neg \lgaback{\lessdot} \top \land \neg \varphi')}]
\).

In this case, we must make up for the fact that DSP in POTL can skip bodies of chains
whose contexts are in the $\lessdot$ relation, while OPSP in OPTL cannot.
In such cases, OPSP in OPTL continue by following the successor edge.
So, we split an OPSP in a path that goes only upwards in the ST
followed by one that can go downwards.
In the latter, $\gamma(\varphi')$ must also hold.
Let $i$ be a position in an OPSP, and let $j_p$, $1 \leq p \leq n$,
be all $n$ positions such that $\chain(i,j_p)$ and $i \lessdot j_p$.
Suppose one of such positions $j_q$, $1 \leq q \leq n$, is also part of the DSP.
Formula $\gamma(\varphi')$, if evaluated in $i_q$,
enforces $\varphi'$ to hold in all positions that the OPSP would span between $i$ and $j_q$.
$\lhuback \lhupglob (\varphi' \land (\lgnext{\doteq} \top \lor \lganext{\doteq} \top \implies \lgnext{\doteq} \delta(\varphi') \lor \lganext{\doteq} \delta(\varphi')))$
enforces $\varphi'$ to hold in all simple chain body starting with $j_k$, $1 \leq k < q$,
(note that $\delta(\varphi')$ only considers the body of the underlying simple chain,
without entering inner chains).
The left side $\lgnext{\doteq} \top \lor \lganext{\doteq} \top$ of the implication
makes sure this formula is required to hold only when such simple chain body continues after $j_k$.
$\lgaback{\lessdot} \lgnext{\lessdot} \delta(\varphi')$
enforces $\varphi'$ in the simple chain body starting in $i+1$.

For other PR combinations containing $\lessdot$,
it suffices to forbid consecutive positions in the wrong relation.
For $\kappa(\luntil{\lessdot \doteq}{\varphi}{\psi})$ and
$\kappa(\luntil{\lessdot \gtrdot}{\varphi}{\psi})$, just substitute $\varphi'$ with, respectively,
$\varphi' \land \neg \lgnext{\gtrdot} \top$ and
$\varphi' \land \neg \lgnext{\doteq} \top$ in
$\kappa(\luntil{\lessdot \doteq \gtrdot}{\varphi}{\psi})$.

The translations for hierarchical operators just need to take into account that,
in OPTL, they are evaluated in the opposite chain context.
\begin{align*}
\kappa(\luntil{\uparrow}{\varphi}{\psi}) &:=
  \lganext{\lessdot} [\neg \lhuback \top \land \lhuuntil{\varphi'}{\psi'}] \\
\kappa(\lsince{\downarrow}{\varphi}{\psi}) &:=
  \lganext{\lessdot} [\neg \lhunext \top \land \lhusince{\varphi'}{\psi'}]
\end{align*}
The translations for the take precedence hierarchical until and since are symmetric.

\end{document}